\documentclass[aps,pre,amsmath,amsfonts,amssymb,superscriptaddress,twocolumn]{revtex4-1}

\usepackage{amsmath}
\usepackage{amsfonts}
\usepackage{enumitem}
\usepackage{yfonts}
\usepackage{mathrsfs}

\usepackage{amssymb}
\usepackage[english]{babel}
\usepackage[pdftex]{graphicx}
\usepackage{tikz}
\usepackage{rotating}
\usepackage[pdftex,hypertexnames=true,linktocpage=true,breaklinks=true]{hyperref} 
\usepackage[hyphenbreaks]{breakurl}
\hypersetup{hidelinks,bookmarksnumbered=false,pdfview=FitB}

\hypersetup{colorlinks=true,linkcolor=blue,anchorcolor=blue,citecolor=magenta,filecolor=blue,urlcolor=blue,bookmarksnumbered=false,pdfview=FitB}

\usepackage[T1]{fontenc}
\usepackage[utf8]{inputenc}

\usepackage{esvect}
\usepackage{xcolor}

\newcommand{\affA}{Aix Marseille University, CNRS, CPT, Marseille, France}
\newcommand{\affB}{CNRS Centre de Physique Th\'eorique UMR7332,
13288 Marseille, France}
 \newcommand{\affG}{Quantum Biology Lab, Howard University, 2400 6th St NW, Washington, DC 20059, USA}

\begin{document}

\title{A Dynamical Spacetime Hypothesis for Quantum Entanglement with a Proposed Experimental Test}

\author{Marco Pettini}
\email{marco.pettini@cpt.univ-mrs.fr}
\email{marco.pettini@howard.edu}
\affiliation{\affA}\affiliation{\affB}\affiliation{\affG}

\date{\today}

\begin{abstract}
Quantum entanglement is a central resource of quantum information science, yet its correlations are ordinarily treated as properties of the quantum state rather than as the outcome of an underlying physical process. Here we explore the hypothesis that these correlations are mediated dynamically by a field propagating locally at finite velocity through an extension of ordinary spacetime,  extending the kinematic proposal of Physical Review Research 7, 013261 (2025) to a constrained bulk-field dynamics. Motivated by the constraints imposed by no-signaling on finite-speed hidden-influence models, we consider a warped five-dimensional spacetime with an additional timelike coordinate. We show that the five-dimensional Einstein equations determine the warp factor within a symmetric ansatz, requiring a positive bulk cosmological constant.
We show that null trajectories propagating through the additional timelike dimension can project onto arbitrarily large spatial distances at the same ordinary time, although these trajectories do not themselves connect separated points on the brane.
Spectral analysis of a minimally coupled bulk scalar reveals a tachyonic four-dimensional continuum. We isolate a normalizable zero-mode sector with stable, retarded effective dynamics in ordinary brane time and obtain an exact solution for a suitably distributed source.
Building on these results, we formulate an equal-time contextual coupling and Bohm–Bub-inspired outcome dynamics as additional hypotheses, subject to Born-rule compatibility and multipartite no-signaling. We propose an experiment involving two independently prepared Bell pairs to test for departures from their expected statistical independence. Under specified assumptions on the field covariance and detector response, we derive a conditional inverse-square decay of correlations between the two pairs' joint measurement outcomes at large separations. Whether such correlations are nonzero, and whether the complete dynamics reproduces Bell-inequality violations while preserving no-signaling and stability, remain open questions.
\medskip

\end{abstract}

\maketitle

\section{Introduction}

Quantum entanglement is mathematically expressed through the non-factorizability of the wavefunction of a composite system. It implies the existence of correlated behaviors between entangled entities even when these are separated by an arbitrarily large distance and in the absence of any known physical connection.  Entanglement is experimentally well established~\cite{aspect1,aspect2,aspect3,nl1,nl2,nlz,nl4}. {From a conceptual viewpoint, however, it remains puzzling because the formalism predicts correlations between spacelike-separated events independently of their distance. If these correlations are attributed to an underlying physical mediator, that mediator would have to produce effectively infinite propagation velocity in ordinary spacetime if it is to remain compatible with operational no-signaling under the corresponding hidden-influence assumptions. This is at odds with conventional physical descriptions of interactions, in which causal influences are mediated by physical degrees of freedom constrained by relativistic locality.}

The uneasiness with  "spooky action at a distance,"   as Einstein called quantum entanglement, had already been expressed, long before, by Isaac Newton with respect to gravity, who argued that matter could not act on matter "without the mediation of something else... through which their action and force may be transmitted" ~\cite{newton}.

 {Newton's remarks highlight a persistent physical intuition: that correlations arising from interactions should admit a mediating physical description.} In this perspective, quantum nonlocality is conceptually
bewildering precisely because it eludes this pattern:
quantum mechanics provides correlation statistics with
striking precision while leaving the causal structure
opaque.
Several attempts have been made to recover a
mediator-based description by hypothesizing superluminal
influences propagating at finite velocity 
$v>c$~\cite{eberhard,scarani,salart,cocciaro,santamaria}.
Experimental bounds on such a signal velocity have been
progressively improved,  reaching $\sim 10^4 c$~\cite{salart}, 
$\sim 5\times 10^6 c$~\cite{cocciaro}, and
$\sim 3.3\times 10^4 c$~\cite{santamaria} in the CMB
preferred frame,  however the programme faces an
obstruction of a different order altogether. These bounds
concern detectability, not principle: quantum correlations
cannot in any case be used to transmit controllable
messages faster than light~\cite{bohm}. The deeper
difficulty was identified by Bancal, Gisin \emph{et al.}~\cite{bancal},
 {who showed that, within the class of finite-velocity hidden-influence models they considered, such mediation leads to operational superluminal signaling in suitable multipartite configurations. Thus, under those assumptions, finite-velocity hidden mediation is incompatible with operational no-signaling.} Their conclusion is: 
\textit{``[\ldots] to keep no-signalling, [\ldots] quantum
nonlocality must necessarily relate discontinuously parts of
the universe that are arbitrarily distant. This gives further
weight to the idea that quantum correlations somehow arise
from outside spacetime, in the sense that no story in space
and time can describe how they occur.''}~\cite{bancal}.
Within those assumptions, this conclusion suggests that ordinary $(3,1)$-dimensional spacetime is insufficient to encode the relevant causal structure. A less problematic interpretation of the statement ``\textit{arise from outside [ordinary] spacetime}'' is therefore to ask whether the relevant dynamics might instead arise from inside an enlarged spacetime. In the construction explored here, the bulk metric admits equal-brane-time null directions, providing a concrete geometrical basis for investigating this possibility.

Hints in this direction come from the long-standing idea of representing Newtonian dynamics as geodesic motion in an extended spacetime, for example resorting to the Eisenhart lift, where the Hamiltonian flow of a mechanical system is encoded in the geodesics of a higher-dimensional Riemannian manifold \cite{CCP,RFMP,pettini-book}. In this geometric formulation, additional coordinates encode effective forces or correlations of the lower-dimensional theory.

More directly relevant to the present proposal are earlier theories
with two timelike coordinates. In Bars's two-time physics, developed
for particle dynamics and subsequently extended to field theory, the
additional timelike direction is governed by a local
$\mathrm{Sp}(2,\mathbb{R})$ gauge symmetry, so that gauge fixing
yields observable dynamics with a single effective
time~\cite{bars4}. By contrast, Wesson investigated noncompactified
five-dimensional relativity with a second time incorporated into
the spacetime geometry. Its signature $(+,-,-,-,+)$ is equivalent,
up to an overall sign, to the convention used
here~\cite{Wesson2002}. Chaichian and Kobakhidze instead formulated a five-dimensional
Randall--Sundrum geometry with a timelike fifth coordinate to
investigate the mass hierarchy and the recovery of effective
four-dimensional gravity on the $(3,1)$-dimensional spacetime 
\cite{ChaichianKobakhidze}. More recently, Strubbe proposed a
formally $4D+\tau$ classical framework in which EPR-type correlations
arise through worldline dynamics evolving with an additional parameter
$\tau$~\cite{Strubbe2026}. In that construction, however, $\tau$ is an
external evolution parameter rather than a fifth spacetime coordinate:
spacetime remains four-dimensional and no five-dimensional metric is
introduced.
None of these earlier approaches, however, was concerned with using
an extra timelike dimension to formulate a dynamical mechanism for
quantum entanglement. Within this broader context,
Ref.~\cite{PRR} proposed extending ordinary $(3,1)$-dimensional
spacetime by an additional temporal coordinate $\tau$ specifically
for this purpose. In that $(3,2)$-dimensional framework, a field
$\mathscr{X}_a({\bf x},t,\tau)$ was hypothesized to propagate at
finite velocity along null geodesics of the enlarged spacetime and
to couple to the wavefunction-collapse dynamics at the endpoints
of entanglement-based experiments. The collapse dynamics was described by a modified version of the nonlinear, nonunitary evolution originally introduced by Bohm and Bub \cite{BB}, with the field $\mathscr{X}_a({\bf x},t,\tau)$ replacing the random hidden variables.

The introduction of a second time dimension raises familiar concerns. Multiple-time theories are commonly associated with
pathologies such as causality violations and ghostlike excitations~\cite{B1,bars4}. These concerns are well founded, but
not necessarily fatal. It has been shown that the initial-value problem for ultrahyperbolic equations, with data posed on a hypersurface of mixed space- and timelike signature, can be rendered well posed after the imposition of a suitable nonlocal constraint on the data~\cite{B4}. Weinstein further argued that deterministic and stable evolution can
exist in theories with more than one time dimension and emphasised that such theories may exhibit ``nonlocality without nonlocality'',
namely nonlocal correlations without nonlocal causation~\cite{weinstein}. That observation captures the central motivation of the present construction. 

 Compared with Ref.~\cite{PRR}, we determine the warp factor within the metric ansatz, derive the bulk-field spectrum and identify its tachyonic continuum; the ensuing restriction to a stable zero-mode sector is an additional assumption. We analyse a warped $(3,2)$-dimensional spacetime hosting the bulk field $\mathscr{X}_a(\mathbf{x},t,\tau)$ and show that, within the metric ansatz adopted here, the five-dimensional vacuum General Relativity equations fix the warp factor uniquely up to sign and an additive constant. This is not a uniqueness theorem in the unrestricted space of five-dimensional solutions, but a determination of the warp function within the adopted warped-product ansatz. We then show that the geometrical framework preserves the standard causal structure on the brane and that the restricted zero-mode sector has a causally retarded response. No-signaling of the hypothesized contextual collapse law is imposed as an independent consistency requirement rather than derived from retardedness.

\paragraph*{Scope of the investigation.}
The framework combines results derived from the five-dimensional
geometry and field equations with additional hypotheses concerning
their application to quantum entanglement. The effective stability
restriction, contextual measurement coupling, and operational
consistency conditions are introduced where they become physically
relevant. The proposed experimental signature is conditional on
these ingredients. 
%Section~\ref{sec:assumptions} distinguishes
%derived results, explicit assumptions, and open questions.

This article is organised as follows.
Section~\ref{four-plus-one} proves the absence of a shortcut
for the \emph{symmetric Randall--Sundrum} extra-spatial metric;
this is not a theorem about all $(4,1)$ geometries.
Section~\ref{extra-time} introduces the warped $(3,2)$
geometry and derives the GR-selected warp factor.
In Sec.~\ref{sec:nullchars} we analyse the null
characteristics of the bulk wave equation and
establish the existence of a null family with
unbounded equal-time brane reach.
Section~\ref{sec:bulkfield} defines the brane-to-brane
response through a $t$-retarded Green function,
together with admissibility conditions in the
extra-time direction.

Section~\ref{sec:Xfield-propagation} develops the
dynamical model in four stages: the single-system
Bohm--Bub collapse model (Sec.~\ref{nutshell}); its
extension to a bipartite entangled system, introducing
the crossed collapse ratios and the channel-normalised
field (Sec.~\ref{sec:bipartite-BB}); the geometrical motivation for a postulated
two-component contextual input, not dynamically derived
from the $E=0$ geodesic kinematics (Sec.~\ref{subsec:meas-exchange}); and a
{ retrospective translation table to the earlier toy model}
(Sec.~\ref{subsec:PRR-continuity}). Contextual
microstates and Born-rule recovery are treated in
Sec.~\ref{subsec:contextuality-lambda}.

Section~\ref{sec:dynmodel} specialises the framework
to a photon pair and develops the two-pair
cross-correlation prediction.
Section~\ref{sec:experimental-signatures} discusses
experimental signatures, including a cross-pair
Bell-test proposal and a parametric estimate of the
signal magnitude. Section~\ref{discussion} gathers the
concluding remarks and gives an explicit account of
what is proved, what is assumed, and what remains open.

\section{Can a (4,1)-dimensional spacetime host a dynamical model of quantum entanglement?}
\label{four-plus-one}
The existence of extra dimensions, beyond the $3+1$ with which we perceive the physical world, has entered theoretical physics already one century ago with the formulation of the five-dimensional Kaluza-Klein theory (KKT), a classical field theory unifying gravitation and electromagnetism.
With a few exotic exceptions \cite{rubakov,visser}, the additional spatial dimension in the KKT is compactified under the so-called cylinder condition, so that the extra dimension remains hidden at macroscopic scales. The Kaluza–Klein theory is considered a precursor of string theory, 
where resorting to extra space dimensions is deemed natural and necessary for mathematical consistency~\cite{rizzo}. Also in these modern theories, the extra dimensions are typically curled up and microscopic, but their conceptual role is central.

Therefore, the idea of introducing an extra spatial dimension is not heretical, and under such an assumption a nonlocal behavior in $(3,1)$ dimensional spacetime can be seen as the projection of fully local phenomena in a higher-dimensional space. Actually, it has been surmised that entanglement and collapse phenomena could involve additional spatial dimensions, even if ordinary (Standard Model) quantum fields remain confined to $(3,1)$ dimensions. The author of Ref. \cite{genovese} affirms  \textit{one may surmise that “the usual fields only ‘live’ in $(3,1)$ dimensions, [whereas] the collapse involves also other dimensions, eventually being induced by ‘some field’ propagating also in these extra-dimensions.’’ In an extended geometry featuring an additional spatial direction entanglement correlations that appear instantaneous or acausal in $(3,1)$ spacetime could instead arise from ordinary, finite-velocity propagation in the higher-dimensional manifold}. 

Independent support for extra dimensions as a natural consequence of quantum mechanics comes from Brody and Graefe~\cite{BrodyGraefe}, who showed that quaternionic quantum mechanics for a spin-$\frac{1}{2}$ particle requires the ambient physical space to have five dimensions, embedding standard three-dimensional dynamics as a canonical reduction. More recently, Furquan, Singh and Wesley~\cite{FurquanSingh} have independently proposed that a $(3,3)$-dimensional spacetime, motivated by a gravi-weak unification programme based on $E_8\otimes E_8$ octonion algebra, can resolve the EPR paradox by reinterpreting apparently spacelike-separated measurement events as timelike-separated events in the extended geometry, their framework also predicts a potential violation of the Tsirelson bound, a further observable distinction between extra-time and standard $(3,1)$ frameworks.

\subsection{Randall--Sundrum geometry in $(4,1)$ dimensions and the
absence of causal shortcuts}
\label{subsec:RS-no-shortcuts}

The central question of this subsection is whether a warped extra
\emph{spatial} dimension can provide a causal mechanism for quantum
correlations between spacelike-separated brane events. The answer is
no: we prove that the Randall--Sundrum warp factor, far from shortening
brane-to-brane travel times, strictly increases them relative to the
brane-confined null geodesic. 

In a flat $(4,1)$-dimensional spacetime with
\begin{equation}
    ds^{2}
    =  g_{\mu\nu} dx^{\mu} dx^{\nu} + d\zeta^{2},
    \label{flatmetric}
\end{equation}
where $g_{\mu\nu}=\mathrm{diag}(-c^{2}, +1, +1, +1)$,
$x^{\mu}=(t,\mathbf{x})$, and $\zeta$ is the extra spatial coordinate,
the brane-to-brane causal structure is identical to that of Minkowski
space and no shortcut is possible. A nontrivial warp factor is therefore
necessary, and we consider the Randall--Sundrum (RS) warped
geometry~\cite{randall1,randall2}
\begin{equation}
    ds^{2}
    = e^{-2f(\zeta)} g_{\mu\nu} dx^{\mu} dx^{\nu} + d\zeta^{2},
    \label{eq:RS-metric-c}
\end{equation}
in which our $(3,1)$ world is a hypersurface (brane) at $\zeta=0$.
In the simplest RS configuration,
\begin{equation}
    f(\zeta)=k\vert\zeta\vert, \qquad k>0,
\end{equation}
with a $\mathbb{Z}_{2}$ symmetry across the brane. Note that
$f(\zeta)\ge 0$ for all $\zeta$, a property used crucially below.

Let $A$ and $B$ be two events on the brane,
\[
A=(t_{A},\mathbf{x}_{A},0), \qquad
B=(t_{B},\mathbf{x}_{B},0),
\]
with spatial separation $|\mathbf{x}_{B}-\mathbf{x}_{A}|$.
A causal influence between $A$ and $B$ must follow a future-directed
causal curve
\[
\gamma(\lambda) = \bigl(t(\lambda),\mathbf{x}(\lambda),\zeta(\lambda)\bigr),
\]
with $\gamma(\lambda_{1})=A$, $\gamma(\lambda_{2})=B$, and $ds^{2}\le 0$
along $\gamma$.

For a null curve $(ds^{2}=0)$, Eq.~\eqref{eq:RS-metric-c} gives
\begin{equation}
    0 = e^{-2f(\zeta)}\bigl(-c^{2}\dot{t}^{2}
    + \dot{\mathbf{x}}^{2}\bigr) + \dot{\zeta}^{2},
\end{equation}
where dots denote differentiation with respect to $\lambda$.
Rearranging,
\begin{equation}
    c^{2}\dot{t}^{2}
    = \dot{\mathbf{x}}^{2} + e^{2f(\zeta)}\dot{\zeta}^{2},
    \label{eq:null-rel-c}
\end{equation}
and for future-directed propagation ($\dot{t}>0$),
\begin{equation}
    \dot{t}
    = \frac{1}{c}\sqrt{\dot{\mathbf{x}}^{2}
    + e^{2f(\zeta)}\dot{\zeta}^{2}}.
\end{equation}
The coordinate travel time between the two brane events along $\gamma$
is therefore
\begin{equation}
    T[\gamma]
    = t_{B}-t_{A}
    = \frac{1}{c}\int_{\lambda_{1}}^{\lambda_{2}}
         \sqrt{\dot{\mathbf{x}}^{2} + e^{2f(\zeta)}\dot{\zeta}^{2}}
         \, d\lambda.
    \label{eq:T-functional-c}
\end{equation}

Since $f(\zeta)\ge 0$ we have $e^{2f(\zeta)}\ge 1$, so the integrand
satisfies
\begin{equation}
    \sqrt{\dot{\mathbf{x}}^{2} + e^{2f(\zeta)}\dot{\zeta}^{2}}
    \;\ge\;
    \sqrt{\dot{\mathbf{x}}^{2} + \dot{\zeta}^{2}}
    \;\ge\;
    |\dot{\mathbf{x}}|,
\end{equation}
which implies
\begin{equation}
    T[\gamma]
    \;\ge\;
    \frac{1}{c}\int_{\lambda_{1}}^{\lambda_{2}} |\dot{\mathbf{x}}|\,
    d\lambda.
    \label{eq:T-lower-1-c}
\end{equation}
The right-hand side is the arc length of the spatial projection
$\mathbf{x}(\lambda)$ in $\mathbb{R}^{3}$, which satisfies the triangle
inequality
\begin{equation}
    \int_{\lambda_{1}}^{\lambda_{2}} |\dot{\mathbf{x}}|\, d\lambda
    \;\ge\;
    \vert\mathbf{x}_{B}-\mathbf{x}_{A}\vert,
\end{equation}
with equality if and only if $\mathbf{x}(\lambda)$ is a straight line
with constant direction. Hence
\begin{equation}
    T[\gamma]
    \;\ge\;
    \frac{|\mathbf{x}_{B}-\mathbf{x}_{A}|}{c}.
    \label{eq:T-lower}
\end{equation}

\noindent\textit{Uniqueness of the saturating path.}
Equality in~\eqref{eq:T-lower} requires simultaneously
that $\dot{\zeta}=0$ almost everywhere, so the curve
stays on the brane, and that $\mathbf{x}(\lambda)$
is a straight segment traversed at constant velocity.
The first condition is necessary because whenever
$\dot\zeta\neq 0$ on a set of positive measure, the
integrand satisfies
$\sqrt{\dot{\mathbf{x}}^2+e^{2f}\dot\zeta^2}
>\sqrt{\dot{\mathbf{x}}^2+\dot\zeta^2}\ge|\dot{\mathbf{x}}|$
strictly (since $e^{2f}\ge 1>0$), so integrating gives
$T[\gamma]>|\mathbf{x}_B-\mathbf{x}_A|/c$.
The unique null path saturating~\eqref{eq:T-lower} is
therefore the brane-confined Minkowski null geodesic
\begin{equation}
\mathbf{x}(\lambda) = \mathbf{x}_{A} +
\frac{\lambda-\lambda_{1}}{\lambda_{2}-\lambda_{1}}
\bigl(\mathbf{x}_{B}-\mathbf{x}_{A}\bigr),
\label{eq:straight-line}
\end{equation}
whose affine parametrisation keeps $\dot{\mathbf{x}}$
at constant direction and magnitude, exactly saturating
the triangle inequality~\eqref{eq:T-lower}. The minimum
travel time is
\begin{equation}
    T_{\min} = \frac{|\mathbf{x}_{B}-\mathbf{x}_{A}|}{c}.
\end{equation}

One might expect a path cutting through the bulk to
save travel time by traversing a shorter coordinate
distance between $A$ and $B$. The warp factor says
otherwise. Any bulk excursion, any segment where
$\dot\zeta\neq 0$, contributes a term
$e^{2f(\zeta)}\dot\zeta^2$ to the integrand of
$T[\gamma]$~\eqref{eq:T-functional-c}, and since
$f(\zeta)=k|\zeta|\ge 0$ this contribution is strictly
positive and grows exponentially with the depth
$|\zeta|$ of the excursion. Dipping into the bulk
costs more time, not less. The brane-confined null
geodesic is the unique path that avoids this penalty
entirely; it is, for that reason, the sole minimiser
of~\eqref{eq:T-lower}.

This lower-bound argument, together with the explicit
mechanism above, proves that in symmetric
Randall--Sundrum warping every 5D null geodesic
projects onto a causal 4D path: the projected velocity
satisfies $|d\mathbf{x}/dt|\le c$ at all times. Two
distant brane points may well look geometrically closer
through the bulk embedding, but the causal structure of
the warped metric turns that apparent proximity into a
strictly greater time cost. The bulk is not a shortcut, 
it is a detour.

 {This result motivates passing, within the class of warped geometries considered here, to a warped
extra \emph{timelike} dimension.}
%%%%%%%%%%%%%%%%%%%%%%%%%%%%%%%%%%%%%%%%%%%%%%%%%%%%%%%%%%%%%%%%%
 \section{Extending Spacetime to (3,2) Dimensions}
 \label{extra-time}
Introducing an additional time dimension raises
well-known theoretical concerns: generic field theories
on a $(3,2)$ background can suffer from ill-posed Cauchy
problems, violations of causality, and ghost
instabilities~\cite{weinstein,B4}. We do not remove the unrestricted-field pathologies in the present work. We isolate a stable reduced mode and treat the additional mediator dynamics as a separate conjecture.
%%%%%%%%%%%%%%%%%%%%%%%%%%%%%%%%%%%%%%%%%%%%%%%%%%%%%%%%%%%%%%%%%%%%
\subsection{Asymmetric warping in the $(3,2)$ case}

An earlier formulation \cite{PRR} briefly remarked that a construction suggested by Visser \cite{visser} for exotic Kaluza--Klein models could be adapted to a five-dimensional spacetime of signature $(-,+,+,+,-)$, i.e.\ a $(3,2)$ geometry with a noncompact extra time dimension. In that formulation, a warped line element was considered in which only the physical time coordinate is multiplied by a warp factor depending on the second temporal coordinate $\tau$, while the spatial part of the induced $(3,1)$ metric remains unwarped. The discussion was intentionally limited to a qualitative indication of how such a geometry might trap ordinary fields on the four-dimensional spacetime while allowing the extra time direction to play a hidden dynamical role. The corresponding Einstein equations, the consistency of the causal structure, and the physical viability of the resulting warp factor were not examined. The Appendix of Ref.~\cite{PRR} concluded by noting that \textit{``how to choose such a metric and how to choose a suitable and physically meaningful function $f(\tau)$ remain far beyond the aim of the present work.''}

The present analysis addresses precisely this open point. The goal is to determine whether a purely time--warped $(3,2)$ metric can provide a consistent and physically meaningful geometric framework for a dynamical mechanism underlying quantum nonlocality.
%%%%%%%%%%%%%%%%%%%%%%%%%%%%%%%%%%%%%%%%%%%%
We therefore analyze the geometry
\begin{equation}
\label{eq:pure-timewarp}
ds^{2}
 = - e^{-2f(\tau)}\,c^{2}dt^{2}
   + d\mathbf{x}^{2}
   - w^{2}d\tau^{2},
\end{equation}
which represents a purely time-warped $(3,2)$ spacetime: the warp factor
modifies only the temporal component of the induced $(3,1)$ metric, while the spatial
directions remain unwarped. However, we show below that
it cannot provide a satisfactory geometric basis for a dynamical account of
entanglement correlations.

\subsubsection*{Curvature and Einstein equations}
For the metric~\eqref{eq:pure-timewarp}, the nonzero metric components and
their inverses are
\begin{eqnarray}
g_{tt}&=&-c^{2}e^{-2f}, \quad
g_{ij}=\delta_{ij},
g_{\tau\tau}=-w^{2},
\nonumber\\
g^{tt}&=&-\frac{e^{2f}}{c^{2}},\quad
g^{ij}=\delta^{ij},\quad
g^{\tau\tau}=-\frac{1}{w^{2}}. \nonumber
\end{eqnarray}
The only nontrivial $\tau$-dependence occurs in $g_{tt}$; using the conventions
of Sec.~\ref{symmwarp3-2} \cite{docarmo}, the only nonvanishing Christoffel
symbols are
\begin{equation}
\Gamma^{t}{}_{t\tau}=\Gamma^{t}{}_{\tau t}
=\frac{1}{2}g^{tt}\,\partial_{\tau}g_{tt}
=\frac{1}{2}\!\left(-\frac{e^{2f}}{c^{2}}\right)\!\left(2c^{2}f'e^{-2f}\right) \nonumber
=-\,f'(\tau),
\end{equation}
\begin{eqnarray}
\Gamma^{\tau}{}_{tt}
&=&-\frac{1}{2}g^{\tau\tau}\,\partial_{\tau}g_{tt}
=-\frac{1}{2}\!\left(-\frac{1}{w^{2}}\right)\!\left(2c^{2}f'e^{-2f}\right)\nonumber \\
&=&+\frac{c^{2}}{w^{2}}\,e^{-2f(\tau)}\,f'(\tau). \nonumber
\label{eq:Gamma_tau_tt}
\end{eqnarray}
The corresponding Ricci tensor components, computed directly from the
Christoffel symbols above, are
\begin{align}
R_{tt} &= \partial_{\tau}\Gamma^{\tau}{}_{tt}
     +\Gamma^{\mu}{}_{\mu\tau}\Gamma^{\tau}{}_{tt}
     -\Gamma^{\mu}{}_{t\tau}\Gamma^{\tau}{}_{\mu t}
  = {\frac{c^{2}e^{-2f}}{w^{2}}\bigl(f''-f'^{2}\bigr)}, \nonumber\\[0.5ex]
R_{ij} &= 0,\nonumber\\[0.5ex]
R_{\tau\tau} &= \bigl(f''-f'^{2}\bigr),\nonumber
\end{align}
and the scalar curvature $R=g^{AB}R_{AB}$ is
\begin{eqnarray}
R = g^{tt}R_{tt}+g^{\tau\tau}R_{\tau\tau}
&=& \left(-\frac{e^{2f}}{c^{2}}\right){\frac{c^{2}e^{-2f}}{w^{2}}(f''-f'^{2})}\nonumber \\
  &+&\left(-\frac{1}{w^{2}}\right)(f''-f'^{2})
{= \frac{2}{w^{2}}\bigl(f'^{2}-f''\bigr).}\nonumber 
\label{eq:R_timewarp}
\end{eqnarray}
The Einstein tensor $G_{AB}=R_{AB}-\tfrac{1}{2}g_{AB}R$ then has
components
\begin{eqnarray}
G_{tt} &=&
  {\frac{c^{2}e^{-2f}}{w^{2}}\bigl(f''-f'^{2}\bigr)
  -\frac{1}{2}(-c^{2}e^{-2f})\!\left(\frac{2}{w^{2}}(f'^{2}-f'')\right)} \nonumber\\
  &= &{0,}\nonumber\\[0.5ex]
G_{ij} &=&
  {-\frac{1}{2}\delta_{ij}\!\left(\frac{2}{w^{2}}(f'^{2}-f'')\right)
  = \frac{f''-f'^{2}}{w^{2}}\,\delta_{ij},}\nonumber\\[0.5ex]
G_{\tau\tau} &=&
  {(f''-f'^{2})
  -\frac{1}{2}(-w^{2})\!\left(\frac{2}{w^{2}}(f'^{2}-f'')\right)
  = 0.} \nonumber
\end{eqnarray}
Both brane-diagonal components $G_{tt}$ and $G_{\tau\tau}$ vanish
identically, independently of the form of $f(\tau)$; this is a special
feature of the pure time-warp ansatz~\eqref{eq:pure-timewarp}.
% and does not occur in the symmetrically warped case treated in Sec.~\ref{symmwarp3-2}.}

Imposing the vacuum Einstein equations with cosmological constant,
\begin{equation}
G_{AB}+\Lambda_{5}\,g_{AB}=0,
\end{equation}
the three independent components yield
\begin{eqnarray}
(tt):& &
 \quad {0
  -\Lambda_{5}\,c^{2}e^{-2f} = 0
  \;\Longrightarrow\;
  \Lambda_{5}=0,}
\label{eq:vac_tt}\\[0.5ex]
(ij):& &
  \quad   {\frac{f''-f'^{2}}{w^{2}}+\Lambda_{5} = 0
  \quad \Longrightarrow\;
  \Lambda_{5}=\frac{f'^{2}-f''}{w^{2}},}
\label{eq:vac_ij}\\[0.5ex]
(\tau\tau):& &
   \quad {0
  -\Lambda_{5}\,w^{2} = 0
  \;\Longrightarrow\;
  \Lambda_{5}=0.}
\label{eq:vac_tautau}
\end{eqnarray}
{The $(tt)$ and $(\tau\tau)$ equations both force $\Lambda_5=0$
identically, for any $f(\tau)$. Consistency with the $(ij)$ equation
then requires}
\begin{equation}
f''(\tau)=f'(\tau)^{2}.
\label{eq:vac_cond_1}
\end{equation}
{This first-order-reducible ODE for $f'$ has the general
nonconstant solution
\begin{equation}
f(\tau)=-\ln\!\left(1-\frac{\tau-\tau_1}{\tau_0}\right),
\qquad \tau<\tau_1+\tau_0,
\label{eq:pure-timewarp-sol}
\end{equation}
with integration constants $\tau_0\neq0$ and $\tau_1$, together with
the trivial constant solution $f'\equiv0$.  The warp
factor generated by~\eqref{eq:pure-timewarp-sol},
\begin{equation}
e^{-2f(\tau)}=\left(1-\frac{\tau-\tau_1}{\tau_0}\right)^{2},
\end{equation}
does not decay monotonically over an infinite range, and it vanishes at the finite coordinate distance $\tau=\tau_1+\tau_0$, where the
chosen coordinates become degenerate (the solution is locally flat,
not curvature-singular). This geometry therefore
provides no mechanism analogous to the Randall--Sundrum exponential
suppression that could localise four-dimensional physics near
$\tau=0$; the construction would have to be cut off, by hand, before
reaching the degeneracy, with no warp-factor decay to motivate
localisation on either side of the cutoff.}
The vacuum Einstein equations {thus admit a nonconstant solution
for the pure time-warp ansatz~\eqref{eq:pure-timewarp}, but only this
locally flat nonconstant family, and only at $\Lambda_5=0$.}
{The branch $f'\equiv0$,} is trivially equivalent (by rescaling $t$) to $f=0$, giving
ordinary five-dimensional Minkowski space.
Consequently, neither branch of pure time warping can generate an exponentially localising geometric
effect of the kind needed to confine Standard Model fields near the brane.
In this sense, pure time warping, even though not forbidden by General Relativity, is unsuited to provide the geometric basis for
the entanglement-correlation mechanism pursued here.

%%%%%%%%%%%%%%%%%%%%%%%%%%%%%%%%%%%%%%%%%%%%%%%%%%%%%%%%%%%%%%%%%%%%%
 \subsection{Modelling the extended spacetime geometry with symmetric warping}
 \label{symmwarp3-2}
 We now turn to a setting where the aim of the present work can be implemented consistently, namely a symmetrically warped $(3,2)$-dimensional spacetime. 
Specifically, we consider a metric which is  a minimal extension of the Randall--Sundrum form 
written as
\begin{equation}
    ds^{2}
    = e^{-2f(\tau)} \eta_{\mu\nu}\, dx^{\mu} dx^{\nu} - w^{2}d\tau^{2},
    \label{eq:32metric}
\end{equation}
where $\eta_{\mu\nu}=\mathrm{diag}(-c^{2},+1,+1,+1)$, $w$ has
dimensions of velocity, $f(\tau)$ is a warping function
depending only on the extra timelike coordinate $\tau$, and sets the scale of propagation in the extra--time
direction.
We take $\tau$ to be of noncompact support, $\tau\in(-\infty,+\infty)$. 
For clarity, we now provide computational details showing that the metric
\eqref{eq:32metric} admits a simple bulk solution of the five-dimensional Einstein
equations with a cosmological constant, and we determine the corresponding analytic
form of $f(\tau)$. We rewrite \eqref{eq:32metric} as
\begin{equation}
\label{eq:metric-def}
ds^{2}
 = g_{AB}\,dx^{A}dx^{B},
\end{equation}
with coordinates $x^{A}=(x^{\mu},\tau)$, $A=0,1,2,3,\tau$.
The nonvanishing components of the metric are
\begin{equation}
g_{\mu\nu} = e^{-2f(\tau)}\,\eta_{\mu\nu},
\qquad
g_{\tau\tau} = -w^{2},
\qquad
g_{\mu\tau} = g_{\tau\mu} = 0,
\end{equation}
and the inverse metric is
\begin{equation}
g^{\mu\nu} = e^{2f(\tau)}\,\eta^{\mu\nu},
\qquad
g^{\tau\tau} = -\frac{1}{w^{2}},
\qquad
g^{\mu\tau} = g^{\tau\mu} = 0.
\end{equation}
The only nonzero derivative of $g_{AB}$ is with respect to $\tau$,
\begin{equation}
\partial_{\tau} g_{\mu\nu}
 = -2 f'(\tau)\,e^{-2f(\tau)}\eta_{\mu\nu}
 = -2 f'(\tau)\,g_{\mu\nu},
\end{equation}
where a prime denotes $d/d\tau$.  All derivatives with respect to $x^{\mu}$
vanish:
\begin{equation}
\partial_{\mu} g_{AB} = 0,\qquad \forall\,A,B,\ \mu=0,1,2,3.
\end{equation}

The Christoffel coefficients are
\begin{equation}
\Gamma^{A}{}_{BC}
 = \frac{1}{2} g^{AD}
   \left(
     \partial_{B} g_{CD}
   + \partial_{C} g_{BD}
   - \partial_{D} g_{BC}
   \right).
\end{equation}
Using the metric above and the fact that $f=f(\tau)$, one finds that the only
nonvanishing components are
\begin{align}
\Gamma^{\rho}{}_{\mu\tau}
 &= \Gamma^{\rho}{}_{\tau\mu}
 = - f'(\tau)\,\delta^{\rho}{}_{\mu},
\label{eq:Gamma-rho-mutau}
\\[0.5ex]
\Gamma^{\tau}{}_{\mu\nu}
 &= -\frac{f'(\tau)}{w^{2}}\,g_{\mu\nu}
 = -\frac{f'(\tau)}{w^{2}}\,e^{-2f(\tau)}\eta_{\mu\nu}.
\label{eq:Gamma-tau-munu}
\end{align}
All other components $\Gamma^{A}{}_{BC}$ vanish.

The Ricci tensor is defined as
\begin{equation}
R_{AB}
 = \partial_{C}\Gamma^{C}{}_{AB}
   - \partial_{B}\Gamma^{C}{}_{AC}
   + \Gamma^{C}{}_{AB}\Gamma^{D}{}_{CD}
   - \Gamma^{C}{}_{AD}\Gamma^{D}{}_{BC}.
\end{equation}
Given that all dependence is on $\tau$ alone, and using
\eqref{eq:Gamma-rho-mutau}--\eqref{eq:Gamma-tau-munu}, the nonvanishing Ricci
components are:
\begin{equation}
R_{\mu\nu}
 = \frac{4 f'(\tau)^{2} - f''(\tau)}{w^{2}}\,g_{\mu\nu}
 = \frac{4 f'^{2} - f''}{w^{2}}\,e^{-2f(\tau)}\eta_{\mu\nu},
\label{eq:Rmunu}
\end{equation}
and
\begin{equation}
R_{\tau\tau}
 = 4 f''(\tau) - 4 f'(\tau)^{2}.
\label{eq:Rtautau}
\end{equation}
The mixed components vanish identically by the warped-product
structure of the metric:
\begin{equation}
R_{\mu\tau} = R_{\tau\mu} = 0,
\qquad
G_{\mu\tau} = G_{\tau\mu} = 0.
\end{equation}
This is an algebraic consequence of the metric ansatz~\eqref{eq:32metric},
not a consequence of the field equations: the Christoffel
symbols~\eqref{eq:Gamma-rho-mutau}--\eqref{eq:Gamma-tau-munu} contain
no mixed $(\mu,\tau)$ terms, so the corresponding Ricci and Einstein
components vanish before any field equation is imposed.

The scalar curvature is
\begin{equation}
R = g^{AB}R_{AB}
  = g^{\mu\nu}R_{\mu\nu} + g^{\tau\tau}R_{\tau\tau}.
\end{equation}
Using $g^{\mu\nu}g_{\mu\nu}=4$ and $g^{\tau\tau}=-1/w^{2}$, and substituting
\eqref{eq:Rmunu} and \eqref{eq:Rtautau}, we obtain
\begin{align}
g^{\mu\nu}R_{\mu\nu}
 & = \frac{16 f'^{2} - 4 f''}{w^{2}},
\\[0.5ex]
g^{\tau\tau}R_{\tau\tau}
 &= -\frac{1}{w^{2}}\,\bigl(4 f'' - 4 f'^{2}\bigr).
\end{align}
Therefore,
\begin{equation}
R = \frac{4}{w^{2}}\bigl(5 f'^{2} - 2 f''\bigr).
\label{eq:Rscalar}
\end{equation}

The Einstein tensor is
\begin{equation}
G_{AB} = R_{AB} - \frac{1}{2} g_{AB} R.
\end{equation}
Using \eqref{eq:Rmunu} and \eqref{eq:Rscalar}, the brane components are
\begin{align}
G_{\mu\nu}
 &= R_{\mu\nu} - \tfrac{1}{2} g_{\mu\nu} R
 = \frac{3}{w^{2}}\bigl(f''(\tau) - 2 f'(\tau)^{2}\bigr)\,g_{\mu\nu},
 \label{eq:Gmunu}
\end{align}
and using \eqref{eq:Rtautau}, \eqref{eq:Rscalar}, and $g_{\tau\tau}=-w^{2}$, the
extra-time component is
\begin{align}
G_{\tau\tau}
 &= R_{\tau\tau} - \tfrac{1}{2} g_{\tau\tau} R
 = 6 f'(\tau)^{2}.
 \label{eq:Gtautau}
\end{align}
The mixed components $G_{\mu\tau}=G_{\tau\mu}=0$ vanish identically, as established above for $R_{\mu\tau}$.

\subsection*{Einstein equations and the warp factor $f(\tau)$}
We now impose the validity of the  five-dimensional Einstein equations with a bulk cosmological
constant $\Lambda_{5}$ and no matter sources in the bulk (i.e.\ away
from any brane-localized stress tensor),
\begin{equation}
G_{AB} + \Lambda_{5}\,g_{AB} = 0.
\label{eq:EE-Lambda}
\end{equation}
From \eqref{eq:Gtautau} and $g_{\tau\tau}=-w^{2}$ we have
\begin{equation}
G_{\tau\tau} + \Lambda_{5} g_{\tau\tau}
 = 6 f'^{2} - \Lambda_{5} w^{2} = 0,
\end{equation}
so
\begin{equation}
f'(\tau)^{2} = \frac{\Lambda_{5} w^{2}}{6}.
\label{eq:fprime2}
\end{equation}
This requires $\Lambda_{5}>0$ and implies that $f'(\tau)$ is constant in the bulk:
\begin{equation}
f'(\tau) = k,
\qquad
k = \pm\, w\sqrt{\frac{\Lambda_{5}}{6}}.
\end{equation}

From \eqref{eq:Gmunu} and $g_{\mu\nu}=e^{-2f}\eta_{\mu\nu}$, the $(\mu\nu)$ component
of \eqref{eq:EE-Lambda} reads
\begin{equation}
\frac{3}{w^{2}}\bigl(f'' - 2 f'^{2}\bigr)\,g_{\mu\nu}
 + \Lambda_{5} g_{\mu\nu} = 0,
\end{equation}
which reduces to
\begin{equation}
f'' - 2 f'^{2} + \frac{\Lambda_{5} w^{2}}{3} = 0.
\label{eq:ODE-f}
\end{equation}
Using \eqref{eq:fprime2}, $\Lambda_{5} w^{2}/3 = 2 f'^{2}$, hence \eqref{eq:ODE-f}
gives
\begin{equation}
f''(\tau)=0,
\end{equation}
consistent with constant $f'$. Integrating in the bulk,
\begin{equation}
f(\tau) = k\,\tau + f_{0},
\qquad
k = \pm\,w\sqrt{\frac{\Lambda_{5}}{6}},
\end{equation}
with $f_{0}$ an integration constant that can be absorbed into a rescaling of the
four-dimensional coordinates. Thus, in the bulk the warped $(3,2)$ metric
\eqref{eq:metric-def} is supported by a positive cosmological constant
$\Lambda_{5}$. 

If a $\mathbb{Z}_{2}$-symmetric thin brane is inserted at $\tau=0$, the Israel
junction condition \cite{israel} fixes the sign of $k$ on either side of the brane, yielding the
standard kink profile
\begin{equation}
f(\tau)=k\,|\tau|,
\qquad
k=w\sqrt{\frac{\Lambda_{5}}{6}},
\label{eq:warp-final}
\end{equation}
so that the warp factor becomes $e^{-2f(\tau)}=e^{-2k|\tau|}$ ($f$ is then only piecewise smooth).

Since $\tau$ is a timelike coordinate with dimensions
of time, the warping constant $k$ has dimensions of
inverse time (s$^{-1}$). It plays the role of the
inverse curvature scale of the bulk, analogous to
$k_{\rm RS}\sim 1/L_{\rm AdS}$ in the
Randall--Sundrum model, with the replacement
$L_{\rm bulk}=w/k$ (which reduces to $c/k$ only when $w=c$).
\paragraph*{Remark.} Let us emphasize that the metric \eqref{eq:32metric} with warp factor \eqref{eq:warp-final} 
is an admissible ansatz for the geometric structure of a $(3,2)$-dimensional spacetime because it satisfies 
the equations of General Relativity.

\paragraph*{Remark on the sign of $\Lambda_5$ and the nature of the bulk.}
The warp function $f(\tau)=k|\tau|$ and the localisation mechanism
are formally identical to RS, but the positive sign of $\Lambda_5$
is a direct consequence of the timelike signature of the extra
dimension and distinguishes the present framework from the
Anti-de~Sitter bulk of the standard RS model.
%%%%%%%%%%%%%%%%%%%%%%%%%%%%%%%%%%%%%%%%%%%%%%%%%%%%%%%%%%%%%%%%%%%%

\subsection{Introducing a massless information-carrying field $\mathscr{X}_a(x,\tau)$}
\label{subsec:vectorX}
As shown in Sec.~\ref{sec:nullchars}, the symmetrically
warped metric~\eqref{eq:32metric} admits bulk null rays
with arbitrarily large spatial projections at fixed $t$.
This geometric feature motivates introducing a massless
bulk field
 $\mathscr{X}_a(x,\tau)$, where $x = (\textbf{x},t)$, with
internal index $a=1,\ldots,N$, propagating in the warped $(3,2)$ geometry~\eqref{eq:32metric}.
``Massless'' here means that no bulk mass term $M^2\mathscr{X}_a^2$
is included in the action. The characteristic structure is
therefore governed by the metric principal part $\Box_{(3,2)}$:
in the geometric--optics limit the characteristics lie on the bulk
null cone. 
Measurement events on the brane act as sources localized at $\tau=0$. 

\paragraph*{Remark.}
The label $a$ stands for internal degrees of freedom and is not
a spacetime index.  One may view $\mathscr{X}_a$ as an ensemble of
scalar fields obeying the same five-dimensional propagation equation.
The multiplicity $N$ has nothing to do with the geometric construction, it just counts
the number of independent collapse-driving channels.
A minimal choice is of course $N=1$ sufficient for a two-outcome
measurement. For a $d$-outcome measurement one may take $N=d$ so that each
component $\mathscr{X}_a$ drives a distinct outcome channel, without qualitative
change of the predicted outcome statistics.
Throughout this paper $\mathscr{X}_a$ is treated as a
classical bulk field that satisfies a classical wave equation,
and quantum correlations enter only through a suitably defined ensemble distribution
$\rho(\lambda)$ over classical field configurations
(Sec.~\ref{sec:bulkfield} and the subsection on response
vs.\ correlations therein). 
%%%%%%%%%%%%%%%%%%%%%%%%%%%%%%%%%%%%%%%%%%%%%%%%%%%%

\subsubsection{Action and field equation}
\label{subsubsec:X_action_matching}

The formal minimally coupled bulk action (used to diagnose the unrestricted spectrum) for the bulk field
$\mathscr{X}_a(x,\tau)$, sourced by a brane-localized current $J_a(x)$ at $\tau=0$, is
\begin{eqnarray}
S[\mathscr{X}]
&=&
-\frac{1}{2}\!\int d^{4}x\, d\tau\, \sqrt{|g|}\,
 g^{AB}\,\partial_{A}\mathscr{X}_a\,\partial_{B}\mathscr{X}_a
\nonumber\\
&&
-\;\kappa\!\int d^{4}x\, d\tau\, \sqrt{|g|}\,
J_a(x)\,\delta(\tau)\,\mathscr{X}_a(x,\tau),
\label{eq:action-vector}
\end{eqnarray}
where summation over the internal index $a$ is understood. 
Varying $S$ yields the sourced wave equation, where $\Box_{5}$ is the
Laplace--Beltrami operator \cite{docarmo}
\begin{equation}
\Box_{5}\mathscr{X}_a
\equiv
\frac{1}{\sqrt{|g|}}\,\partial_{A}\!\left(\sqrt{|g|}\,g^{AB}\,\partial_{B}\mathscr{X}_a\right)
=
+\,\kappa\,J_a(x)\,\delta(\tau).
\label{eq:EOM-vector}
\end{equation}
For the warped metric
$ds^{2}=e^{-2f(\tau)}\eta_{\mu\nu}dx^{\mu}dx^{\nu}-w^{2}d\tau^{2}$, Eq.~\eqref{eq:EOM-vector}
becomes
\begin{equation}
-\frac{1}{w^{2}}\partial_{\tau}\!\left(e^{-4f}\,\partial_{\tau}\mathscr{X}_a\right)
+e^{-2f}\,\Box_{4}\mathscr{X}_a
=
+\,\kappa\,e^{-4f}\,J_a(x)\,\delta(\tau),
\label{eq:EOM-vector-explicit}
\end{equation}
with $\Box_{4}=\eta^{\mu\nu}\partial_{\mu}\partial_{\nu}$.
The positive sign on the right-hand side of
\eqref{eq:EOM-vector-explicit} follows from the
$-\kappa J_a\mathscr X_a$ source term with the standard
negative kinetic sign in the action~\eqref{eq:action-vector}.
Integrating Eq.~\eqref{eq:EOM-vector-explicit} across
a narrow interval around $\tau=0$ gives the brane
matching (jump) condition
\begin{eqnarray}
\Bigl[\partial_{\tau}\mathscr{X}_a(x,\tau)
\Bigr]_{0^-}^{0^+}
&\equiv&
\partial_{\tau}\mathscr{X}_a(x,0^+)
-\partial_{\tau}\mathscr{X}_a(x,0^-)
\nonumber\\
&=&
-\,w^{2}\kappa\,J_a(x).
\label{eq:jump_condition_Xa}
\end{eqnarray}
Equation~\eqref{eq:jump_condition_Xa} makes precise
in what sense the brane source injects a bulk
disturbance: under standard regularity assumptions the
field $\mathscr{X}_a$ itself is continuous at $\tau=0$,
while its normal derivative has a discontinuity fixed
by the source strength.

\subsubsection{Separation along the brane and spectral
resolution in $\tau$}
\label{subsubsec:X_spectral_resolution}

Since the background depends only on $\tau$, the brane
directions $t$ and $\mathbf{x}$ are Killing directions
and a Fourier decomposition along the brane is natural,
\begin{equation}
\mathscr{X}(x,\tau)=
\int \frac{d\omega\,d^{3}\mathbf{k}}{(2\pi)^{4}}\,
 e^{-i\omega t+i\mathbf{k}\cdot\mathbf{x}}\,
\psi_{\omega,\mathbf{k}}(\tau),
\label{eq:fourier_decomp_Xa}
\end{equation}
which reduces the bulk wave equation (away from
$\tau=0$) to a second-order ODE in $\tau$ with
$(\omega,\mathbf{k})$ as parameters,
\begin{equation}
\psi''-4f'\,\psi'
+w^{2}e^{+2f(\tau)}\!\left(\mathbf{k}^{2}-
\frac{\omega^{2}}{c^{2}}\right)\psi=0,
\quad (\tau\neq 0).
\label{eq:tau_eq}
\end{equation}
The bulk wave equation is diagonal in the internal
index $a$, so the geometry does not mix different
components; Eq.~\eqref{eq:tau_eq} holds for each
component of $\mathscr{X}_a$ separately. Throughout
this section the index $a$ is suppressed: $\mathscr{X}$,
$\psi$, and $\phi$ denote any single component, and
the full field is recovered by restoring $a$.

A plane-wave ansatz $e^{iq\tau}$ does not diagonalize
the $\tau$-dependence, because the warp factor $f(\tau)$
breaks $\tau$-translation invariance and the
coefficients in~\eqref{eq:tau_eq} depend explicitly on
$\tau$. The standard approach is to recast the problem as an eigenvalue equation for the 
self-adjoint Sturm--Liouville operator in $\tau$
induced by the warp factor, whose normalizable eigenfunctions 
play the role of plane waves in the flat case.

Concretely, inserting the separated ansatz
$\mathscr{X}(x,\tau)=\phi(x)\,\psi(\tau)$ into
the bulk wave equation~\eqref{eq:EOM-vector-explicit}
and dividing through by $\phi(x)\psi(\tau)$ splits the
left-hand side into a function of $x$ alone and a
function of $\tau$ alone. The two sides must therefore
each equal the same constant, the separation
constant, written as $-\mu^2$ with the sign chosen so that $\mu^2\geq0$ for the positive $\tau$-operator below.
This yields two decoupled equations: the $\tau$-equation
\begin{equation}
-\,\frac{1}{w^{2}}\frac{d}{d\tau}\!\left(e^{-4f(\tau)}\,
\frac{d\psi}{d\tau}\right)
=\mu^{2}\,e^{-2f(\tau)}\,\psi,
\label{eq:tau_eigenvalue_again}
\end{equation}
and the four-dimensional equation
$(\Box_4+\mu^2)\phi=0$. In the mostly-plus signature the standard massive equation is $(\Box_4-m^2)\phi=0$, so a mode with $\mu^2>0$ has $m^2=-\mu^2<0$: it is tachyonic, not massive (see the spectral stability warning below).

Equation~\eqref{eq:tau_eigenvalue_again} is of
Sturm--Liouville type~\cite{courant},
\begin{eqnarray}
-\frac{d}{d\tau}\!\bigl(p(\tau)\psi'(\tau)\bigr)
&=&\lambda\,W(\tau)\psi(\tau),
\nonumber\\
p(\tau)&=&\frac{1}{w^2}e^{-4f(\tau)},
\quad \lambda=\mu^2,
\label{sturm}
\end{eqnarray}
and therefore fixes the warped weight as
\begin{equation}
W(\tau)=e^{-2f(\tau)}.
\label{eq:warped_weight}
\end{equation}
This weight is the function with respect to which the
$\tau$-operator is self-adjoint. Equivalently, it is
the measure that appears in the bulk kinetic term when
one inserts a separated ansatz into the action. Indeed,
starting from the bulk kinetic term
\begin{equation}
S_{\rm kin}=-\frac{1}{2}\int d^4x\,d\tau\,\sqrt{|g|}
\left[g^{\mu\nu}\,\partial_\mu\mathscr{X}\,
\partial_\nu\mathscr{X}
+g^{\tau\tau}(\partial_\tau\mathscr{X})^2\right],
\end{equation}
inserting the separated ansatz
$\mathscr{X}(x,\tau)=\phi(x)\psi(\tau)$
and using $\sqrt{|g|}=w e^{-4f(\tau)}$,
$g^{\mu\nu}=e^{2f(\tau)}\eta^{\mu\nu}$,
and $g^{\tau\tau}=-w^{-2}$, we obtain
\begin{align}
S_{\rm kin}
={}&-\frac{w}{2}
\left[\int d\tau\,e^{-2f(\tau)}\psi^2(\tau)\right]
\int d^4x\,\partial_\mu\phi\,\partial^\mu\phi
\nonumber\\
&+\frac{1}{2w}
\left[\int d\tau\,e^{-4f(\tau)}
(\partial_\tau\psi)^2\right]
\int d^4x\,\phi^2 .
\end{align}

The second term contributes to the effective
four-dimensional mass term, whereas the first
determines the normalization of the four-dimensional
kinetic term. Writing $W(\tau)=e^{-2f(\tau)}$,
we therefore have
\begin{equation}
\begin{aligned}
S_{\rm kin}^{(4)}
&=-\frac{w}{2}
\left[\int d\tau\,W(\tau)\psi^2(\tau)\right]
\int d^4x\,\partial_\mu\phi\,\partial^\mu\phi\\
&=-\frac{w}{2}
\int d^4x\,\partial_\mu\phi\,\partial^\mu\phi,
\end{aligned}
\end{equation}
where the last equality follows from the weighted
normalization
$\int d\tau\,W(\tau)\psi^2(\tau)=1$.

Thus, defining the canonically normalized
four-dimensional field by
$\phi_{\rm can}=\sqrt{w}\,\phi$,
we recover the standard kinetic term
\begin{equation}
S_{\rm kin}^{(4)}
=-\frac{1}{2}\int d^4x\,
\partial_\mu\phi_{\rm can}
\partial^\mu\phi_{\rm can}.
\end{equation}
which is therefore the weighted mode norm (the proper-measure
kinetic coefficient includes the constant factor $w$).

\paragraph*{Remark on dimensions.}
With the standard five-dimensional kinetic term, each
component $\mathscr{X}_a$ has dimension
$[\mathscr{X}_a]\sim L^{-3/2}$ in $\hbar=c=1$ units.
\smallskip

If $\psi_n$ and $\psi_m$ are eigenfunctions with
eigenvalues $\mu_n^2$ and $\mu_m^2$, they satisfy the
Sturm--Liouville equations
\begin{eqnarray}
-\frac{d}{d\tau}\!\bigl(p(\tau)\,\psi_n'(\tau)\bigr)
&=&\mu_n^{2}\,W(\tau)\,\psi_n(\tau),
\nonumber\\
-\frac{d}{d\tau}\!\bigl(p(\tau)\,\psi_m'(\tau)\bigr)
&=&\mu_m^{2}\,W(\tau)\,\psi_m(\tau).
\label{eq:SL_pair}
\end{eqnarray}
Multiplying the first equation in~\eqref{eq:SL_pair}
by $\psi_m$ and the second by $\psi_n$ gives
\begin{eqnarray}
-\psi_m\,(p\psi_n')'
&=&\mu_n^{2}\,W\,\psi_n\psi_m,
\nonumber\\
-\psi_n\,(p\psi_m')'
&=&\mu_m^{2}\,W\,\psi_n\psi_m.
\label{eq:SL_multiplied}
\end{eqnarray}
Subtracting the two relations
in~\eqref{eq:SL_multiplied} yields
\begin{equation}
-\psi_m\,(p\psi_n')' + \psi_n\,(p\psi_m')'
= (\mu_n^{2}-\mu_m^{2})\,W\,\psi_n\psi_m.
\label{eq:SL_subtracted}
\end{equation}
The left-hand side is a total derivative
\begin{equation}
-\psi_m\,(p\psi_n')' + \psi_n\,(p\psi_m')'
=\frac{d}{d\tau}\Bigl[p(\tau)\bigl(\psi_n\psi_m'
-\psi_m\psi_n'\bigr)\Bigr].
\label{eq:SL_total_derivative}
\end{equation}
Integrating~\eqref{eq:SL_subtracted} over
$\tau\in(-\infty,+\infty)$ therefore gives
\begin{eqnarray}
(\mu_n^2-\mu_m^2)&&\int_{-\infty}^{\infty}
d\tau\,W(\tau)\,\psi_n(\tau)\psi_m(\tau)
\nonumber\\
&=&
\Bigl[p(\tau)\bigl(\psi_n\psi_m'
-\psi_m\psi_n'\bigr)\Bigr]_{-\infty}^{+\infty}.
\label{eq:SL_identity}
\end{eqnarray}
Thus, if the boundary term on the right-hand side
vanishes, eigenfunctions with distinct eigenvalues are
orthogonal with respect to the weighted inner product
$\langle\psi,\varphi\rangle_W\equiv
\int d\tau\,W\,\psi\,\varphi$. This is the warped
counterpart of the familiar plane-wave orthogonality
in a translation-invariant direction.

\paragraph*{Spectral stability warning.}
For the transverse Sturm--Liouville problem,
self-adjointness guarantees that $\mu^2$ is real,
while positivity of the associated quadratic form gives
$\mu^2\geq0$. Because the extra dimension is timelike,
the projected equation nevertheless has the sign
$(\Box_4+\mu^2)\phi_\mu=\kappa\,\psi_\mu(0)\,J_a(x)$ and dispersion
$\omega^2=c^2(|\mathbf p|^2-\mu^2)$.
For every $\mu>0$, Fourier components with
$|\mathbf p|<\mu$ admit exponentially growing solutions
in ordinary time. Thus, weighted normalizability and
vanishing boundary terms do \emph{not} imply dynamical
stability. A generic brane-localized $\delta(\tau)$
current also excites the continuum.

\paragraph*{(P1) Effective stability restriction.}
We therefore retain the normalizable massless zero-mode
sector in subsequent calculations, with sources restricted
or projected consistently onto this sector. The resulting
effective field has stable brane-time evolution.

We consider $\mathbb{Z}_2$-even bulk configurations
under $\tau\mapsto-\tau$ and impose matching and
asymptotic boundary conditions that make the weighted
transverse operator
$\mathcal L_\tau=W^{-1}\mathcal D_\tau$ self-adjoint,
where $W=e^{-2f}$ and
$\mathcal D_\tau=-w^{-2}\partial_\tau
(e^{-4f}\partial_\tau)$.
These conditions ensure that the boundary terms in
Eq.~\eqref{eq:SL_identity} vanish.
The resulting spectral decomposition includes
normalizable bound modes and a continuum of
generalized eigenfunctions. With these domain conditions,
the spectral theorem provides a real, complete spectral
resolution~\cite{reed}. The nonnegativity of the
quadratic form follows explicitly by integration by parts:
\begin{equation}
\langle\psi,\mathcal L_\tau\psi\rangle_W
=\frac{1}{w^2}\int_{-\infty}^{\infty}d\tau\,
e^{-4f}|\partial_\tau\psi|^2\geq0.
\label{eq:tau_positive_form}
\end{equation}
Here the boundary term is assumed to vanish as part of
the selected self-adjoint domain.

On the noncompact domain $\tau\in\mathbb{R}$ the
spectrum need not be purely discrete: in general it
consists of any normalizable bound mode(s) (if
present) plus a continuum of generalized eigenmodes.
We therefore write the spectral expansion in the form
\begin{equation}
\mathscr{X}(x,\tau)
=
\sum_{n} \phi_{n}(x)\,\psi_n(\tau)
\;+\;
\int_{0}^{\infty} d\mu\;\phi_{\mu}(x)\,\psi_{\mu}(\tau),
\label{eq:X_spectral}
\end{equation}
where the bound components are normalized with respect
to the weighted inner product,
\begin{equation}
\int_{-\infty}^{\infty} d\tau\,W(\tau)\,
\psi_n(\tau)\psi_m(\tau)=\delta_{nm},
\end{equation}
while the continuum components satisfy the
corresponding delta-normalization,
\begin{equation}
\int_{-\infty}^{\infty} d\tau\,W(\tau)\,
\psi_\mu(\tau)\psi_{\mu'}(\tau)=\delta(\mu-\mu').
\end{equation}
These weighted normalizations are the analogue of
plane-wave normalization. Together with completeness
of the adopted self-adjoint spectral resolution, they
make the expansion~\eqref{eq:X_spectral} invertible and
allow its components to be projected independently
onto their four-dimensional equations.

\subsubsection{Effective 4D equations and mode
propagation}
\label{subsubsec:X_effective4D}

We now consider the brane-localized source term
in~\eqref{eq:EOM-vector-explicit} and project the
equation onto a single $\tau$-eigenfunction
$\psi_\mu(\tau)$. Inserting~\eqref{eq:X_spectral}
into~\eqref{eq:EOM-vector-explicit}, multiplying by
$\psi_\mu(\tau)$, and integrating over $\tau\in\mathbb{R}$
we get
\begin{eqnarray}
&&\int_{-\infty}^{\infty} d\tau\,\psi_\mu(\tau)\,
\Bigl[\mathcal{D}_\tau
+\;e^{-2f(\tau)}\Box_4\Bigr]\mathscr{X}(x,\tau)
\nonumber\\
&=&
\kappa\,J_a(x)\int_{-\infty}^{\infty} d\tau\,
\psi_\mu(\tau)\,e^{-4f(\tau)}\delta(\tau),
\label{eq:projection_start}
\end{eqnarray}
where $\mathcal D_\tau$ is the transverse differential
expression defined above and appearing in
Eq.~\eqref{eq:tau_eigenvalue_again}. The index $a$ reappears on the right-hand side because
the source $J_a(x)$ carries the internal structure of
the coupling: different components $J_a$ drive
different outcome channels of the collapse dynamics,
and this physical differentiation is not visible to the
bulk geometry. The left-hand side, by contrast, is
diagonal in $a$, the wave operator $\Box_5$ acts
identically on each component, so the suppression
of $a$ remains valid there.

More precisely, $\mathcal L_\tau=W^{-1}\mathcal D_\tau$
is self-adjoint in the weighted space with inner product
$\langle\psi,\varphi\rangle_W\equiv
\int d\tau\,W(\tau)\,\psi^*\varphi$,
for the chosen self-adjoint domain. For real
mode functions, integration by parts then gives,
for any admissible $\varphi$,

\begin{eqnarray}
\int_{-\infty}^{\infty} d\tau\,\psi_\mu\,
(\mathcal{D}_\tau\varphi)
&=&
\int_{-\infty}^{\infty} d\tau\,\varphi\,
(\mathcal{D}_\tau\psi_\mu)
\nonumber\\
&=&
\mu^2\int_{-\infty}^{\infty} d\tau\,W(\tau)\,
\varphi\,\psi_\mu,
\label{eq:Dtau_selfadj}
\end{eqnarray}
where the last step uses the eigenvalue equation
$\mathcal{D}_\tau\psi_\mu = \mu^2\,W(\tau)\,\psi_\mu$.
Applying~\eqref{eq:Dtau_selfadj} with
$\varphi = \mathscr{X}$ and using the spectral
expansion~\eqref{eq:X_spectral} together with the
weighted orthogonality
\begin{equation}
\int_{-\infty}^{\infty} d\tau\,W(\tau)
\psi_\mu\psi_{\mu'}
=\delta(\mu-\mu'),
\label{eq:SL_use}
\end{equation}
the $\mathcal{D}_\tau$ part of the left-hand side
of~\eqref{eq:projection_start} diagonalizes to
$\mu^2\phi_{\mu}(x)$, the $e^{-2f}\Box_4$ part
diagonalizes by the same weighted orthogonality, and since
$\Box_4$ acts only on $x$, it can be brought outside
the $\tau$-integral. The remaining $\tau$-integral is
\begin{eqnarray}
&&\int_{-\infty}^{\infty} d\tau\, \psi_\mu(\tau)\,
e^{-2f(\tau)}\, \psi_{\mu'}(\tau)\nonumber\\
&=& \int_{-\infty}^{\infty} d\tau\, W(\tau)\,
\psi_\mu(\tau)\, \psi_{\mu'}(\tau)
= \delta(\mu-\mu'),
\end{eqnarray}
which selects $\phi_{\mu}(x)$ and yields the
contribution $\Box_4\phi_{\mu}(x)$. The left-hand
side of~\eqref{eq:projection_start} therefore gives
$(\Box_4+\mu^2)\phi_{\mu}(x)$.

The source projection is supported at $\tau=0$ and
gives
\begin{eqnarray}
\int_{-\infty}^{\infty} d\tau\,\psi_\mu(\tau)\,
e^{-4f(\tau)}\delta(\tau)
&=&
\Bigl[\psi_\mu(\tau)\,e^{-4f(\tau)}\Bigr]_{\tau=0}
\nonumber\\
&=& \psi_\mu(0)\,e^{-4f(0)},
\label{eq:delta_projection}
\end{eqnarray}
so the right-hand side of~\eqref{eq:projection_start}
becomes $\kappa\,\psi_\mu(0)\,e^{-4f(0)}\,J_a(x)$.
The overall factor $e^{-4f(0)}$ (evaluated at the
brane position $\tau=0$) is a fixed constant that can
be absorbed into $\kappa$ by convention, yielding the
$\mu$-mode equation
\begin{equation}
\left(\Box_4+\mu^2\right)\phi_{\mu}(x)
=
\kappa\,\psi_\mu(0)\,J_a(x),
\label{eq:projected_4D_eq}
\end{equation}
so each spectral component couples to the brane
current only through its brane overlap $\psi_\mu(0)$.

The same result applies to discrete bound modes,
\begin{equation}
\left(\Box_{4}+\mu_{n}^{2}\right)\phi_{n}(x)
=\kappa\,\psi_{n}(0)\,J_a(x),
\label{eq:phi4D_continuum}
\end{equation}
so in both cases a brane-localized interaction excites
a given bulk mode only through its brane overlap
$\psi_\mu(0)$ (or $\psi_n(0)$). Modes with small
brane overlap couple weakly to the brane source and
therefore contribute little to brane observables.

For every $\mu>0$, the dispersion relation
$\omega^2=c^2(|\mathbf p|^2-\mu^2)$
admits exponentially growing solutions in ordinary
time whenever $|\mathbf p|<\mu$.
The positive-$\mu$ continuum is therefore
excluded from the restricted effective model
by assumption (P1), rather than being
dynamically suppressed.

By contrast, as $f(\tau)\to+\infty$ for
$|\tau|\to\infty$, the warped measure
$W(\tau)=e^{-2f(\tau)}$ can make the constant profile
$\psi_{0}=\mathrm{const.}$ normalizable, yielding a
$\tau$-bound massless mode with $\mu_{0}=0$. For the
warp factor $f(\tau)=k|\tau|$ derived in
Sec.~\ref{symmwarp3-2} this normalizability is
explicit: setting $\psi_0=c_0$ (constant),
\begin{equation}
\int_{-\infty}^{+\infty}d\tau\,W(\tau)\,|\psi_0|^2
=c_0^2\int_{-\infty}^{+\infty}e^{-2k|\tau|}\,d\tau
=\frac{c_0^2}{k}<\infty\quad ,
\label{eq:zero_mode_norm}
\end{equation}
$\forall\,k>0$, so the zero mode normalized with respect to $d\tau\,W(\tau)$
is $\psi_0=\sqrt{k}$; with the proper measure $w\,d\tau\,W$
it is $\psi_0=\sqrt{k/w}$.

\paragraph*{Meaning and limitations of the normalizable zero mode.}
The zero mode has a constant profile along the extra dimension,
$\psi_0(\tau)=\sqrt{k}$. Nevertheless, it is normalizable
because the normalization integral involves the warped weight
$W=e^{-2k|\tau|}$, giving
$\int W\psi_0^2\,d\tau=1$.
Restricting the field to this mode in the action before
variation yields a conventional massless four-dimensional
field. Its retarded response propagates along the ordinary
four-dimensional light cone: a constant profile in $\tau$
does not imply instantaneous communication between different
points on the brane.

A separate question is whether such a zero-mode configuration
can be generated by a physical source. The brane-localized
current $J_a(x)\delta(\tau)$ appearing in
Eq.~\eqref{eq:EOM-vector} generally excites the continuum
as well, so that a pure zero-mode configuration is not an
exact solution of the full sourced equation.
An exact zero-mode solution can instead be sourced by a
current distributed along $\tau$. In the weighted field
equation its profile is $\rho_Z(\tau)=ke^{-2f(\tau)}$,
giving $\mathscr X_a=\sqrt{k}\phi_a$ with
$\Box_4\phi_a=\kappa\sqrt{k}J_a$.
Equivalently, the covariant equation reads
$\Box_5\mathscr X_a=\kappa k e^{2f(\tau)}J_a(x)$;
$\rho_Z$ denotes the weighted forcing, not the
covariant source. This exact construction uses a
distributed rather than a brane-localized current.
Accordingly, the zero-mode calculations developed below
are understood as a restricted effective description
under assumption (P1).

\paragraph*{A preferred-foliation route beyond P1.}
The zero-mode restriction is not the only conceivable way to address
the tachyonic instability. An Einstein--aether-inspired preferred
direction, aligned with ordinary brane time, suggests a different
mediator kinetic operator~\cite{JacobsonMattingly2001,DonnellyJacobson2010}.
On the background~\eqref{eq:32metric}, consider the unit field and
orthogonal projector
\begin{equation}
\begin{aligned}
u^A&=\frac{e^{f(\tau)}}{c}\,\delta^A_t,
&u^Au_A&=-1,\\
h^{AB}&=g^{AB}+u^Au^B.
\end{aligned}
\label{eq:pf_projector}
\end{equation}
Here $\delta^A_t$ denotes the unit coordinate vector
along ordinary time $t$, while $u^A$ is normalized.
The tensor $h^{AB}$ projects
orthogonally to $u^A$, removing the ordinary-time
direction while retaining the spatial directions
and the additional timelike coordinate $\tau$.
At the level of a fixed-background investigation, a candidate
quadratic action for each mediator component is
\begin{equation}
S_{\rm pf}^{(2)}
=-\frac12\int d^5x\,\sqrt{|g|}\,
h^{AB}\partial_A\mathscr X_a\partial_B\mathscr X_a.
\label{eq:pf_action}
\end{equation}
Treating $u^A$ as a prescribed background field at this stage,
its Euler--Lagrange equation, away from brane sources, becomes
\begin{equation}
\begin{aligned}
\mathcal D_{\rm pf}\mathscr X_a
&\equiv \frac{1}{\sqrt{|g|}}\partial_A\!\left(
\sqrt{|g|}\,h^{AB}\partial_B\mathscr X_a\right)=0,\\
\Longleftrightarrow\quad
\bigl[\partial_\tau^2-4f'\partial_\tau
&-w^2e^{2f}\Delta_3\bigr]\mathscr X_a=0.
\end{aligned}
\label{eq:pf_operator}
\end{equation}
Its principal symbol contains no ordinary-time frequency.
More explicitly, spatial Fourier decomposition at fixed $t$
gives
\begin{equation}
\partial_\tau\!\left(
e^{-4f}\partial_\tau\widehat{\mathscr X}_a
\right)
+w^2e^{-2f}|\mathbf q|^2
\widehat{\mathscr X}_a=0.
\label{eq:pf_fourier}
\end{equation}
Thus, the spatial Fourier contribution enters the extra-time
evolution equation with a nonnegative coefficient, rather than
producing the brane-time tachyonic dispersion of the minimally
coupled theory.

The remaining operator is hyperbolic in $\tau$, with
characteristics
\begin{equation}
\begin{aligned}
p_\tau^2&=w^2e^{2f(\tau)}|\mathbf p|^2,\\
\frac{dt}{d\tau}&=0,\qquad
\left|\frac{d\mathbf x}{d\tau}\right|=we^{f(\tau)}.
\end{aligned}
\label{eq:pf_characteristics}
\end{equation}
%Thus, the equal-$t$ spatial reach is retained without imposing 
%the original zero-mode restriction.
These equations identify a promising, action-based route 
to avoiding the specific brane-time tachyonic dispersion, 
while retaining the equal-\(t\) spatial reach without imposing
 the zero-mode restriction (P1).
They do not yet constitute a proof of complete stability.
Since the projected equation describes evolution in
the extra time $\tau$, its coupling to detectors evolving
in ordinary time $t$ remains to be specified.
Establishing a fully coupled aether--mediator theory,
including source and boundary conditions, interacting
stability, physical brane response, and operational
no-signaling, constitutes a natural next stage of
this research program and is reserved for a separate
investigation.
The present calculations continue to use P1.

\paragraph*{Scope of the spectral restriction.}
The unrestricted $(3,2)$ Klein--Gordon model is not established to be
stable: its spectral resolution contains a tachyonic four-dimensional
continuum that is not eliminated by weighted normalizability, vanishing
boundary flux, $\mathbb Z_2$ parity, or the retarded prescription.
The zero-mode construction is an effective four-dimensional restriction,
not a dynamically selected sector of a generic brane-sourced theory.
A stable five-dimensional realization of the separate equal-time
contextual coupling, together with a nonsignaling multiparty collapse
law, remains to be constructed; neither a ghost-free quantization nor
a well-posed unrestricted ultrahyperbolic Cauchy problem is claimed.

\section{Null characteristics: monotonicity in $\tau$
and equal-time reach}
\label{sec:nullchars}

In the geometric-optics (WKB) regime, wavefront
propagation of a minimally coupled massless bulk field
is governed by the characteristics of the bulk wave
operator. For the bulk d'Alembertian
$\Box_{5}=\nabla^{A}\nabla_{A}$, the principal symbol
is $P(x,k)=g^{AB}(x)\,k_A k_B$, so characteristic
covectors satisfy $P(x,\partial S)=0$, i.e.\
the eikonal equation
$g^{AB}\partial_A S\,\partial_B S=0$.
In a $(3,2)$-signature bulk, $\Box_5$ is
\emph{ultrahyperbolic}, with two timelike directions,
and is therefore not hyperbolic in the standard
one-time PDE sense. The characteristic set is
nonetheless the metric null cone, and the associated
bicharacteristics project onto null geodesics of the
bulk metric. In what follows, ``null characteristic
ray'' and ``null geodesic'' are used interchangeably.

For a pseudo-Riemannian metric of signature $(3,2)$,
the null cone at a generic point $p$ is
\begin{equation}
\mathcal{N}_p^{(3,2)}=\Bigl\{\,v\neq 0:\;
g_{AB}(p)\,v^{A}v^{B}=0 \Bigr\},
\label{eq:null_cone_32}
\end{equation}
with timelike and spacelike sets defined by the sign
of $g_p(v,v)$,
\begin{equation}
\mathcal{T}_p^{(3,2)}
=\{\,v:\; g_p(v,v)<0\,\},\quad
\mathcal{S}_p^{(3,2)}
=\{\,v:\; g_p(v,v)>0\,\}.
\end{equation}
With two timelike directions, $\mathcal{T}_p^{(3,2)}$
is connected, admitting no canonical future-past
split; in signature $(3,1)$ it splits into two
disconnected components.

\subsection{Conserved quantities and monotonicity
in $\tau$}
\label{subsec:tau_monotone}

Parametrize a null ray by an affine parameter $s$ and
write $\dot{x}^{A}=(\dot t,\dot{\mathbf x},\dot\tau)$.
The ray equations follow from the geodesic equations
$\ddot x^{A}+\Gamma^{A}{}_{BC}\dot x^{B}\dot x^{C}=0$
or, equivalently, from the Euler--Lagrange equations
of $L=\tfrac12 g_{AB}\dot x^{A}\dot x^{B}$,
\begin{equation}
\frac{d}{ds}\!\left(g_{AB}\dot{x}^{B}\right)
-\frac12\,\partial_A g_{BC}\,
\dot{x}^{B}\dot{x}^{C}=0,
\label{eq:geo-eq-general-clean}
\end{equation}
subject to the null constraint
$g_{AB}\dot x^{A}\dot x^{B}=0$.

Since the metric~\eqref{eq:32metric} is independent
of $t$ and $\mathbf{x}$, the corresponding canonical
momenta are conserved:
\begin{equation}
E \equiv e^{-2f(\tau)}c^{2}\dot t,
\qquad
p_i \equiv e^{-2f(\tau)}\dot x^{i},
\qquad
p^{2}\equiv \delta^{ij}p_ip_j.
\label{eq:conserved}
\end{equation}
The null constraint gives
\begin{equation}
\begin{aligned}
0&=e^{-2f}\!\left(-c^{2}\dot t^{2}
+\dot{\mathbf{x}}^{2}\right)-w^{2}\dot\tau^{2}\\
&=e^{2f(\tau)}\!\left(
p^{2}-\frac{E^{2}}{c^{2}}\right)-w^{2}\dot\tau^{2},
\end{aligned}
\label{eq:null_geod}
\end{equation}
or equivalently,
\begin{equation}
w^{2}\dot\tau^{2}=e^{2f(\tau)}\,\alpha,
\qquad
\alpha\equiv p^{2}-\frac{E^{2}}{c^{2}}
=\mathrm{const}.
\label{eq:tau_dot}
\end{equation}

Since $e^{2f(\tau)}>0$, real null rays require
$\alpha\ge 0$. For $\alpha>0$,
Eq.~\eqref{eq:tau_dot} gives
$|\dot\tau|=(e^{f(\tau)}/w)\sqrt{\alpha}>0$
everywhere along the ray. Thus $\dot\tau$ cannot
change sign on a smooth geodesic, and $\tau(s)$
is strictly monotone, with its direction fixed
by the initial conditions. The borderline case
$\alpha=0$ yields $\dot\tau\equiv 0$, corresponding
to propagation confined to a fixed
$\tau=\mathrm{const.}$ slice.

\paragraph*{Single-ray kinematics vs.\ brane-to-brane
propagation.}
\label{par:single_ray_vs_kernel}

The monotonicity of $\tau(s)$ for $\alpha>0$
implies that a single smooth null ray leaving
the brane at $\tau=0$ cannot return to it by
free geodesic motion. This does not invalidate
the equal-time spatial projection of bulk null
characteristics, but the latter should not
be identified with a physical brane-to-brane
signal.

The field response must instead be determined
from the bulk wave equation and its source and
boundary conditions. As discussed in
Sec.~\ref{subsec:mode_kernel}, the unrestricted
field admits a formal spectral decomposition
into normalizable discrete modes and generalized
continuum eigenfunctions $\psi_\mu(\tau)$.
The corresponding four-dimensional amplitudes
satisfy Klein--Gordon-type equations, with the
positive-$\mu$ continuum exhibiting the tachyonic
instability identified above.

Under the effective stability restriction (P1),
only the normalizable massless zero mode is
retained. Its retarded four-dimensional response
has support on the ordinary brane light cone
and does not itself generate an instantaneous
measurement-induced interaction between two
spatially separated parties (conventionally
denoted Alice and Bob). A generic brane-localized
source can also excite the excluded continuum,
so the restricted model requires a compatible
source prescription.
Thus, the geometrical equal-time reach and
the restricted spectral dynamics are distinct
but compatible results. The preferred-foliation
construction discussed above offers an alternative
dynamical framework for investigating equal-time
correlations beyond the zero-mode restriction.

\subsection{A $t$-stationary null family and unbounded equal-time reach}
\label{subsec:E0}
A useful limiting class is obtained by setting
\begin{equation}
E=0,\qquad p^{2}>0.
\label{eq:E0choice}
\end{equation}
Then Eqs.~\eqref{eq:conserved} and~\eqref{eq:null_geod}
imply
\begin{equation}
\dot t=0,\qquad
\dot\tau=\pm \frac{e^{f(\tau)}}{w}\,|p|,\qquad
\dot{\mathbf x}=e^{2f(\tau)}\,\mathbf p,
\label{eq:E0_family}
\end{equation}
so these characteristics are everywhere null and satisfy
$dt\equiv 0$. Since $\alpha=p^{2}>0$, $\tau(s)$ is monotone on
each such curve, with the sign fixed by initial
conditions. Dividing the last two relations
in~\eqref{eq:E0_family} gives the brane displacement
per unit $\tau$,
\begin{equation}
\frac{d\mathbf x}{d\tau}
=\pm\,w\,e^{f(\tau)}\,\hat{\mathbf n},
\qquad \hat{\mathbf n}\equiv \frac{\mathbf p}{|\mathbf p|},
\label{eq:dx_dtau_E0}
\end{equation}
so a one-way excursion from $\tau=0$ to depth $\tau$
produces an equal-time brane projection
\begin{equation}
\Delta\mathbf x\,(\tau)
=\pm\,w\,\hat{\mathbf n}
\int_{0}^{\tau} e^{f(\bar\tau)}\,d\bar\tau.
\label{eq:equal_time_shift}
\end{equation}
For the warp factor $f(\tau)=k|\tau|$ derived in
Sec.~\ref{symmwarp3-2}, the integral
in~\eqref{eq:equal_time_shift} diverges as
$\tau\to\infty$, so the equal-time brane reach of
the $E=0$ family is unbounded. Explicitly, for $\tau>0$,
\begin{equation}
|\Delta\mathbf x(\tau)|=\frac{w}{k}\left(e^{k\tau}-1\right),
\qquad
\tau_*(L)=\frac1k\ln\!\left(1+\frac{kL}{w}\right),
\label{eq:tau_reach}
\end{equation}
so that a brane separation $L$ is reached at depth $\tau_*(L)$; the flat estimate $\tau_*\simeq L/w$ holds only for $kL/w\ll1$.
Consequently, for any prescribed brane separation
$L=|\mathbf x_B-\mathbf x_A|$ and any fixed brane time
$t_0$, one can find a depth $\tau_\ast$ and a direction
$\hat{\mathbf n}$ such that an everywhere-null
characteristic with $\dot{t}\equiv 0$ emitted from
$A=(t_0,\mathbf x_A,0)$ reaches the bulk point
\[
(t_0,\mathbf x_A+L\,\hat{\mathbf n},\tau_\ast),
\]
whose brane projection has spatial displacement $L$.
This concerns the equal-time reach of null
characteristics in the bulk; it should not be read as
the existence of a single null geodesic connecting two
brane events $A=(t_0,\mathbf x_A,0)$ and
$B=(t_0,\mathbf x_B,0)$.
It is nevertheless useful to summarise this kinematics
by saying that the equal-time projection of null
propagation has an unbounded effective correlation
velocity,
\begin{equation}
v_{\rm eff}^{(\mathrm{corr})}\equiv
\frac{|\Delta\mathbf x|}{\Delta t},
\end{equation}
in the sense that for any prescribed brane separation
$L$ one can find a bulk depth $\tau_*$ such that the
corresponding null characteristic with $\dot{t}\equiv 0$
reaches $|\Delta\mathbf{x}|=L$
(Eq.~\eqref{eq:equal_time_shift}); hence
$\lim_{\Delta t\to 0}|\Delta\mathbf{x}|/\Delta t$ is
unbounded. This is shorthand for ``unbounded
equal-time reach'': the limit is taken at fixed,
arbitrarily large $|\Delta\mathbf{x}|$, not at a point
where $\Delta t$ has already been set to zero. 
This geometrical reach is distinct from source-driven
brane-to-brane propagation. Under P1, the response is
governed by $\mathcal K_{\rm br}^{(\mathrm{P1})}$
[Eq.~\eqref{eq:brane_kernel_P1}] and vanishes for
brane-spacelike separation.

\textit{Summary of Sec.~\ref{sec:nullchars}.}
The $E=0$ null geodesic family is a structural feature
of the warped $(3,2)$ metric: $\dot{t}\equiv 0$ and
the equal-time brane reach is unbounded for any
asymptotically growing warp factor, in particular for
$f(\tau)=k|\tau|$. Each null ray with $\alpha>0$ is
monotone in $\tau$ and never returns to the brane.
The absence of a returning single ray leaves open a
collective Green-function mechanism, but a stable source-to-brane
kernel of this kind is not constructed for the unrestricted field. 
The $E=0$ family identifies geometrically allowed equal-time null
directions, while a power-law equal-time covariance may be assigned
to a classical ensemble of zero-mode solutions. Neither observation
proves that $E=0$ rays contribute a well-defined stationary-phase
saddle to the physical two-brane source response, or that the
measurement event creates an instantaneous distant response.
Causal response and controllable
signaling are governed by the $t$-retarded Green
function, which is strictly light-cone limited on the
brane. The two channels, equal-time correlations
and retarded response, are operationally distinct
and coexist without contradiction.
%=====================================================
\section{Bulk field equation and the brane-to-brane Green function}
\label{sec:bulkfield}

We first distinguish the unrestricted bulk equation from the
zero-mode effective dynamics adopted under P1. For a single
component of the mediator (suppressing the index $a$), the
minimally coupled equation~\eqref{eq:EOM-vector} with a
brane-localized source reads
\begin{equation}
\Box_5\mathscr X=\kappa J(x)\delta(\tau).
\label{eq:waveeq}
\end{equation}
The corresponding covariantly normalized bulk Green-function
equation would be
\begin{equation}
\Box_5G_5(x,\tau;x',\tau')=
\frac{\delta^{(4)}(x-x')\delta(\tau-\tau')}
{\sqrt{|g(x,\tau)|}}.
\label{eq:green_def}
\end{equation}
We do not construct a stable, unrestricted bulk fundamental
solution retarded in ordinary time $t$. In particular, a
retarded prescription does not eliminate the exponentially
growing positive-$\mu$ components excited by a generic
brane-localized source.

Under the effective restriction P1, the field is instead
confined to its normalizable massless zero mode,
$\mathscr X_a(x,\tau)=\psi_0(\tau)\phi_a(x)$, with
$\psi_0(\tau)=\sqrt{k}$. A compatible projected source then
gives
$\Box_4\phi_a=\kappa\sqrt{k}\,J_a$. Writing
$G^{(0)}_{\rm ret}\equiv G^{(4)}_{{\rm ret},0}$ for the
ordinary four-dimensional retarded Green function, we require
\begin{equation}
G^{(0)}_{\rm ret}(x-x')=0\qquad\text{for }t<t'.
\label{eq:retarded_support}
\end{equation}
The resulting response is
\begin{equation}
\phi_a(x)=\kappa\sqrt{k}\int d^4x'\,
G^{(0)}_{\rm ret}(x-x')J_a(x'),
\label{eq:brane_response_Xa}
\end{equation}
so that the brane field itself satisfies
$\mathscr X_a(x,0)=\sqrt{k}\,\phi_a(x)$ and its
source-response kernel is $\kappa kG^{(0)}_{\rm ret}$.
This restriction is not a solution of
Eq.~\eqref{eq:waveeq} for an arbitrary delta-function
source: such a source generally has nonzero overlap with
the excluded continuum. P1 must therefore be implemented
through an effective projection or an explicitly compatible
source prescription. The zero-mode response is brane-causal;
it cannot itself transmit a measurement-generated signal at
equal $t$ between separated brane points.

\subsection{Mode expansion and the brane kernel}
\label{subsec:mode_kernel}

The spectral resolution described in
Sec.~\ref{subsubsec:X_spectral_resolution} contains
normalizable discrete modes $\psi_n(\tau)$ and generalized
continuum eigenfunctions $\psi_\mu(\tau)$, with spectral
normalization based on the weight $W(\tau)=e^{-2f(\tau)}$.
The corresponding four-dimensional amplitudes formally obey
operators $\Box_4+\mu^2$, as derived in
Sec.~\ref{subsubsec:X_effective4D}. For the unrestricted
brane source, separation of variables suggests the formal
brane response kernel
\begin{equation}
\begin{aligned}
\mathcal K_{\rm br}^{\rm formal}(x-x')
={}&\sum_n\psi_n(0)^2\,
G^{(4)}_{{\rm ret},\mu_n}(x-x')\\
&+\int_0^\infty d\mu\;\varrho(\mu)\,
G^{(4)}_{{\rm ret},\mu}(x-x'),
\end{aligned}
\label{eq:mode_sum_mixed}
\end{equation}
where $\varrho(\mu)\geq0$ incorporates the continuum
spectral measure and brane overlap. With delta-normalized
continuum modes it is proportional to $|\psi_\mu(0)|^2$.
The overlaps quantify the coupling of individual modes to
brane-localized sources; they do not imply that the continuum
is dynamically suppressed. Equation~\eqref{eq:mode_sum_mixed}
is a \emph{formal spectral diagnostic}, not an established
stable $t$-retarded propagator for the ultrahyperbolic bulk
operator. It is normalized as a response kernel for the
coordinate-delta source of Eq.~\eqref{eq:waveeq}, rather
than as the direct restriction of the invariant Green
function in Eq.~\eqref{eq:green_def}.

Under P1, the physical effective brane kernel contains only
\begin{equation}
\mathcal K_{\rm br}^{(\mathrm{P1})}(x-x')
=\psi_0(0)^2G^{(4)}_{{\rm ret},0}(x-x')
=k\,G^{(0)}_{\rm ret}(x-x').
\label{eq:brane_kernel_P1}
\end{equation}
With $\Box_4=-c^{-2}\partial_t^2+\Delta_3$, the
massless retarded Green function is defined by
\begin{equation}
\Box_4G^{(0)}_{\rm ret}(x-x')
=\delta^{(4)}(x-x'),
\end{equation}
and explicitly, for $r=|\mathbf x-\mathbf x'|>0$,
\begin{equation}
G^{(0)}_{\rm ret}(\Delta t,r)
=-\frac{1}{4\pi r}\,
\delta\!\left(\Delta t-\frac{r}{c}\right),
\qquad\Delta t=t-t'.
\label{eq:massless_retarded_4D}
\end{equation}
Its support lies on the future brane light cone
\cite{Jackson}.

\subsubsection*{Response vs.\ correlations in a classical field}

A source-driven response and a statistical correlation are
different objects. The first describes how changing a source
changes the field; the second describes joint fluctuations
within a specified ensemble of configurations. 
Retarded propagation restricts the source-response
channel but does not, by itself, guarantee operational
no-signaling, which remains an independent requirement
on the complete measurement statistics (P3).
\paragraph*{(i) Response: the causal zero-mode kernel.}
For comparison with the unrestricted spectral analysis, a
positive-$\mu$ amplitude driven by the brane source would
formally satisfy
\begin{equation}
(\Box_4+\mu^2)\phi_\mu(x)
=\kappa\psi_\mu(0)J(x),
\end{equation}
with the individual mode response
\begin{equation}
\phi_\mu^{\rm resp}(x)
=\kappa\psi_\mu(0)\int d^4x'\,
G^{(4)}_{{\rm ret},\mu}(x-x')J(x').
\label{eq:resp_def_classical}
\end{equation}
Each fixed four-dimensional operator $\Box_4+\mu^2$
admits a retarded fundamental solution supported in the
future brane light cone, even though its $\mu>0$ solutions
include exponentially growing modes. In particular,
\begin{equation}
G^{(4)}_{{\rm ret},\mu}(\Delta t,r)=0
\quad\text{for }r>c|\Delta t|
\  \text{(brane-spacelike interval)}.
\label{eq:ret_spacelike_zero}
\end{equation}
Here brane-spacelike means
$(x-x')^2=r^2-c^2\Delta t^2>0$ in the present
metric convention. The existence of such four-dimensional
retarded kernels is standard for normally hyperbolic
operators on Minkowski spacetime
\cite{BaerWave2010,DerezinskiPropagators2024};
it does not establish a stable retarded Green operator for
the unrestricted $(3,2)$ bulk. In the physical P1 sector,
only the $\mu=0$ response of
Eq.~\eqref{eq:brane_response_Xa} is retained.

\paragraph*{(ii) Correlations: an ensemble of zero-mode
solutions.}
Within P1, one may specify a probability ensemble of
classical zero-mode initial data on a brane-time Cauchy
surface $t=t_0$, possibly supplemented by contextual
variables $\lambda$. For a normalized measure $P(\lambda)$,
write
\begin{equation}
\langle F\rangle_\lambda
=\int d\lambda\,P(\lambda)F[\lambda].
\label{eq:ensemble_average_def}
\end{equation}
The connected covariance of the zero-mode amplitude is
\begin{equation}
\begin{aligned}
C_0(x,x')={}&\langle\phi_0(x)\phi_0(x')\rangle_\lambda\\
&-\langle\phi_0(x)\rangle_\lambda
\langle\phi_0(x')\rangle_\lambda.
\end{aligned}
\label{eq:classical_covariance}
\end{equation}
Unlike a retarded response kernel, this covariance need
not vanish at equal-time, spatially separated points.
Such correlations reflect the statistical ensemble and
cannot, by themselves, be interpreted as a dynamically
transmitted signal.

\paragraph*{Brane-restricted correlation kernel.}
An unrestricted spectral expansion could also be used to
write a \emph{formal} diagonal-mode covariance when the
distinct spectral amplitudes have vanishing connected
cross-covariances:
\begin{equation}
\begin{aligned}
C_{\rm br}^{\rm formal}(x,x')
={}&\sum_n\psi_n(0)^2 C_{\mu_n}(x,x')\\
&+\int_0^\infty d\mu\;\varrho(\mu)
C_\mu(x,x').
\end{aligned}
\label{eq:brane_corr_mode_sum_classical}
\end{equation}
This expression is an algebraic diagnostic, not a
construction of a stable ensemble containing the excluded
tachyonic continuum. In the P1 effective model, the
physical brane covariance is instead simply
\begin{equation}
\begin{aligned}
C_{\rm br}^{(\mathrm{P1})}(x,x')
&=\psi_0(0)^2 C_0(x,x')\\
&=k\,C_0(x,x').
\end{aligned}
\label{eq:brane_corr_P1}
\end{equation}

Thus, within P1, the brane covariance is entirely
determined by the statistical ensemble of zero-mode
configurations. It may be nonzero at equal time for
spatially separated points, despite the light-cone
restriction on source-driven response. Neither its
specific form nor its dynamical generation follows
from the zero-mode restriction alone.

\paragraph*{Field correlators versus Bell correlators.}

Under the zero-mode restriction (P1), the connected
two-point covariance of the mediator field on the
brane is
$C_{\mathscr X}(x,x')=k\,C_0(x,x')$, where
$C_0$ is determined by the statistical ensemble
of zero-mode configurations. It may be nonzero
at equal time for spatially separated points,
even though the source-driven response is
restricted to the ordinary brane light cone.
Neither the magnitude nor the spatial decay of
this covariance is determined by P1 alone.

A Bell--CHSH correlator,
$E(\mathbf a,\mathbf b)
=\langle A_{\mathbf a}B_{\mathbf b}\rangle$,
instead concerns measurement outcomes for
detector settings $\mathbf a$ and $\mathbf b$.
For comparison, a strictly Bell-local classical
readout would assign outcomes through
\begin{equation}
A_{\mathbf a}=\mathcal F_A(\mathbf a;\lambda),
\qquad
B_{\mathbf b}=\mathcal F_B(\mathbf b;\lambda),
\label{eq:local_readout_maps}
\end{equation}
where $\lambda$ includes the relevant shared
field variables. The corresponding correlator is
\begin{equation}
E(\mathbf a,\mathbf b)
=\int d\lambda\,P(\lambda)\,
\mathcal F_A(\mathbf a;\lambda)\,
\mathcal F_B(\mathbf b;\lambda).
\label{eq:bell_from_lambda}
\end{equation}
Here $\mathcal F_A$ and $\mathcal F_B$ are local
readout functions taking values $\pm1$,
determining the outcomes at Alice's and Bob's
detectors from their respective analyzer settings
and the hidden variables $\lambda$, which include
the relevant shared mediator-field variables.
For dichotomic outcomes and a setting-independent
$P(\lambda)$, this factorization obeys the
Bell--CHSH bound $|S_{\rm CHSH}|\leq 2$.
Consequently, a shared classical mediator field,
combined with strictly local readout, does not
by itself reproduce Bell-inequality violations.

Reproducing such violations requires additional
measurement dynamics beyond this Bell-local
factorization, while operational no-signaling
remains an independent requirement (P3).
The relation between mediator-field correlations
and observable Bell correlators must therefore
be established from the complete detector
coupling and outcome statistics. In particular,
neither the distance dependence of Bell
correlations nor a geometry-dependent cross-pair
correlation follows directly from the zero-mode
covariance.

\subsection{Normalizability and ``vanishing flux'' at
$|\tau|\to\infty$}
\label{subsec:vanishing_flux}

Here ``vanishing flux'' means that integration by
parts in the transverse Sturm--Liouville expression
produces no boundary term. In the weighted inner
product $\langle\psi,\varphi\rangle_W
=\int d\tau\,W\psi^*\varphi$, symmetry requires
\begin{equation}
\Bigl[p(\tau)\bigl(\psi^*(\tau)\varphi'(\tau)
-\psi'^*(\tau)\varphi(\tau)\bigr)
\Bigr]_{-\infty}^{+\infty}=0.
\label{eq:SL_boundary_form}
\end{equation}
The vanishing boundary form makes the operator symmetric.
A self-adjoint realization also specifies matching at
$\tau=0$ and the asymptotic conditions. Normalizable
discrete modes, including the zero mode, satisfy
\begin{equation}
\int_{-\infty}^{\infty}d\tau\,W(\tau)
|\psi(\tau)|^2<\infty,
\label{eq:normalizability_W}
\end{equation}
whereas continuum modes are generalized,
delta-normalized eigenfunctions. Completeness is
understood with respect to the chosen self-adjoint
spectral resolution.

\subsection{WKB: geometric-optics view}
\label{subsec:wkb_view}

In the short-wavelength 
regime, the minimally coupled massless
bulk wave equation \eqref{eq:EOM-vector} (with $J_a=0$) 
admits standard geometrical-optics (eikonal or
WKB) approximation, we write the bulk field as
 (see, e.g., Ref.~\cite{TaylorPDE2})
\begin{equation}
\mathscr{X}_a(x,\tau)\;\sim\;
\mathrm{Re}\!\left\{\mathscr{A}_a(x,\tau;\varepsilon)\,
\exp\!\Bigl[\tfrac{i}{\varepsilon}S(x,\tau)
\Bigr]\right\},
\label{eq:wkb_ansatz_simple}
\end{equation}
with $\varepsilon\sim\lambda/L\ll 1$ the ratio of
the field wavelength $\lambda$ to the geometric
variation length $L\sim w/k$ of the warp factor,
where $S$ is the rapidly varying phase and
$\mathscr{A}_a$ is a slowly varying amplitude.
Substituting \eqref{eq:wkb_ansatz_simple} into the
wave equation and keeping the leading term in
$\varepsilon$ gives the eikonal equation
\begin{equation}
g^{AB}\,\partial_A S\,\partial_B S=0,
\qquad
k_A\equiv\partial_A S,
\label{eq:eikonal_simple}
\end{equation}
so the wave covector $k_A$ is null, and the
corresponding ray trajectories are therefore null
characteristics of the bulk wave equation which,
for a minimally coupled nondispersive massless
field, coincide with null geodesics up to
reparametrization.
Away from caustics, a local bulk WKB wave kernel
has the schematic null-ray expansion
%[Gutzwiller 1990; Littlejohn 1992], with the sum running over contributing null-ray families γ\gamma , S_\gamma
% the associated phase, and A_\gamma
%​ a geometrical-optics amplitude."
%
(see, e.g., Refs.~\cite{DuistermaatFIO,SoggeFIO}),
\begin{equation}
\mathscr{K}_{\rm WKB}(X,X')\ \sim\ \sum_{\gamma}
A_{\gamma}(X,X';\varepsilon)\,
\exp\!\left[\frac{i}{\varepsilon}
S_{\gamma}(X,X')\right],
\label{eq:wkb_generic_simple}
\end{equation}
Here $X=(x,\tau)$ and $X'=(x',\tau')$ denote
bulk endpoints, and $\gamma$ labels connecting rays.
This local high-frequency expression describes
wavefront propagation, not a global retarded Green
function or an ensemble covariance. By $\tau$
monotonicity, a ray that leaves the brane into the
bulk cannot return to it; brane-confined null rays
($\alpha=0$) are a separate case.

\paragraph*{Remark: retarded response versus ensemble covariance.}
The representation~\eqref{eq:wkb_generic_simple} is a
geometric-optics description of a Green kernel; it
is not an established expansion of the independently
prescribed ensemble covariance. For a retarded kernel,
the contributing branches are additionally subject to
causal support conditions.
Under P1, causal response is instead given by
$G^{(0)}_{\rm ret}$ [Eq.~\eqref{eq:massless_retarded_4D}].
The $E_\gamma=0$ bulk rays provide geometrical motivation
for equal-time correlations, whose values depend on the
zero-mode ensemble [Eq.~\eqref{eq:brane_corr_P1}].

\paragraph*{Geometric-optics link to the $E=0$
equal-time null family.}
In the WKB description of the unrestricted bulk
field, $E=-\partial_t S$, so the $E=0$ family
corresponds to null rays propagating at fixed
ordinary time $t$. These rays provide geometrical
motivation for the proposed equal-time correlations,
whose covariance depends on the mediator dynamics
and statistical ensemble.
\subsection{Causality and operational consistency}
\label{subsec:causality_operational}

In a $(3,2)$ signature, the bulk admits timelike
loops as abstract curves in the $(t,\tau)$ plane.
The physically relevant question, however, is
whether such loops can be exploited by brane observers
to signal into their own past.

Within P1, the brane response is governed by
$G^{(0)}_{\rm ret}$ [Eq.~\eqref{eq:retarded_support}],
or equivalently by $\mathcal K_{\rm br}^{(\mathrm{P1})}$
[Eq.~\eqref{eq:brane_kernel_P1}].
Its support is confined to the future brane light
cone, excluding closed causal-response loops
returning signals to the brane past.

Independently, the bulk null-geodesic analysis gives,
for $E>0$ [Eq.~\eqref{eq:conserved}],
\begin{equation}
\dot{t}(s)=\frac{E}{c^{2}}e^{2f(\tau(s))}>0,
\label{eq:tdotpos}
\end{equation}
while $E=0$ implies $\dot t=0$. Thus, along null
rays restricted to $E\geq0$, ordinary time never
decreases, excluding propagation back into the
brane past through such rays.

The restriction $E\geq0$ does not, however,
eliminate the tachyonic transverse modes.
The $E=0$ null family provides the geometrical
basis for the proposed equal-time propagation.

\paragraph*{Limits of the causal-response guarantee.}
The above results concern the restricted brane
response and the selected null-ray families,
rather than global causality of the unrestricted
bulk. Operational no-signaling of the complete
detector statistics remains the independent
requirement P3.

\paragraph*{Coexistence of causal response and
equal-time correlations.}
Two distinct structures coexist in the theory.
Within P1, the source-driven response
$\delta\mathscr X_a(x,0)=\kappa\int d^4x'\,
\mathcal K_{\rm br}^{(\mathrm{P1})}(x-x')J_a(x')$
is confined to the future brane light cone.
At the same time, the preparation ensemble
can support nonzero equal-time spatial correlations,
$C_{\rm br}^{(\mathrm{P1})}(x,x')=kC_0(x,x')$,
without implying an instantaneous source response.
The $E=0$ bulk null family provides complementary
geometrical motivation for the proposed
equal-time contextual mediation.
%=====================================================
\section{Role of the field $\mathscr{X}_a(x,\tau)$ in the Bohm--Bub collapse model}
\label{sec:Xfield-propagation}
This section develops the dynamical model in four
steps. First (Sec.~\ref{nutshell}), we recall the
single-system Bohm--Bub (BB) collapse model and
identify the object that must be replaced by a field.
Second (Sec.~\ref{sec:bipartite-BB}), we extend it
to a bipartite entangled system, introducing the
channel projections and crossed collapse ratios that
drive the two-wing dynamics. Third
(Sec.~\ref{subsec:meas-exchange}), we geometrise the
information-exchange mechanism by specifying the two
bulk sources $J_a^{(\rm prep)}$ and $J_a^{(\rm meas)}$
and motivating a postulated two-component contextual input
from the $E=0$ geodesic kinematics.
Fourth (Sec.~\ref{subsec:PRR-continuity}), we provide a retrospective
translation table showing how the present framework
{ relates the construction to the assumptions of the
earlier toy model~\cite{PRR}, without an exact derivation.} Contextual microstates, equivariance,
and Born-rule recovery are treated separately in
Sec.~\ref{subsec:contextuality-lambda}.
\paragraph*{(P2) Contextual mediation hypothesis.}
In the BB model, $\xi(t)$ (defined below) is an abstract
run-dependent auxiliary. Here the detector inputs
are hypothesized to depend on the brane value
$\mathscr X_a(\mathbf x,t,0;\lambda)$ of a propagating
mediator. P1 supplies its retarded zero-mode response,
while preparation can correlate the initial field data.
P2 introduces a distinct, proposed equal-$t$
measurement coupling, motivated by the $E=0$ bulk
characteristics (Sec.~\ref{sec:nullchars}). The
collapse evolution is deterministic for fixed $\lambda$;
statistics follow by averaging over $\rho(\lambda)$,
subject to Born compatibility and P3.

\paragraph*{(P3) Operational no-signaling condition.}
The contextual mediation hypothesis (P2) must satisfy a separate
operational requirement: for any freely chosen analyser settings,
the outcome statistics of any group of observers must be independent
of the settings chosen by the other observers. This applies to joint
as well as single-detector marginals after averaging over
$\rho(\lambda)$. Neither the retarded field response nor single-system
Born-rule compatibility implies this multipartite condition; its
dynamical realization remains open.
%----------------------------------------------------------------------
%----------------------------------------------------------------------
\subsection{The Bohm--Bub collapse model for a single system}
\label{nutshell}

We recall the BB model~\cite{BB} for a single quantum
system interacting with a macroscopic measuring apparatus
(see also the refined derivation in Ref.~\cite{tutsch}).
Bohm and Bub supplement the Schr\"odinger equation by a
nonunitary term $\mathscr{B}$:
\begin{equation}
\frac{\partial\Psi(\mathbf{x},t)}{\partial t}
=\mathscr{B}(\mathbf{x},t)
-\frac{i}{\hbar}\hat{H}\,\Psi(\mathbf{x},t),
\label{schroedinger}
\end{equation}
where $\mathscr{B}$ is negligible outside measurement
interactions but dominates during a measurement interval,
so that the Hamiltonian term $-i\hat{H}/\hbar$ is a
subleading correction on the collapse timescale.

Expanding $\Psi(\mathbf{x},t)=\sum_i\psi_i(t)\phi_i(\mathbf{x})$
in the measurement eigenbasis, one takes
\begin{equation}
\mathscr{B}(\mathbf{x},t)
=\gamma\sum_i\psi_i(t)\phi_i(\mathbf{x})
\sum_j|\psi_j(t)|^2\,(R_i-R_j),
\label{settoreB}
\end{equation}
with
\begin{equation}
R_i(t)=\frac{|\psi_i(t)|^2}{|\xi_i(t)|^2},
\end{equation}
where $\xi_i(t)$ are the components of a
run-dependent auxiliary hidden vector $\xi(t)$
expanded in the same orthonormal basis
$\{|i\rangle\}$ as $\psi_i(t)=\langle i|\psi(t)\rangle$,
so that $R_i$ depends on the chosen measurement
basis.
Equation~\eqref{schroedinger} then yields the amplitude
dynamics
\begin{equation}
\frac{d\psi_i(t)}{dt}
=\gamma\,\psi_i(t)\sum_j|\psi_j(t)|^2\,(R_i-R_j)
-\frac{i}{\hbar}\sum_j H_{ij}\psi_j(t).
\label{eq:BB-single}
\end{equation}
The nonlinear BB term drives $\psi_i(t)$
deterministically toward a single outcome once $\xi(t)$
is fixed: if $R_i > R_j$ for all $j\neq i$, i.e.\ if
$|\xi_i|<|\psi_i|$ relative to all other components,
the amplitude $\psi_i$ is amplified at the expense of
the others and the state collapses to the $i$-th
outcome. The rate $\gamma>0$ sets the collapse
timescale; the Hamiltonian term is negligible on that
timescale but restores unitary evolution between
measurements. The statistical character of quantum
mechanics is recovered, in the usual way, by allowing
$\xi$ to fluctuate between runs: for a uniform
distribution of $\xi$ on the unit sphere the Born rule
$P_i=|\psi_i|^2$ follows directly~\cite{BB}. 

Among the available nonlinear collapse
models~\cite{bassi,genovese2005,BrodyHughston2023},
the BB framework is adopted here for three reasons,
each rooted in structural compatibility rather than
mere convenience. First, it already contains a
run-dependent hidden variable $\xi(t)$ with no
intrinsic spacetime character, making it, in fact,
the minimal framework admitting replacement by the
brane-restricted bulk field $\mathscr{X}_a$ without further structural modification. 
Second, Born-rule recovery
in the single-system case is explicit and
basis-independent: for a uniform distribution of $\xi$
on the unit sphere, $P_i=|\psi_i|^2$ holds in every
measurement basis~\cite{BB}, which is precisely the
property the bipartite extension must inherit. Third,
the nonlinear term vanishes identically between
measurements, so the model reduces to standard unitary
evolution outside the collapse window; no unphysical
spontaneous collapse of isolated systems occurs, and
the theory remains well-behaved in the absence of
measurement. 
%----------------------------------------------------------------------
\subsection{Extension to a bipartite entangled system}
\label{sec:bipartite-BB}

Consider two quantum subsystems with Hilbert spaces
$\mathcal{H}_1$ and $\mathcal{H}_2$, without
committing to a specific observable beyond the choice
of local detector basis. Let
\begin{equation}
\mathcal{B}_k=\{|\phi^{(k)}_i\rangle\}_{i=1}^{d_k},
\qquad k=1,2,
\end{equation}
be orthonormal measurement bases. Any pure state of
the composite system expands as
\begin{equation}
|\Psi(t)\rangle
=\sum_{i=1}^{d_1}\sum_{j=1}^{d_2}C_{ij}(t)\,
   |\phi^{(1)}_i\rangle\otimes|\phi^{(2)}_j\rangle,
\label{eq:general-bipartite-state}
\end{equation}
with $C_{ij}(t)$ the (in general non-diagonal)
coefficient matrix in this basis.
The Schmidt decomposition theorem guarantees
that, for any bipartite pure state, there exists a pair
of orthonormal local bases, the \emph{Schmidt bases},
one for each subsystem, in which the coefficient matrix
is diagonal \cite{nielsen}. For the illustrative reduction below we choose these
Schmidt bases at preparation, giving
\begin{equation}
C_{ij}(t_0)=c_i(t_0)\,\delta_{ij},\qquad d_1=d_2\equiv d,
\label{eq:schmidt-diagonal}
\end{equation}
for the occupied Schmidt sector. A diagonal form at later
times requires evolution preserving this sector. More
importantly, independent Bell-analyser settings generally
rotate the local measurement bases, in which $C_{ij}$
need not be diagonal. The resulting local-amplitude
collapse equations are therefore a restricted illustrative
model; a complete treatment of arbitrary Bell settings
must use the joint channels of Eq.~\eqref{eq:general-bipartite-state}.
Within the restricted Schmidt-basis description, the state reads
\begin{eqnarray}
|\Psi(t)\rangle
&=&\sum_{i=1}^{d}c_{i}(t)\,
   |\phi^{(1)}_i\rangle\otimes|\phi^{(2)}_i\rangle,
\nonumber\\
\Psi(\mathbf{x}_1,\mathbf{x}_2,t)
&=&\sum_{i}c_{i}(t)\,
   u^{(1)}_i(\mathbf{x}_1)\,u^{(2)}_i(\mathbf{x}_2),
\label{eq:bipartite-state-PRD}
\end{eqnarray}
where
\begin{equation}
u^{(k)}_i(\mathbf{x})
\equiv\langle\mathbf{x}|\phi^{(k)}_i\rangle
\label{eq:detector-channel-profiles}
\end{equation}
are the detector-basis wavefunctions (channel profiles)
on the brane. The single coefficient vector
$c_i(t)$ encodes the full entanglement structure of the
prepared state in this basis, since the off-diagonal
entries $C_{ij}$, $i\ne j$, vanish identically by
construction; it is, in fact, the object from which both
the local channel amplitudes and the non-factorising
bulk field configuration are derived.

Define the single-station amplitude
$\psi^{(k)}_i(t)\equiv c_i(t)$ for $k=1,2$ (the two
stations share the same coefficient by the diagonal
structure~\eqref{eq:schmidt-diagonal}, although the
local readout maps $\mathcal F_{ij}$ that act on it at
each station, introduced below, need not coincide); by
the same diagonal structure,
$\sum_i|\psi^{(1)}_i(t)|^2=\sum_i|\psi^{(2)}_i(t)|^2
=\sum_i|c_i(t)|^2=1$ identically, so the single-station
norm coincides with the norm of the full bipartite
state~\eqref{eq:bipartite-state-PRD} at every instant.
\label{eq:marginal-norm-consistency}
This
is what allows us to treat $\psi^{(k)}_i(t)$ as obeying
its own self-contained, separately normalised BB-type
equation of motion at each station, Eq.~\eqref{eq:BB-entgl}
below, rather than as a marginal whose dynamics would,
for a non-diagonal state, only be implicitly fixed by
some more fundamental joint equation.

\paragraph*{Bulk-mediated contextual input and the
crossed projections.}
The collapse at each station is driven by the
projection of the brane-restricted bulk field onto the
local detector channels. We define the
channel-resolved overlaps
\begin{equation}
\mathscr{X}^{(2\to 1)}_{a,i}(t_0;\lambda)
=\int_{\Omega_1}d^3x_{(1)}\;
u^{(1)\,*}_i(\mathbf{x}_{(1)})\,
\mathscr{X}_a(\mathbf{x}_{(1)},t_0,0;\lambda),
\label{eq:Xproj-2to1}
\end{equation}
and analogously
\begin{equation}
\mathscr{X}^{(1\to 2)}_{a,j}(t_0;\lambda)
=\int_{\Omega_2}d^3x_{(2)}\;
u^{(2)\,*}_j(\mathbf{x}_{(2)})\,
\mathscr{X}_a(\mathbf{x}_{(2)},t_0,0;\lambda).
\label{eq:Xproj-general}
\end{equation}
The superscript ``$2\to1$'' labels the projection
at station~$1$ of the bulk field whose non-factorising
preparation source encodes the correlations of both
stations; ``$1\to2$'' is defined analogously at
station~$2$. These quantities are detector-basis
overlaps of the brane-evaluated bulk field serving
as contextual inputs for the local collapse dynamics,
not components of the bipartite state
$\Psi(\mathbf{x}_1,\mathbf{x}_2,t)$; the arrows do
not assert a brane-spacelike retarded influence from
one wing to the other.
Moreover, let us remark that
the integration variables in
\eqref{eq:Xproj-2to1}--\eqref{eq:Xproj-general} are
brane-space coordinates local to each detector, not
the two independent configuration-space coordinates
$\mathbf{x}_1,\mathbf{x}_2$ of $\Psi$, a
distinction that matters when interpreting the crossed
structure of the collapse drive.
The crossed projections reflect the shared
preparation field and the contextual measurement
contribution hypothesized in P2. The collapse ratio
at station~$1$ depends on the projection
of the field onto outcome channel~$i$. Its correlation
with the projection at station~$2$, in the detector's
chosen analyser basis, is a property to be imposed on
or derived from the preparation-field ensemble.
The shared contextual microstate $\lambda$ allows
non-factorising field projections, but their particular
channel dependence and the resulting Bell correlations
require the preparation ensemble and detector readout
to satisfy the conditions described in
Sec.~\ref{subsec:meas-exchange}.

\paragraph*{Channel normalisation and distance
independence.}
Because P1 alone does not fix the spatial covariance
or the preparation-field amplitude, we introduce the
channel-normalised field
\begin{equation}
\hat{\mathscr{X}}^{(k)}_{a,i}
\equiv
\frac{\mathscr{X}^{(k)}_{a,i}}{|\mathscr{X}^{(k)}_a|},
\qquad
|\mathscr{X}^{(k)}_a|
=\Bigl(\sum_i|\mathscr{X}^{(k)}_{a,i}|^2\Bigr)^{1/2},
\label{eq:channel-norm}
\end{equation}
where $|\hat{\mathscr{X}}^{(k)}_{a,i}|^2\equiv
\delta^{ab}\hat{\mathscr{X}}^{(k)}_{a,i}
(\hat{\mathscr{X}}^{(k)}_{b,i})^*$.
If the channel amplitudes have a common distance-dependent
envelope, that factor cancels in this normalisation.
The normalised input then encodes their relative weights;
its distance independence is conditional on this common-envelope
assumption and is not implied by the covariance alone. When no
ambiguity with spacetime indices arises we write
$\hat{\mathscr{X}}^{(k)}_i$ for the
$\delta_{ab}$-contracted shorthand.

\paragraph*{Crossed BB ratios and the bipartite collapse equation.} 
The crossed BB ratios are
\begin{equation}
\widetilde{R}^{(1)}_i(t_0;\lambda) =\frac{|\psi^{(1)}_i(t_0)|^2}
{|\hat{\mathscr{X}}^{(2\to 1)}_{a,i}(t_0;\lambda)|^2},
\label{eq:R-general-PRD}
\end{equation}
\begin{equation}
\widetilde{R}^{(2)}_j(t_0;\lambda) =\frac{|\psi^{(2)}_j(t_0)|^2}
{|\hat{\mathscr{X}}^{(1\to 2)}_{a,j}(t_0;\lambda)|^2}.
\label{eq:R-general-PRDa}
\end{equation}
The internal norm is $|\hat{\mathscr{X}}_{a,i}|^2\equiv
\delta^{ab}\hat{\mathscr{X}}_{a,i}\hat{\mathscr{X}}^*_{b,i}$, with
the index $a$ summed implicitly when not displayed. The full
bipartite BB equation including the Hamiltonian then reads
\begin{eqnarray}
\frac{d\psi^{(k)}_i}{dt} &=&\gamma_k\,\psi^{(k)}_i(t)
\sum_{m=1}^{d_k}|\psi^{(k)}_m(t)|^2
\bigl(\widetilde{R}^{(k)}_i(t_0;\lambda)
-\widetilde{R}^{(k)}_m(t_0;\lambda)\bigr) \nonumber\\
&&-\;\frac{i}{\hbar}\sum_m H^{(k)}_{im}\psi^{(k)}_m(t),
\label{eq:BB-entgl}
\end{eqnarray}
for $k=1,2$, where $H^{(k)}_{im}$ is the local Hamiltonian matrix
at station $k$. As in the single-system case~\eqref{eq:BB-single},
the nonlinear BB drive dominates during the collapse timescale; the
Hamiltonian term is negligible on that same timescale, and between
measurements the nonlinear term vanishes identically, restoring
unitary evolution. For fixed $\lambda$ the dynamics is fully
deterministic, driving each station to a definite outcome; quantum
statistics emerge only upon averaging over $\rho(\lambda)$, as
detailed in Sec.~\ref{subsec:contextuality-lambda} and in the
Appendix. For independent analyzer bases, Born statistics are
required of the contextual equilibrium ensemble; their
basis-independent recovery from the proposed bipartite
dynamics has not been established.

\paragraph*{From a bulk field to two detector readouts.}
A brane-localized source event at
$A=(\mathbf x_1,t_0)$ enters the unrestricted equation as
\begin{equation}
\Box_5\mathscr X_a(x,\tau)
=J_a^{(1)}(x)\delta(\tau),
\label{eq:single_source_bulk_eq}
\end{equation}
Such a source generally excites the tachyonic
continuum. Under P1 its projected brane response is
\begin{equation}
\mathscr X_a^{(0)}(x,0)
=\int d^4x'\,
\mathcal K_{\rm br}^{(\mathrm{P1})}(x-x')
J_a^{(1)}(x'),
\label{eq:single_source_bulk_sol}
\end{equation}
with the coupling absorbed into $J_a^{(1)}$.
A response at $B=(\mathbf x_2,t_0+\Delta t)$ requires
$\Delta t\ge|\mathbf x_2-\mathbf x_1|/c$;
no source-driven response occurs at equal brane time
for distinct detector positions.
For spacelike-separated events at a common brane time
$t_0$, the readouts
$\mathscr{X}_a(\mathbf{x}_1,t_0,0;\lambda)$ and
$\mathscr{X}_a(\mathbf{x}_2,t_0,0;\lambda)$ both sample
the same contextual microstate $\lambda$.
Their possible correlation is attributed to an assumed
non-factorising preparation ensemble, geometrically
motivated by the $E=0$ null family, and not to a
retarded signal between the stations.
The bulk wave equation for $\mathscr{X}_a$ is linear,
but the brane field depends sensitively on the
microscopic structure of $J_a(x)$ through the
superposition of propagation channels with different
phases. This dependence, together with the nonlinear
BB collapse map and environmental fluctuations,
justifies treating $\lambda$ as an effective
run-dependent stochastic variable, as developed in the
Appendix.

\subsection{Two bulk sources: preparation and measurement}
\label{subsec:meas-exchange}
\paragraph*{The two sources and their physical roles.}
Two physically distinct classes of events source the
bulk field $\mathscr{X}_a$, each with a different
function.
The \emph{preparation source}
$J_a^{(\rm prep)}(\mathbf{x},t)$, active at the
creation event $x_{\rm prep}$, is brane-localised
($\tau=0$) and generates the shared contextual bulk
configuration associated with the prepared quantum
state. The \emph{measurement sources}
$J_a^{(k,\rm meas)}(\mathbf{x},t)$, active at the
measurement time $t_0$ at each detector $k=A,B$,
are distributed according to the detector
$\tau$-profiles $\eta_k(\tau)$ (assumption H1
below). These sources encode the local
measurement events and their associated
outcome-channel information.

For clarity, we first write the unrestricted
two-source equation formally; its P1 projection follows:
\begin{eqnarray}
\Box_5\,\mathscr{X}_a(\mathbf{x},t,\tau)
&=& J^{(\rm prep)}_a(\mathbf{x},t)\,\delta(\tau)\nonumber\\
&+& \sum_{k=A,B}J^{(k,\rm meas)}_a(\mathbf{x},t)\,\eta_k(\tau),
\label{eq:full-sourced}
\end{eqnarray}
where $\Box_5$ is the five-dimensional Laplace--Beltrami
operator on the warped background.
A well-defined stable \emph{effective} version is obtained by
projecting the source onto the normalisable zero mode \emph{before}
evolution. With $f=k|\tau|$ and $\psi_0=\sqrt{k}$, the
projection of Eq.~\eqref{eq:full-sourced} gives

\begin{equation}
\begin{aligned}
\mathscr{X}^{(0)}_a(x,\tau)
&=\sqrt{k}\,\phi_{0,a}(x),\\
\Box_4\phi_{0,a}&=\sqrt{k}\,J_a^{\rm eff},\\
J_a^{\rm eff}&=J_a^{(\rm prep)}
+\sum_s b_sJ_a^{(s,\rm meas)},\\
b_s&=\int d\tau\,e^{-4k|\tau|}\eta_s(\tau).
\end{aligned}
\label{eq:projected-two-source}
\end{equation}

Here the coupling constant is absorbed into the
currents as in Eq.~\eqref{eq:full-sourced}. This is the zero-mode
projection of the sourced equation, not an exact solution of the
unrestricted equation with delta-like or Gaussian sources. The
$\tau$-independent zero mode has a retarded four-dimensional response:
$b_s$ changes the effective coupling but produces neither an
$E=0$ pulse nor the distance-dependent overlap of
Eq.~\eqref{eq:eps-meas}. Those ingredients require an additional
stable source-generated mediator dynamics, not constructed here;
these ingredients enter the contextual mediation hypothesis (P2).
Here $s$ ranges over the activated detectors.

\begin{figure}[h!]
% \centering
\includegraphics[scale=0.30,keepaspectratio=true,angle=0]{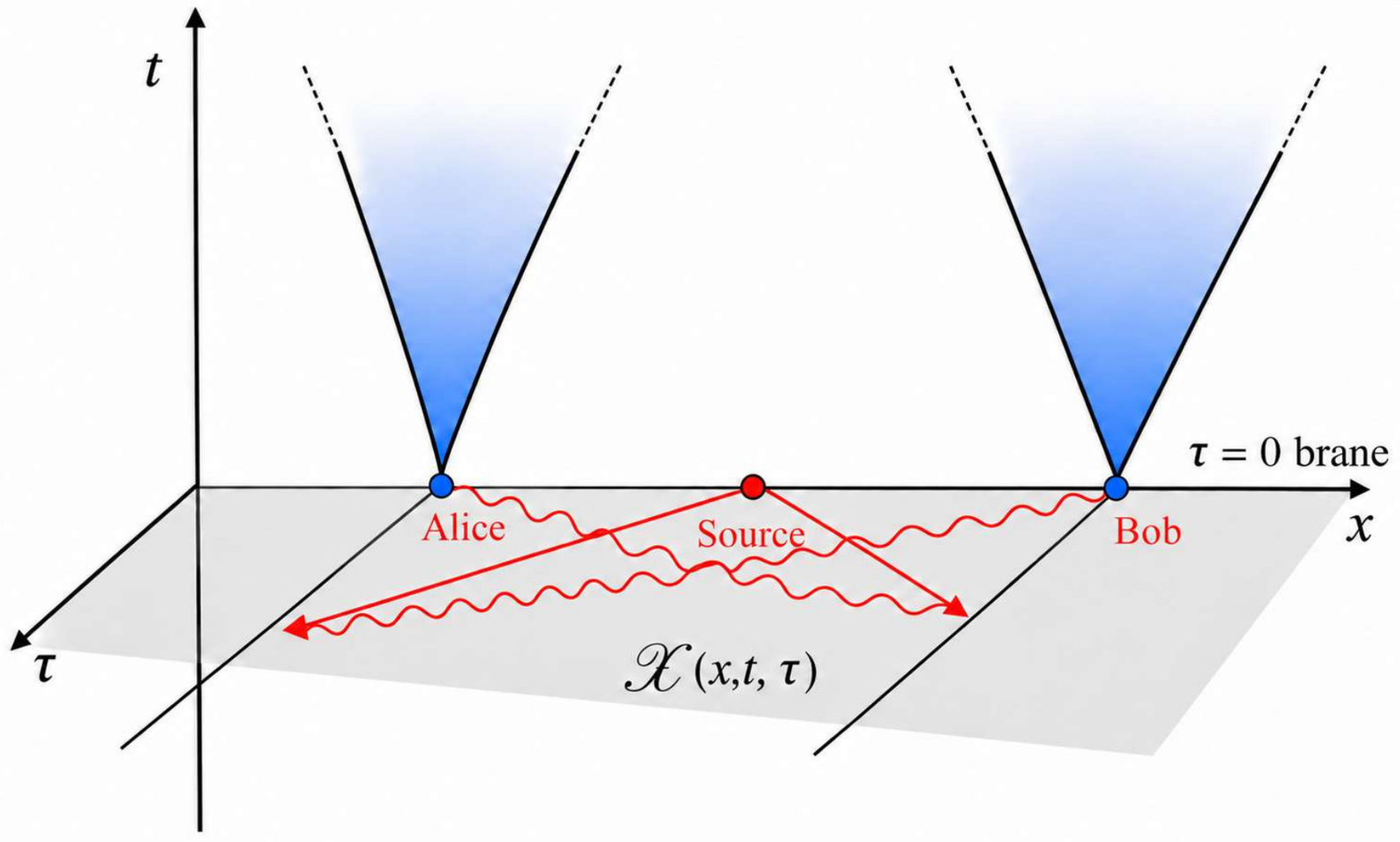}
\vskip -1truecm
  \caption{Schematic representation of the proposed bulk-mediated coupling between distant measurement events in a (3,2)-dimensional spacetime. Alice and Bob are located on the $\tau=0$ brane, where their measurement events are spacelike separated. The source generates a shared contextual configuration of the bulk field $\mathscr{X}(x,t,\tau)$, represented schematically by the straight red arrows. The red wavy lines illustrate the hypothesized reciprocal contributions associated with the two measurement events, mediated through the additional timelike dimension. Their projection onto the brane may connect events at equal ordinary time $t$, without requiring a signal propagating between the detectors within the brane. The diagram is conceptual.}
\label{fig1}
\end{figure}

For a generic brane-delta source the equation is formal: P1
requires zero-mode projection or the compatible distributed
source of Sec.~\ref{subsubsec:X_effective4D}.
Superposition holds for the linear field equation, not automatically
for the subsequent nonlinear detector readout. The preparation term supplies the shared
contextual background inherited from the entangled
source; the measurement term describes the local
detector response in a given run. The two source
terms produce additive contributions to the
contextual inputs $\xi^{(A)}_i$ and $\xi^{(B)}_j$
at each detector; their decomposition into
preparation and measurement components is given in
Eq.~\eqref{eq:xi-B-decomp}.
The physical interpretation of the two-source construction is schematically illustrated in Fig. \ref{fig1}.  The straight red arrows represent the preparation contribution. The wavy lines represent the postulated equal-time measurement coupling, not a source-to-source response derived from the stable field equation.

\paragraph*{Preparation source and the shared
mediator background.}
We represent the preparation of a bipartite
quantum state $|\Psi\rangle$ by the effective
source
\begin{equation}
J_a^{(\rm prep)}(\mathbf{x},t;\Psi,\lambda)
=\mathcal J_a(\mathbf{x};\Psi,\lambda)\,
s_{\rm prep}(t),
\label{eq:Ja-prep-ansatz}
\end{equation}
where $\mathcal J_a$ describes the spatial
and channel structure of the preparation source,
$s_{\rm prep}(t)$ its temporal profile, and
$\lambda$ the run-dependent variables.
The dependence on $|\Psi\rangle$ allows the
source to encode the joint preparation of the
two particles, including their entangled-state
structure. The coupling strength is absorbed
into $J_a^{(\rm prep)}$.

The P1 response, including homogeneous initial
data, is then
\begin{equation}
\begin{aligned}
\mathscr X_a^{(0)}(\mathbf x,t_0,0;\lambda)
={}&\int d^3x'\,dt'\,
\mathcal K_{\rm br}^{(\mathrm{P1})}
(\mathbf x-\mathbf x',t_0-t')\\
&\times J_a^{(\rm prep)}
(\mathbf x',t';\Psi,\lambda)
+\mathscr X_{a,\rm hom}^{(0)}
(\mathbf x,t_0;\lambda),
\end{aligned}
\label{eq:Xa-prep-integral}
\end{equation}
where $\mathcal K_{\rm br}^{(\mathrm{P1})}$ is the
retarded zero-mode brane kernel.
The preparation profile $s_{\rm prep}(t)$
may describe either a localized preparation
pulse or a source acting over a finite time
interval. The mediator configuration at the
detector time $t_0$ also includes the
contribution of the homogeneous initial data.

\paragraph*{Complementarity of the prepared field.}
For a common analyzer basis, we consider the
following complementary condition on the
positive-channel field intensities:
\begin{equation}
|\mathscr{X}_{a,+}(\mathbf{x}_A,t_0,0;\lambda)|^2
+|\mathscr{X}_{a,+}(\mathbf{x}_B,t_0,0;\lambda)|^2
=\mathcal N_a,
\label{eq:anticorr-field}
\end{equation}
where $\mathcal N_a$ is independent of the
run-dependent variables $\lambda$.
Here $\mathscr X_{a,+}$ denotes the
positive-channel component of the mediator
field in the specified basis. This condition
expresses complementary fluctuations of
the field intensities at the two detectors
and provides a possible preparation-level
ingredient for singlet anticorrelation.
For the singlet state,
\begin{equation*}
|\Psi^-\rangle=
\frac{1}{\sqrt{2}}
\left(|+\rangle_A|-\rangle_B
-|-\rangle_A|+\rangle_B\right),
\end{equation*}
the preparation source depends on the joint,
non-factorising state through
$\mathcal J_a(\mathbf{x};\Psi^-,\lambda)$.

For independently oriented polarisation
analysers, the quantum-mechanical joint
probabilities of an ideal singlet are
\begin{equation}
P_{ij}^{(AB)}(\theta_A,\theta_B)
=\frac{1-ij\cos[2(\theta_A-\theta_B)]}{4},
\qquad i,j=\pm1.
\label{eq:singlet-born-target}
\end{equation}
These probabilities provide the benchmark
for the subsequent contextual detector dynamics.

\paragraph*{Assumption (H1): detectors have finite
$\tau$-thickness.}
In the strict brane limit, physical objects
are confined to $\tau=0$. For the proposed
bulk-mediated measurement exchange, we assume
that detector couplings extend over a finite
range in the extra-time direction, described by
\begin{equation}
\eta(\tau)\geq0,
\qquad
\int_{-\infty}^{+\infty}
\eta(\tau)\,d\tau=1,
\label{eq:detector-profile}
\end{equation}
with characteristic width $\delta\tau>0$.
For example, a symmetric Gaussian profile is
\begin{equation*}
\eta(\tau)=
\frac{1}{\sqrt{2\pi}\,\delta\tau}
\exp\left(-\frac{\tau^2}
{2\delta\tau^2}\right).
\end{equation*}
Such a profile allows the detector to couple
to bulk configurations extending beyond
$\tau=0$. In the limit $\delta\tau\to0$,
the profile approaches $\delta(\tau)$,
suppressing the proposed equal-time bulk
exchange between spatially separated
detectors, while leaving preparation-encoded
correlations available to their readout.

\paragraph*{Assumption (H2): measurement interactions
source the bulk field throughout the detector
$\tau$-profile.}
When a particle arrives at detector $k$ at brane time $t_0$,
the measurement interaction sources $\mathscr{X}_a$
throughout the profile $\eta_k(\tau)$ as written in
Eq.~\eqref{eq:full-sourced}.
 H3 restricts the explicit source dependence to the
outcome-channel variables; it does not prove the absence
of setting dependence in their distribution.

\paragraph*{The $E=0$ pulse and its equal-time propagation.}

In the postulated geometric-optics extension, a
measurement at $A$ is associated with a bulk disturbance; no such
spacelike response is generated by the stable projected theory
\eqref{eq:projected-two-source}.
Assumption (P2) models its equal-$t$ exchange by a pulse
associated with the $E=0$ null-geodesic family; the
existence of such a stable source-generated pulse is not
derived from the bulk wave equation.
From the geodesic equations~\eqref{eq:E0_family},
with $\dot{t}=0$, a pulse originating at
$(\mathbf{x}_A,t_0,0)$ reaches the bulk point
\begin{equation}
\bigl(\mathbf{x}_A+\hat{\mathbf n}\,\tfrac{w}{k}\bigl(e^{k\Delta\tau}-1\bigr),\;
t_0,\;\Delta\tau\bigr)
\label{eq:E0-pulse-travel}
\end{equation}
at parameter depth $\Delta\tau$, with the brane time $t$
unchanged along the geometrical ray. This reach relation
is kinematic and does not establish transport by a stable
physical mediator.
The condition for the pulse to reach Bob's detector at $\mathbf{x}_B$, with $\ell=|\mathbf{x}_B-\mathbf{x}_A|$, gives $\Delta\tau=\tau_*(\ell)=k^{-1}\ln(1+k\ell/w)$ [Eq.~\eqref{eq:tau_reach}], and the pulse arrives at bulk depth $\tau_*$ at the same brane time $t_0$ at which it was emitted. For $k\ell/w\ll1$ this reduces to the propagation time $\Delta\tau=\ell/w$ of Ref.~\cite{PRR}, which is thus recovered from the geodesic equations as their near-brane limit.

\paragraph*{The two-component contextual input at Bob's
detector.}
Bob's detector, with profile $\eta_B(\tau)=
(\sqrt{2\pi}\,\delta\tau)^{-1}
\exp(-\tau^2/2\delta\tau^2)$, a Gaussian of width
$\delta\tau$ centred on the brane, probes
the bulk field throughout its $\tau$-extent. The
contextual input driving Bob's collapse is the
$\tau$-weighted overlap
\begin{equation}
\xi^{(B)}_j(t_0;\lambda)
\equiv
\int_{-\infty}^{+\infty}d\tau\;\eta_B(\tau)\;
\mathscr{X}_{a,j}(\mathbf{x}_B,t_0,\tau;\lambda).
\label{eq:xi-B-full}
\end{equation}
Using the field decomposition~\eqref{eq:full-sourced},
the bulk field at Bob's position splits into two
additive contributions:
\begin{eqnarray}
\mathscr{X}_{a,j}(\mathbf{x}_B,t_0,\tau;\lambda)
&=&\underbrace{
\mathscr{X}^{(\rm prep)}_{a,j}(\mathbf{x}_B,t_0,\tau;\lambda)
}_{\text{preparation term}}\nonumber\\
&+&\underbrace{
\mathscr{X}^{(A\to B)}_{a,j}(\mathbf{x}_B,t_0,\tau;\lambda)
}_{\text{postulated measurement contribution}},
\label{eq:field-decomp}
\end{eqnarray}
Here $\mathscr{X}^{(A\to B)}$ denotes the source-induced
measurement contribution, not a separately normalised field.
We parameterise its detector overlap using an effective
channel amplitude $\xi_j^{(B,\rm meas)}$ and a dimensionless
suppression factor $\varepsilon_{\rm meas}(\ell)$; the
factorisation of the overlap is a narrow-pulse modelling
assumption rather than an exact consequence of the bulk
field equation.

The following profile-overlap estimate is a
phenomenological part of P2 and must not be read as the response of
the zero-mode equation~\eqref{eq:projected-two-source}.
The two terms are treated differently on physical grounds.
The measurement pulse is normalised so that
$\varepsilon_{\rm meas}(\ell)$, the bulk-travel
suppression factor defined below in
Eq.~\eqref{eq:eps-meas}, alone controls its
weight in the denominator of $\widetilde{R}^{(B)}_j$;
normalising the preparation term independently would
introduce a second free scale and destroy the
physically meaningful competition between the two
contributions. The preparation term therefore retains
its source-dependent amplitude at
$r_{\rm prep}=|\mathbf{x}_B-\mathbf{x}_{\rm prep}|$.
If the channel amplitudes share a common distance envelope,
that envelope does not affect their channel-to-channel
ratio $|\mathscr{X}^{(\rm prep)}_{a,+}|/
|\mathscr{X}^{(\rm prep)}_{a,-}|$,
which varies with $\lambda$ and carries the
entanglement structure~\eqref{eq:anticorr-field};
only the overall scale of the preparation term
decays with distance.
Substituting into~\eqref{eq:xi-B-full} gives
\begin{eqnarray}
\xi^{(B)}_j
&=&\underbrace{
\int d\tau\;\eta_B(\tau)\;
\mathscr{X}^{(\rm prep)}_{a,j}
(\mathbf{x}_B,t_0,\tau;\lambda)
}_{\xi^{(B,\rm prep)}_j}\nonumber\\
&+&\underbrace{
\int d\tau\;\eta_B(\tau)\;
\mathscr{X}^{(A\to B)}_{a,j}
(\mathbf{x}_B,t_0,\tau;\lambda)
}_{\varepsilon_{\rm meas}(\ell)\,\xi^{(B,\rm meas)}_j}.
\label{eq:xi-B-decomp}
\end{eqnarray}
Alice's collapse ratio follows by the identical
construction with $A\leftrightarrow B$; the framework
is symmetric under this exchange.

\paragraph*{The two components evaluated.}

The preparation and measurement contributions
have different physical origins. Under P1,
the preparation field is represented by its
$\tau$-independent zero mode. For
$\delta\tau\ll 1/k$, its detector overlap is
therefore well approximated by the brane value,
$\xi^{(B,\rm prep)}_j\approx
\mathscr{X}^{(\rm prep)}_{a,j}
(\mathbf{x}_B,t_0,0;\lambda)$.
For an ideal normalized detector profile, this
reduction is exact within the zero-mode sector.

The measurement contribution
$\xi_j^{(B,\rm meas)}$, instead, describes
Bob's coupling to the bulk disturbance
generated by Alice. At Bob's spatial position,
the corresponding equal-time null trajectory
reaches the extra-time depth
$\tau_*(\ell)=k^{-1}\ln(1+k\ell/w)$.

The profile $\eta_B(\tau)$ describes Bob's
sensitivity to the incoming field at depth
$\tau$. If the pulse is narrow compared with
the variation scale of this profile, its
overlap with the detector is proportional
to $\eta_B(\tau_*)$. For the Gaussian profile,
normalization to $\eta_B(0)$ gives the
dimensionless suppression factor
\begin{equation}
\varepsilon_{\rm meas}(\ell)
\equiv
\exp\!\left(
-\frac{[\ln(1+k\ell/w)]^2}
{2k^2\delta\tau^2}
\right).
\label{eq:eps-meas}
\end{equation}

This factor decreases from unity at $\ell=0$
toward zero as the separation increases.
A characteristic exchange length is defined
by $\tau_*(\ell_{\rm exch})=\delta\tau$, giving
$\ell_{\rm exch}=(w/k)(e^{k\delta\tau}-1)$.
For $k\delta\tau\ll1$, this reduces to
$\ell_{\rm exch}\simeq w\delta\tau$, and
Eq.~\eqref{eq:eps-meas} becomes
$\exp(-\ell^2/2w^2\delta\tau^2)$.

More generally, the suppression is Gaussian
in $\ln(1+k\ell/w)$ rather than in $\ell$.
For $k\delta\tau\gtrsim1$, its variation near
$\ell_{\rm exch}$ can resemble a weak power law,
with a gradual crossover rather than a sharp
cutoff.

The Gaussian profile is a convenient explicit
choice. Other localized detector profiles lead
to analogous suppression when
$\tau_*(\ell)\gg\delta\tau$, although the precise
distance dependence changes.

The exchange length depends exponentially on
$k\delta\tau$. For illustration, taking
$w/k=1$\,mm and $k\delta\tau=5,10,15$
gives respectively
$\ell_{\rm exch}\approx0.15$\,m, $22$\,m,
and $3.3$\,km.
Here $\delta\tau$ characterizes the detector's
extension in the extra-time direction; it is
distinct from the ordinary-time coincidence
window and would ultimately be determined by
the microscopic detector coupling.

The two contributions enter the contextual
ratio at Bob's detector as
\begin{equation}
\widetilde{R}^{(B)}_j(t_0;\lambda)
=\frac{|\psi^{(B)}_j(t_0)|^2}
      {\bigl|\xi^{(B,\rm prep)}_j(t_0;\lambda)
       +\varepsilon_{\rm meas}(\ell)\,
        \xi^{(B,\rm meas)}_j(t_0;\lambda)\bigr|^2}.
\label{eq:R-full-B}
\end{equation}

This is the raw, unnormalised ratio.
Defining
$\xi^{(B)}_j=\xi_j^{(B,\rm prep)}+
\varepsilon_{\rm meas}\xi_j^{(B,\rm meas)}$
and $N_B^2=\sum_j|\xi_j^{(B)}|^2>0$,
the corresponding normalized Bohm--Bub
ratios satisfy
\begin{equation}
\widetilde R^{(B),\rm norm}_j
= N_B^2\widetilde R^{(B),\rm raw}_j.
\label{eq:raw-norm-relation}
\end{equation}

Normalization multiplies all channel ratios
and their differences by the same positive
factor. It therefore preserves their ordering
and the selected outcome channel, while
rescaling the collapse rate within a run.
Finite-time outcome statistics also depend
on the collapse-time prescription.

\paragraph*{Non-vanishing of the denominator.}

The denominator of
$\widetilde R^{(B)}_j$ must remain nonzero.
Exact cancellation between preparation and
measurement contributions is excluded for
runs in which the preparation term strictly
dominates and is itself nonzero.
Possible field nodes must be handled by
the contextual ensemble, which is taken
to assign zero probability to singular
denominators.

\paragraph*{The preparation-dominance condition.}

We assume the hierarchy
\begin{equation}
|\mathscr{X}^{(\rm prep)}_{a,j}
(\mathbf{x}_B,t_0,0;\lambda)|
\;\gg\;
\varepsilon_{\rm meas}(\ell)\,
|\xi_j^{(B,\rm meas)}|,
\label{eq:prep-dominance}
\end{equation}
with finite, dimensionally consistent
channel amplitudes.

This condition expresses the physical
assumption that the preparation field is
stronger than the detector-generated
contribution, including its transverse
overlap suppression. The exchange term
then acts as a perturbation of the
preparation-controlled collapse dynamics.

For a Bell experiment with the preparation
source approximately midway between the
detectors, $r_{\rm prep}\approx\ell/2$.
The actual hierarchy depends on the
preparation-source profile and detector
couplings. In particular, the ensemble
covariance in Eq.~\eqref{eq:C0_powerlaw}
does not determine the amplitude of an
individual field realization.

Cancellation of a common overall field
amplitude is implemented through the
channel normalization of
Eq.~\eqref{eq:channel-norm}. The ensemble
must also provide the appropriate Born
statistics, while operational no-signaling
is required by P3.

\paragraph*{Two dynamical regimes.}

The separation dependence of
$\varepsilon_{\rm meas}(\ell)$ identifies
two regimes.

For $\ell\lesssim\ell_{\rm exch}$, the
detector overlap is appreciable and the
measurement contribution can influence
the contextual ratios. Under
Eq.~\eqref{eq:prep-dominance}, this
influence remains perturbative.
Its observable consequences depend on
the detector readout and the contextual
ensemble, particularly for correlations
between independently prepared pairs.

For $\ell\gg\ell_{\rm exch}$, the
measurement contribution becomes strongly
suppressed, and the ratio reduces to
\begin{equation}
\widetilde{R}^{(B)}_j(t_0;\lambda)
\;\approx\;
\frac{|\psi^{(B)}_j(t_0)|^2}
     {|\xi^{(B,\rm prep)}_j(t_0;\lambda)|^2}.
\label{eq:R-prep-dominated}
\end{equation}

In this preparation-dominated regime, the
contextual evolution approaches the original
Bohm--Bub structure. Born statistics are
recovered when the preparation ensemble
provides the corresponding Born basin
weights. The complementary field condition
\eqref{eq:anticorr-field} can contribute to
the required pairwise anticorrelation, in
conjunction with the complete channel
relations of the detector dynamics.

Thus, suppressing the measurement exchange
leaves the preparation-controlled mechanism
as the dominant contribution. The separation
at which this occurs depends on
$k\delta\tau$ and the detector profile.

\paragraph*{Relation to the standard quantum prediction.}

At separations
$\ell\lesssim\ell_{\rm exch}$, the additional
measurement contribution may modify the
collapse dynamics. The model is intended
to preserve standard single-pair Bell
statistics through an appropriate
contextual equilibrium.

Its proposed experimentally distinguishable
effect is instead a cross-pair correlation
between two independently prepared Bell
pairs, absent for independent pairs in
standard quantum mechanics
(Sec.~\ref{subsec:proposal-C-nonzero}).

The range $\ell\lesssim\ell_{\rm exch}$
is particularly relevant experimentally,
because the measurement-exchange contribution
is least suppressed there. Its extent can
vary substantially with $k\delta\tau$.

\paragraph*{Assumption (H3): outcome-channel
dependence of the measurement source.}

The measurement source is taken to encode
local outcome-channel amplitudes without
an independent explicit dependence on the
freely chosen analyzer setting. For Alice,
this assumption reads
\begin{equation}
J^{(A,\rm meas)}_a(\mathbf{x},t;\lambda)
=J^{(A,\rm meas)}_a\bigl(\mathbf{x},t;\,
\xi^{(A)}_\pm(\lambda)\bigr),
\label{eq:H3}
\end{equation}
where $\xi^{(A)}_\pm(\lambda)$ vary across
individual realizations.

Although the source has no independent
explicit dependence on $\theta_A$, the
channel amplitudes themselves may depend
statistically on the analyzer setting.
Consequently, H3 must be supplemented by
the operational no-signaling requirement
P3 for the complete multipartite
outcome statistics.

\paragraph*{Roles of the two mechanisms and
singlet anticorrelation.}

The preparation and measurement contributions
play complementary roles. The preparation
source~\eqref{eq:Ja-prep-ansatz} establishes
the shared contextual background, while
measurement-generated disturbances provide
the proposed exchange contribution.

The field-amplitude condition
\eqref{eq:anticorr-field} describes
complementary positive-channel intensities.
The actual detector outcomes depend on
both positive- and negative-channel
amplitudes through their Bohm--Bub ratios.

For equal analyzer bases, a necessary
ordering condition for deterministic
opposite-channel selection, away from
ties, is
\begin{equation}
(\widetilde R^{(A)}_+-\widetilde R^{(A)}_-)
(\widetilde R^{(B)}_+-\widetilde R^{(B)}_-)<0.
\label{eq:singlet-ordering}
\end{equation}

This condition connects the complementary
preparation field to the ordering of the
detector readout channels. For independently
oriented analyzers, the full joint-channel
statistics must recover the singlet
probabilities of
Eq.~\eqref{eq:singlet-born-target}.

In the contextual measurement-exchange
hypothesis (P2), contributions from Alice
and Bob enter both readouts at the same
brane time. Their combined outcome
statistics are subject to the operational
no-signaling condition P3.

%----------------------------------------------------------------------
\subsection{Contextuality, Born probabilities,
and $\mathscr{X}_a(x,\tau)$}
\label{subsec:contextuality-lambda}

\paragraph*{Born weights in the Bohm--Bub framework.}
In the BB model the nonlinear collapse equations depend
on the ratios $R_i=|\psi_i|^2/|\xi_i|^2$, where the
auxiliary vector $\xi$ varies from run to run. Bohm
and Bub showed that averaging over the ensemble of
$\xi$ recovers the Born rule $P_i=|\psi_i|^2$~\cite{BB}.
Probabilities do not therefore enter as a postulate at
the level of a single run; they emerge from the
run-to-run fluctuations of $\xi$.

\paragraph*{Contextuality from the bulk field.}
The Kochen--Specker theorem motivates the importance
of contextuality, although it does not by itself derive
the collapse parameters used here~\cite{KS}. In the present framework this
context dependence is encoded in the bulk field
$\mathscr{X}_a$, shaped by the global experimental
arrangement. The relevant contextual information is
coarse-grained into a single effective parameter
$\lambda$ (in standard Bell notation),
\begin{equation}
\lambda\equiv\lambda\!\left[\mathscr{X}_a\right]
\quad(\text{evaluated at }\tau=0).
\end{equation}
For fixed $\lambda$ the collapse is deterministic;
statistical predictions arise upon averaging:
\begin{equation}
P_i=\int_{\Lambda_i}\rho(\lambda)\,d\lambda,
\end{equation}
where $\Lambda_i$ is the basin leading to outcome $i$
and $\rho(\lambda)$ is the distribution of contextual
microstates. In Appendix~\ref{app:roles_X_lambda} we
exhibit a minimal drift--diffusion model in which
$\rho$ relaxes to a unique stationary distribution
reproducing $P_i=|\psi_i|^2$.

\paragraph*{Structure of $\Lambda_i$, Born compatibility,
and the Born rule.}
The basin $\Lambda_i$ is the set of microstates
$\lambda$ for which the BB dynamics drives the
wavefunction to outcome $i$; its boundary is
determined by the ordering of the ratios
$\widetilde{R}^{(k)}_i(\lambda)$. The distribution
$\rho(\lambda)$ is \emph{Born-compatible} at the measurement onset if
\begin{equation*}
\int_{\Lambda_i}\rho(\lambda)\,d\lambda=|\psi_i(t_0)|^2
\end{equation*}
for every measurement basis and every outcome $i$:
the ensemble weight of each basin must match the
corresponding Born probability. The Born rule then
follows from averaging the deterministic outcome over
the Born-compatible ensemble; preservation of that
ensemble under a physical evolution remains open.

Equivariance is not a freely adjustable assumption; it
singles out the class of distributions $\rho(\lambda)$
compatible with quantum predictions. Whether generic initial distributions relax
dynamically toward this class is the bipartite
analogue of a result due to
Valentini~\cite{valentini1991,valentini2005}, that is, 
arbitrary sub-quantum distributions
$\rho(\lambda)\neq|\psi|^2$ are driven toward the
Born-rule distribution by the guidance dynamics,
in analogy with Boltzmann's $H$-theorem for
classical gases. This remains an open question here; it is
identified in Sec.~\ref{discussion}, and the Appendix exhibits
an illustrative stationary ensemble with prescribed
Born basin weights for a fixed measurement context.

\paragraph*{Consistency of the channel normalisation
with equivariance.}
The ordering $\widetilde{R}^{(k)}_i>\widetilde{R}^{(k)}_m$
determining the basin boundary is equivalent to
$|\psi^{(k)}_i|^2/|\mathscr{X}^{(k)}_{a,i}|^2>
|\psi^{(k)}_m|^2/|\mathscr{X}^{(k)}_{a,m}|^2$,
from which the common factor $|\mathscr{X}^{(k)}_a|^2$
cancels; hence $\Lambda_i^{\hat{\mathscr{X}}}
=\Lambda_i^{\mathscr{X}}$ and the fixed-context Born-basin condition
is unaffected by a common channel normalisation.
The collapse-driving differences
$\widetilde{R}^{(k)}_i-\widetilde{R}^{(k)}_m$ are
rescaled by the positive factor
$|\mathscr{X}^{(k)}_a|^2$, which changes the collapse timescale but not the
ratio ordering or deterministic winning channel;
no statement about all dynamical observables follows.

\paragraph*{Distinct roles of $\mathscr{X}_a$ and $\lambda$.}
The bulk field $\mathscr{X}_a$ is the physical mediator
entering the deterministic nonlinear collapse equations.
The parameter $\lambda$ is a coarse-grained label for
the run-dependent microstate of $\mathscr{X}_a$; it 
determines which collapse trajectory is realised in a
given run and, through $\rho(\lambda)$, fixes the
outcome statistics.
%======================================================================
\subsection{{ Continuity with the earlier toy model}}
\label{subsec:PRR-continuity}

With the bipartite BB model and the two-source bulk
mechanism fully established, we now make explicit the
continuity with Ref.~\cite{PRR}, identifying which
elements of that earlier construction were ad hoc
assumptions and which are here derived.

\paragraph*{{ Earlier collapse equations.}}
{ In the earlier formulation, the collapse equations for two}
spacelike-separated detectors at positions
$\mathbf{x}_1$ (Alice) and $\mathbf{x}_2$ (Bob)
read [Eqs.~(23)--(26) of Ref.~\cite{PRR}]:
\begin{align}
\frac{d\psi^{(1)}_{+}(t)}{dt}
&=\gamma(t)\bigl(R^{(1)}_{+}-R^{(1)}_{-}
               +\widetilde{R}^{(2)}_{-}-\widetilde{R}^{(2)}_{+}\bigr)
  \psi^{(1)}_{+}(t)\,|\psi^{(1)}_{-}(t)|^2,
\nonumber\\
\frac{d\psi^{(1)}_{-}(t)}{dt}
&=\gamma(t)\bigl(R^{(1)}_{-}-R^{(1)}_{+}
               +\widetilde{R}^{(2)}_{+}-\widetilde{R}^{(2)}_{-}\bigr)
  \psi^{(1)}_{-}(t)\,|\psi^{(1)}_{+}(t)|^2,
\label{eq:PRR-23-26}
\end{align}
and their $(-)$ counterparts, with
\begin{align}
R^{(k)}_\pm
&=\frac{|\psi^{(k)}_\pm|^2}
       {|\mathscr{X}_{a,\pm}(\mathbf{x}_k,t_0,\tau_0+\Delta\tau)|^2},
\label{eq:PRR-28}
\\[4pt]
\widetilde{R}^{(1)}_\pm
&=\frac{|\psi^{(1)}_\pm|^2}
       {|\mathscr{X}_{a,\pm}(\mathbf{x}_2-\mathbf{w}\Delta\tau,
         t_0,\tau_0+\Delta\tau)|^2},
\label{eq:PRR-27a}
\\[4pt]
\widetilde{R}^{(2)}_\pm
&=\frac{|\psi^{(2)}_\pm|^2}
       {|\mathscr{X}_{a,\pm}(\mathbf{x}_1-\mathbf{w}\Delta\tau,
         t_0,\tau_0+\Delta\tau)|^2},
\label{eq:PRR-27b}
\end{align}
and, in the earlier flat-reach approximation,
$\Delta\tau=|\mathbf{x}_2-\mathbf{x}_1|/|\mathbf{w}|$.
The complementary field condition
\eqref{eq:anticorr-field} provides a possible
preparation-level contribution to anticorrelated
detector readouts. In the present two-channel
dynamics, the term
$\widetilde{R}^{(2)}_-
-\widetilde{R}^{(2)}_+$
drives Alice's contextual evolution, while
the corresponding
$\widetilde{R}^{(1)}_\pm$ terms enter Bob's.
The resulting outcome correlations depend on
the complete channel-amplitude configuration
and its evolution.

For comparison with Ref.~\cite{PRR}, the qubit
reduction of Eq.~\eqref{eq:BB-entgl} is
illustrated for $i=+$, $k=1$, with the
Hamiltonian term omitted. This reduction
establishes the two-channel algebraic form,
while the separate-ratio decomposition is
discussed below.

Setting $d_1=2$ and expanding $\sum_{m\in\{+,-\}}$ explicitly,
\begin{equation*}
\frac{d\psi^{(1)}_{+}}{dt}
= \gamma\,\psi^{(1)}_{+}
  \Bigl[
    \underbrace{|\psi^{(1)}_{+}|^{2}
      \bigl(\widetilde{R}^{(1)}_{+}-\widetilde{R}^{(1)}_{+}\bigr)}_{=\,0}
  + |\psi^{(1)}_{-}|^{2}
      \bigl(\widetilde{R}^{(1)}_{+}-\widetilde{R}^{(1)}_{-}\bigr)
  \Bigr].
\end{equation*}
The $m=i$ term vanishes identically; the remaining term brings 
the factor $|\psi^{(1)}_{-}|^{2}$ visible in { the earlier equation.}
 
In the general equation~\eqref{eq:BB-entgl} the collapse driver
for station~$1$ involves only $\widetilde{R}^{(1)}_\pm$, defined
via the channel-normalised crossed projection
$|\hat{\mathscr{X}}^{(2\to 1)}_{a,\pm}|^{2}$
[Eq.~\eqref{eq:R-general-PRD}].
{ The earlier equations}~\eqref{eq:PRR-23-26} involve instead two
distinct ratio families, $R^{(1)}_\pm$ and $\widetilde{R}^{(2)}_\pm$,
defined via the bare field $\mathscr{X}_{a,\pm}$ evaluated at two
different spacetime points as shown in Eqs.~\eqref{eq:PRR-28}--\eqref{eq:PRR-27b}.

The present contextual input does contain additive
preparation and measurement \emph{field amplitudes}
[Eq.~\eqref{eq:xi-B-decomp}]. However, the collapse
ratio involves the inverse squared modulus of their sum:
\begin{equation}
\widetilde R^{(B)}_j
=\frac{|\psi^{(B)}_j|^2}
{|\xi^{(B,\rm prep)}_j+\varepsilon_{\rm meas}
\xi^{(B,\rm meas)}_j|^2}.
\end{equation}
This is not the sum of two collapse ratios. The two
ratio families used in Ref.~\cite{PRR} can therefore be
related to the preparation- and exchange-dependent inputs
only as a \emph{phenomenological correspondence}, not as
an exact algebraic reduction of Eq.~\eqref{eq:BB-entgl}.
In particular, the earlier crossed collapse equations
are recovered only if an additional effective approximation
for the complete ratio differences is specified and checked.
No such approximation is established here.

\paragraph*{Correspondences.}
 The earlier formulation used two phenomenological inputs:
\textit{(a)} the field propagates at finite velocity
$\mathbf{w}$ in $\tau$ but at infinite velocity in
ordinary $(3,1)$ spacetime; and
\textit{(b)} the collapse-sourcing term
$F[\mathbf{x},\psi^{(j)}_\pm,\tau;t_0,\tau_0]$ in the
field equation was left unspecified.
 Here the first receives a geometric reach law; the second
is specified by explicit, though not microscopically derived, sources.

\textit{i)} The $E=0$ null-geodesic reach in the warped
metric is $\tau_*(d)=k^{-1}\ln(1+kd/w)$
[Eq.~\eqref{eq:tau_reach}], reducing to $d/w$ only
for $kd/w\ll1$. It is a kinematic property of the
null cone; the stability and no-signaling of a physical
measurement-induced exchange are not implied by it.

\textit{ii)} { The earlier cross-ratios} $\widetilde{R}^{(1,2)}_\pm$
correspond to $\widetilde{R}^{(B,A)}_j|_{\text{zero mode, meas}}$
of Eq.~\eqref{eq:R-full-B}, with
$\varepsilon_{\rm meas}\,\xi^{(B,\rm meas)}_j$ playing
the role of
$\mathscr{X}_{a,\pm}(\mathbf{x}_{1,2}-\mathbf{w}\Delta\tau,
t_0,\tau_0+\Delta\tau)$.
The proposed $\varepsilon_{\rm meas}(d)$ is obtained
by evaluating the detector profile at
$\tau_*(d)=k^{-1}\ln(1+kd/w)$ under the narrow-pulse
approximation; it is not a derived stable bulk propagator.

\textit{iii)} The local ratios $R^{(k)}_\pm$ { of the earlier formulation} correspond
to the zero-mode brane projection
$\widetilde{R}^{(k)}_\pm|_{\text{zero mode}}
=|\psi^{(k)}_\pm|^2/|\hat{\mathscr{X}}^{(k)}_{a,\pm}|^2$.

\textit{iv)} Assumption~$(b)$ { of the earlier formulation} is replaced by the bulk
wave equation~\eqref{eq:full-sourced} with sources
$J_a^{(\rm prep)}$ and $J_a^{(k,\rm meas)}$ given
in~\eqref{eq:Ja-prep-ansatz} and~\eqref{eq:H3}.

\textit{v)} The ordinary zero-mode retarded response
is brane-causal. Operational no-signaling for the separate
contextual measurement exchange does \emph{not} follow
from (H3) and must satisfy the full condition (P3).

The present formulation retains the geometric
motivation and provides a conditional contextual
extension of the earlier effective collapse
structure in the regime
$\ell\lesssim\ell_{\rm exch}$.
The complementary field condition
\eqref{eq:anticorr-field} can be incorporated
into the preparation ensemble, while the
two-channel detector dynamics determines
the resulting outcome correlations.
Operational no-signaling remains the
independent requirement P3.

%----------------------------------------------------------------------
\section{Photon-pair specialisation and cross-pair prediction}
\label{sec:dynmodel}
The general bipartite BB framework and the two-source
bulk mechanism were developed in
Secs.~\ref{sec:bipartite-BB}--\ref{subsec:meas-exchange}.
This section specialises them to a photon pair,
 introduces the effective contextual field
$\mathscr{Y}^{(k)}_a$ as the $\tau$-weighted detector input,
and formulates a conditional cross-pair experimental test.
%----------------------------------------------------------------------
\subsection{Single entangled photon pair and its
measurement basis}
\label{subsec:single-pair}

Consider two photons $A$ and $B$, each in the
two-dimensional Hilbert space
$\mathcal{H}=\mathrm{span}\{|H\rangle,|V\rangle\}$,
prepared in the singlet Bell state
\begin{equation}
|\psi_{\rm Bell}\rangle
=\frac{1}{\sqrt{2}}
\bigl(|H\rangle_A|V\rangle_B-|V\rangle_A|H\rangle_B\bigr).
\end{equation}
Local measurements are performed along analyser
orientations $\theta_A$ and $\theta_B$, with eigenstates
\begin{align}
|+\rangle_\theta
&=\cos\theta\,|H\rangle+\sin\theta\,|V\rangle,
\\
|-\rangle_\theta
&=-\sin\theta\,|H\rangle+\cos\theta\,|V\rangle.
\end{align}
At the onset of measurement the wavefunction is
\begin{equation}
|\psi(t_0)\rangle
=\sum_{i,j=\pm}\psi_{ij}(t_0)\,
|i\rangle_{\theta_A}|j\rangle_{\theta_B},
\end{equation}
where $\psi_{ij}(t_0)=
{}_{\theta_A}\langle i|\,{}_{\theta_B}\langle j|
\psi_{\rm Bell}\rangle$ are the singlet projections
onto the local eigenstates.

%----------------------------------------------------------------------
\subsection{Effective contextual field and detector-channel projections}
\label{subsec:bulk-field-detector}

The general framework of Sec.~\ref{subsec:meas-exchange}
defines the contextual input $\xi^{(k)}_j$ at each
detector as the $\tau$-weighted bulk-field
overlap~\eqref{eq:xi-B-full}, with the two-component
decomposition~\eqref{eq:xi-B-decomp}.
For the photon-pair specialisation it is convenient to
introduce the \emph{effective contextual field}
\begin{equation}
\mathscr{Y}^{(k)}_a(\mathbf{x},t_0;\lambda)
\equiv
\int d\tau\;\eta_k(\tau)\;
\mathscr{X}_a(\mathbf{x},t_0,\tau;\lambda),
\label{eq:Y-def-full}
\end{equation}
which is linear in the individual bulk-field components;
internal norms are taken only after channel projection.
In the brane limit we have $\eta_k\to\delta(\tau)$,
$\mathscr{Y}^{(k)}_a\to\mathscr{X}_a(\mathbf{x},t_0,0;\lambda)$ 
and only the preparation-encoded mechanism survives.
The detector-channel projections are then
\begin{align}
\mathscr{X}^{(A)}_{a,i}(t_0;\lambda)
&=\int_{\Omega_A}d^3x\;
  u^{(A)*}_i(\mathbf{x})\,
  \mathscr{Y}^{(A)}_a(\mathbf{x},t_0;\lambda),
\label{eq:XA-full}
\\
\mathscr{X}^{(B)}_{a,j}(t_0;\lambda)
&=\int_{\Omega_B}d^3x\;
  u^{(B)*}_j(\mathbf{x})\,
  \mathscr{Y}^{(B)}_a(\mathbf{x},t_0;\lambda),
\label{eq:XB-full}
\end{align}
where $u^{(k)}_i(\mathbf{x})$ are the detector-basis
channel profiles of Eq.~\eqref{eq:detector-channel-profiles}.
 Under P2, these projections sample the
preparation field and the hypothesized
measurement-exchange contribution.

The channel-normalised intensity amplitudes are
$\hat{\mathscr{X}}^{(k)}_i\equiv
\left(\delta^{ab}\mathscr{X}^{(k)}_{a,i}
\mathscr{X}^{(k)*}_{b,i}\right)^{1/2}/|\mathscr{X}^{(k)}|$,
with $|\mathscr{X}^{(k)}|^2=\sum_i\delta^{ab}
\mathscr{X}^{(k)}_{a,i}\mathscr{X}^{(k)*}_{b,i}$,
Under the common-envelope assumption of
Sec.~\ref{sec:bipartite-BB}, the common distance factor
cancels from these normalized channel amplitudes.
 These amplitudes use the channel norm of
Eq.~\eqref{eq:channel-norm}. The older crossed labels $2\to1$ and $1\to2$
refer to \emph{where} a shared preparation field is projected,
not to two instantaneous one-wing propagators. In the strict brane
limit $\eta_k\to\delta(\tau)$ the station-$A,B$ projections
agree with Eqs.~\eqref{eq:Xproj-2to1} and
\eqref{eq:Xproj-general}, respectively. For finite detector
profiles, Eqs.~\eqref{eq:XA-full}--\eqref{eq:XB-full}
\emph{replace} those brane-only projections; literal equality
between the two overlap integrals is not assumed.
The effective crossed collapse ratios for the
photon-pair specialisation are then
\begin{equation}
\widetilde{R}^{(A)}_i
=
\frac{|\psi^{(A)}_i|^2}
     {|\hat{\mathscr{X}}^{(A)}_i|^2},
\qquad
\widetilde{R}^{(B)}_j
=
\frac{|\psi^{(B)}_j|^2}
     {|\hat{\mathscr{X}}^{(B)}_j|^2},
\label{eq:Rij-physical-unified}
\end{equation}
The photon-pair collapse equation is obtained from
Eq.~\eqref{eq:BB-entgl} by taking $k\in\{A,B\}$
and $d_k=2$.

\noindent\textit{Remark.} The retarded zero-mode response
is brane-causal; multipartite no-signaling for the
postulated contextual readout remains condition (P3)
(Sec.~\ref{item:D}).

%----------------------------------------------------------------------
\subsection{Two independent Bell pairs}
\label{subsec:two-pairs}
\paragraph*{(P4) Conditional experimental signature.}
We now consider a possible experimental signature
of the contextual bulk mediator: correlations
between two independently prepared Bell pairs.
Unlike an ordinary Bell test, which probes
correlations within an entangled pair, the
proposed experiment searches for correlations
between initially independent pairs. Their
magnitude and distance dependence are governed
by the contextual field ensemble and detector
couplings.

Consider two independent sources emitting
the Bell pairs $(A,B)$ and $(E,T)$, denoted
pair~1 (Alice and Bob) and pair~2 (Eve and Tom).
Their initial quantum state is a product:
\begin{equation}
|\Psi(t_0)\rangle
=|\psi^{(AB)}_{\rm Bell}\rangle
\otimes|\psi^{(ET)}_{\rm Bell}\rangle.
\end{equation}
In the product measurement basis, the general
four-party state is written
\begin{equation}
|\Psi(t)\rangle
=\sum_{i,j,k,l=\pm}
\Psi_{ijkl}(t)\,
|i\rangle_{\theta_A}|j\rangle_{\theta_B}
|k\rangle_{\theta_E}|l\rangle_{\theta_T},
\label{eq:Psi-four-index-expansion}
\end{equation}
where the four indices label the respective
detector channels.

Each detector samples the effective contextual
field $\mathscr{Y}^{(k)}_a$ defined previously.
For Alice and Bob, the channel amplitudes
follow from the spatial and basis projections
in Eqs.~\eqref{eq:XA-full}--\eqref{eq:XB-full};
the same construction applies to Eve and Tom.
In particular, the Bob amplitudes are obtained
from Eq.~\eqref{eq:XB-full}.

The two preparation ensembles are independent
in the absence of contextual coupling. To
describe the detector readout, we introduce
a simple pair-level map formed from products
of the normalized single-detector amplitudes:
\begin{align}
\hat{\mathscr{X}}^{(1)}_{ij}
&\equiv
\hat{\mathscr{X}}^{(A)}_i\,
\hat{\mathscr{X}}^{(B)}_j,
\label{eq:X1-projected-two-pairs}
\\
\hat{\mathscr{X}}^{(2)}_{kl}
&\equiv
\hat{\mathscr{X}}^{(E)}_k\,
\hat{\mathscr{X}}^{(T)}_l.
\label{eq:X2-projected-two-pairs}
\end{align}

These products preserve normalization:
if each single-detector amplitude has unit
summed squared modulus, so does the
corresponding pair-level amplitude.
They define the contextual readout used
in the four-index Bohm--Bub evolution
\eqref{eq:BB-two}.

The linear projections of
Eqs.~\eqref{eq:Xproj-AB}--\eqref{eq:Xproj-ET}
serve instead as overlap diagnostics; they
do not replace the product amplitudes in
the collapse denominators. The choice
of detector map is physically relevant
because it also determines how the
outcome correlations depend on the
analyzer orientations.

For each pair, the contextual ratios are
\begin{equation}
\widetilde{R}^{(p)}_{ij}
=\frac{|\psi^{(p)}_{ij}|^2}
      {|\hat{\mathscr{X}}^{(p)}_{ij}|^2},
\qquad p=1,2,
\end{equation}
where the indices $ij$ stand for the two
channels of the corresponding pair.

The four-party contextual drive is then
\begin{equation}
\widetilde{R}_{ijkl}(t_0;\lambda)
\equiv
\widetilde{R}^{(1)}_{ij}(t_0;\lambda_1)
+\widetilde{R}^{(2)}_{kl}(t_0;\lambda_2)
+\Delta_{ijkl}(t_0;\lambda),
\label{eq:Rijkl-def}
\end{equation}
where $\Delta_{ijkl}$ represents the
additional contextual coupling between
the two pairs. It vanishes when their
contextual configurations are independent,
with $\lambda=(\lambda_1,\lambda_2)$
distributed according to a factorized
ensemble.

The corresponding four-index collapse
equation becomes
\begin{equation}
\frac{d\Psi_{ijkl}}{dt}
=\gamma\,\Psi_{ijkl}
\sum_{i',j',k',l'}|\Psi_{i'j'k'l'}|^2
\bigl(\widetilde{R}_{ijkl}
     -\widetilde{R}_{i'j'k'l'}\bigr),
\label{eq:BB-two}
\end{equation}
with $\widetilde{R}_{ijkl}$ given by
Eq.~\eqref{eq:Rijkl-def}.

%----------------------------------------------------------------------

\subsection{Factorisation vs.\ induced cross-pair
correlations}
\label{subsec:factorization}

The central question is whether the
four-index dynamics~\eqref{eq:BB-two}
preserves the independence of the two pairs
or introduces correlations between them.

Independence is preserved when the contextual
drive separates into a contribution depending
only on $(ij)$ and another depending only
on $(kl)$. Additive constants are irrelevant
because they cancel in the differences
entering the collapse equation.

A sufficient decoupling condition is
$\Delta_{ijkl}=0$, together with a factorized
contextual distribution
$\rho(\lambda)=\rho_1(\lambda_1)
\rho_2(\lambda_2)$:
\begin{equation}
\widetilde{R}_{ijkl}(t_0;\lambda)
=\widetilde{R}^{(1)}_{ij}(t_0;\lambda_1)
+\widetilde{R}^{(2)}_{kl}(t_0;\lambda_2).
\label{eq:Rijkl-separable}
\end{equation}
With factorized initial conditions, this
structure preserves the product evolution
$\Psi_{ijkl}(t)=
\psi^{(1)}_{ij}(t)\psi^{(2)}_{kl}(t)$.
The joint outcome distribution then
factorizes as
$P_{ijkl}=P^{(1)}_{ij}P^{(2)}_{kl}$,
in agreement with standard quantum
mechanics for independent pairs.

A cross-pair coupling appears when
$\Delta_{ijkl}$ introduces a genuinely
nonseparable dependence on the two index
pairs. A term that can be written as a
sum of separate functions of $(ij)$ and
$(kl)$ merely modifies the individual
pair drives and does not, by itself,
couple their dynamics.

Within P2, $\Delta_{ijkl}$ represents
the proposed measurement-generated
contextual exchange. Its strength depends
on the spatial arrangement and on the
underlying field and detector couplings.

For a phenomenological estimate in the
large-separation regime, we introduce
the scaling
\begin{equation}
|\Delta_{ijkl}|\sim
\varepsilon\equiv\left(\frac{r_{AB}}{r_{AE}}\right)^2,
\label{eq:eps-scaling}
\end{equation}
where $r_{AB}$ is the intra-pair baseline
and $r_{AE}$ the separation between
Alice and Eve.

This provides a small parameter when
$r_{AE}\gg r_{AB}$. It should be
distinguished from the detector-overlap
suppression $\varepsilon_{\rm meas}(\ell)$
introduced earlier. In particular,
the near-Alice--Eve configuration of
Fig.~\ref{fig2} lies outside the
controlled large-separation regime
of this estimate.

To identify a measurable cross-pair
effect, we associate the outcomes
$\pm$ with the numerical values $\pm1$.
For the two pairs, define the parity
observables $O^{(1)}=ij$ and
$O^{(2)}=kl$. Their connected
cross-correlation is
\begin{equation}
C_{\rm pair}=\langle O^{(1)}O^{(2)}\rangle
 -\langle O^{(1)}\rangle\langle O^{(2)}\rangle.
\label{eq:crosscorr-def}
\end{equation}

Standard quantum mechanics predicts
$C_{\rm pair}=0$ for independently
prepared pairs. The same result follows
in the factorized contextual sector
of Eq.~\eqref{eq:Rijkl-separable}.

A genuinely nonseparable contextual
coupling can instead generate
$C_{\rm pair}\neq0$, depending on
the detector map and the ensemble
statistics. After conventional
experimental correlations have been
excluded, such a signal would provide
a candidate signature of contextual
coupling between otherwise independent
quantum systems.

This test differs from Bell--CHSH
measurements. CHSH probes outcome
correlations within a single entangled
pair, whereas $C_{\rm pair}$ probes
statistical dependence between the
outcomes of two independently
prepared pairs.

%----------------------------------------------------------------------

\paragraph*{Born-rule consistency.}

The four-party extension retains the
Born-rule requirements of the individual
pairs. A nonseparable contextual drive
must be accompanied by a joint ensemble
that reproduces the appropriate
single-pair Born probabilities.
Operational multipartite no-signaling
remains the additional requirement P3.

\section{Possible Experimental Tests}
\label{sec:experimental-signatures}
The hypothesis explored here attributes quantum
correlations to a bulk mediator $\mathscr{X}_a(x,\tau)$.
The controlled zero-mode response is causal in ordinary
brane time; a separate measurement-generated equal-time
interaction is postulated, not derived. Section~\ref{sec:nullchars}
establishes only the kinematic $E=0$ null family with
equal-$t$ reach, not an instantaneous field response or
a proof of operational no-signaling. Two qualitatively
distinct experimental tests arise, differing in
physical meaning and empirical strength.

\subsection{Experimental idea based on asymmetric
detector geometry}
\label{subsec:asym-test-idea}
An experimental test was proposed in Ref.~\cite{PRR},
based on an asymmetric detector geometry. Two identical
EPR sources simultaneously emit one entangled pair
each; the four apparatuses are arranged so that the
distance between Alice and Eve is much shorter than
Alice--Bob and Eve--Tom.
Within the conditional detector model of
Sec.~\ref{sec:dynmodel}, a covariance-based ratio
$(r_{AB}/r_{AE})^2$ motivates stronger cross-pair
sensitivity at smaller separations, but it does not
follow from a source-generated propagator. The warped
$E=0$ geodesic has reach depth
$\tau_*(r)=k^{-1}\ln(1+kr/w)$, reducing to $r/w$
only for $kr/w\ll1$; it is not a measured propagation
delay in ordinary brane time.

Consider two independent Bell experiments operated
simultaneously, producing photon pairs $(A_1,B_1)$
and $(A_2,B_2)$ in identical Bell states. The
apparatuses are arranged so that Alice and Eve are
close to each other while their respective partners
Bob and Tom are at larger distances:
\begin{equation}
\Delta\tau(A\!\leftrightarrow\!E)\;\ll\;
\Delta\tau(A\!\leftrightarrow\!B),
\quad
\Delta\tau(A\!\leftrightarrow\!E)\;\ll\;
\Delta\tau(E\!\leftrightarrow\!T).
\end{equation}
If the bulk-mediated contextual input does not preserve
source identity at the detector level, Alice's and
Eve's particles may couple contextually with greater
weight than they couple to their respective remote
partners.
Standard quantum mechanics predicts complete statistical
independence between the two pairs:
\begin{equation}
P_{ij,kl}^{\mathrm{QM}}=P^{(1)}_{ij}\,P^{(2)}_{kl}.
\end{equation}
In the present framework a nonseparable contextual
coupling between the $(A_1,B_1)$ and $(A_2,B_2)$
systems is in principle allowed. The principal experimental targets are the direct
Alice--Eve covariance and the pair-parity covariance.
 A CHSH test sharing Alice's settings with $AB$ is
subject to nonsignaling monogamy.

For a standard bipartite CHSH test, define
\begin{equation}
S_{AB}\equiv
\left|
E(\theta_A,\theta_B)
+E(\theta_A',\theta_B)
-E(\theta_A,\theta_B')
+E(\theta_A',\theta_B')
\right|,
\label{eq:CHSH-def}
\end{equation}
where
\begin{equation}
E(\theta_A,\theta_B)
\equiv
\sum_{i,j=\pm 1} ij\,P_{ij}^{(AB)}(\theta_A,\theta_B)
=
\langle a\,b\rangle_{\theta_A,\theta_B},
\label{eq:E-AB-def}
\end{equation}
with $i,j\in\{\pm1\}$ the dichotomic outcomes at the
two stations for analyser settings $(\theta_A,\theta_B)$
and $P_{ij}^{(AB)}(\theta_A,\theta_B)$ the
corresponding joint outcome probabilities. This is the
CHSH form~\cite{bell,chsh} of Bell's inequality. Any
locally causal theory satisfies $S_{AB}\le 2$; quantum
mechanics permits values up to $2\sqrt{2}$ for suitable
entangled states (Tsirelson bound~\cite{tsirelson}).

The standard CHSH inequality, $|S_{XY}|\leq2$, can be violated independently by the two disjoint pairs $AB$ and $ET$. An Alice–Eve CHSH analysis, however, constitutes a separate and more demanding diagnostic. If the $AB$ and $AE$ tests share Alice's two measurement settings, multipartite nonsignaling imposes the monogamy bound $|S_{AB}|+|S_{AE}|\leq 4$, thereby excluding simultaneous CHSH violations for these overlapping pairs. This restriction does not exclude nonzero cross-pair correlations $C_{AE}^{\rm dir}$. The primary tests of the present model remain 
$C_{AE}^{\rm dir}$ and $C_{\rm pair}$; any additional CHSH protocol requires an explicit specification of measurement settings and a separate verification of P3 compatibility.

%------------------------------------------------------------
\subsubsection{Loophole controls and distance-dependence test}
\label{subsubsec:control-tests}
%------------------------------------------------------------
Of course, to perform a loophole-free test of this
kind the polariser settings at each of the four
stations must be changed randomly while the photons
are in flight between the sources and the
polarisers, so that no setting is correlated with
the source emission or with the orientations at the
other stations,  and each measurement event must be
spacelike separated from all others, thus closing
the locality loophole for the full four-station
configuration; the two Bell setups must moreover be
suitably electromagnetically shielded in order to
avoid any conventional communication channel
between them.
Beyond these standard precautions, the distance
$d$ between Alice and Eve must be varied across
experimental runs, to test a possible distance dependence; the
$(\ell/d)^2$ phenomenological expansion is restricted
to $d\gg\ell$ and does not apply to the near
Alice--Eve configuration. Varying $d$ alone does not
rule out conventional sources of correlations.

 The primary tests are $C_{AE}^{\rm dir}$ and
$C_{\rm pair}$, not an Alice--Eve Bell violation.
Original-pair CHSH tests do not replace controls for
channel mixing, pairing errors, common-source noise
or timing artefacts.

%%%%%%%%%%%%%%%%%%%%%%%%%%%%%%%%%%%%%%%%%%%%%%%%%%%%%%%%%%%%
\subsection{Cross-pair correlations: a conditional test of nonfactorization}
\label{subsec:proposal-C-nonzero}
\begin{figure}[h!]
 \centering
\includegraphics[width=\columnwidth,keepaspectratio=true]{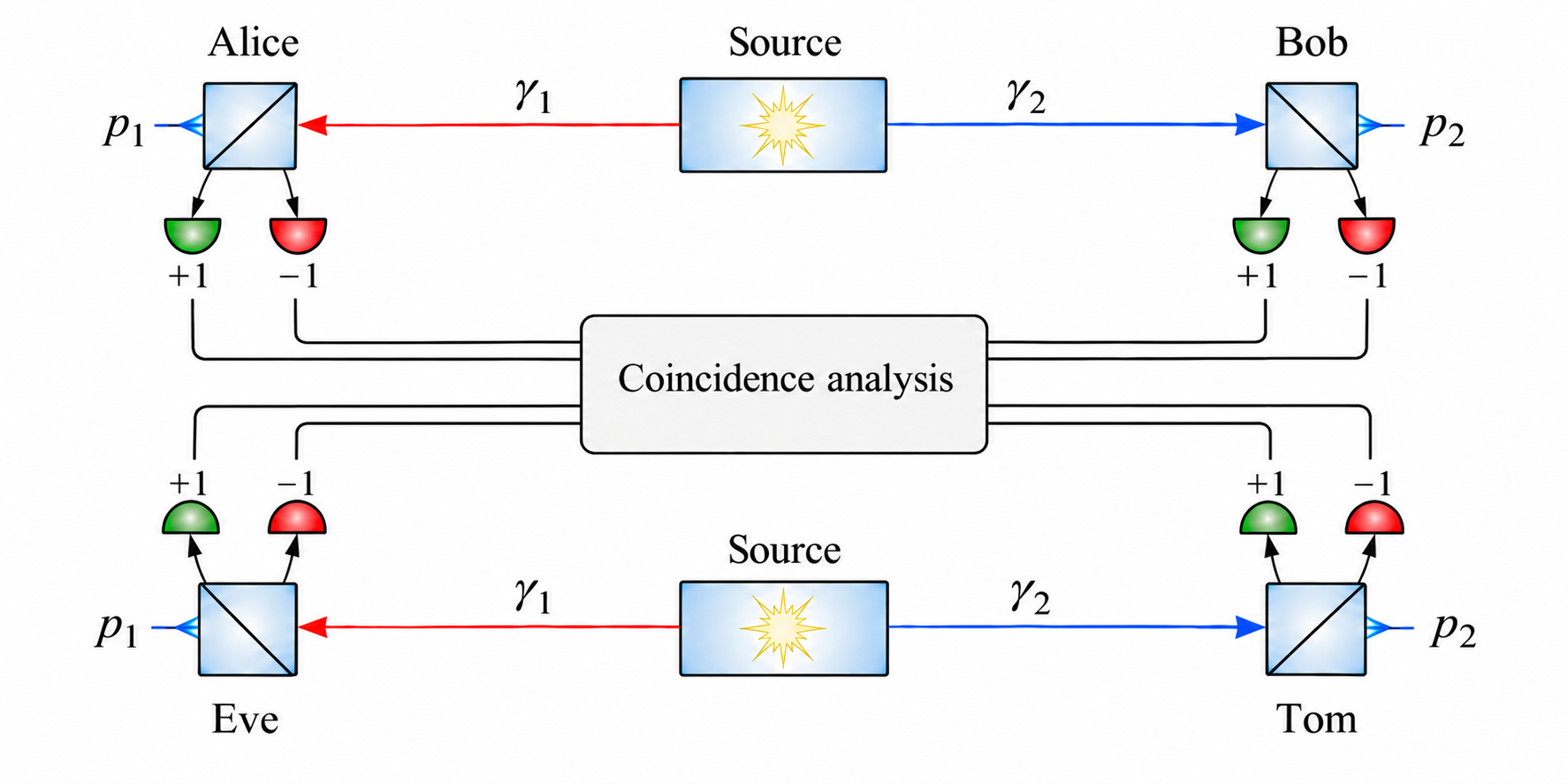}
\caption{Cross-pair test adapted from Ref.~\protect\cite{PRR}.
Two nominally independent Bell-pair experiments are operated
simultaneously, with Alice and Eve close together
($d_{AE}\ll d_{AB},d_{ET}$). After controls for conventional
cross-talk and pairing artifacts, the direct Alice--Eve
covariance $C_{AE}^{\rm dir}$ and, separately, the
pair-parity covariance $C_{\rm pair}$ can be measured.
An Alice--Eve CHSH-type analysis is a separate option,
subject to the nonsignaling monogamy constraint discussed
in Sec.~\ref{subsubsec:control-tests}. The scheme applies as well
to spin systems.}
\label{fig2}
\end{figure}

\paragraph*{Two distinct cross-pair observables.}
We distinguish the direct Alice--Eve outcome covariance
$C_{AE}^{\rm dir}=\langle ik\rangle-\langle i\rangle\langle k\rangle$
from the covariance between Bell-pair parities
$C_{\rm pair}=\langle(ij)(kl)\rangle-\langle ij\rangle\langle kl\rangle$.
These quantities test different forms of statistical dependence:
neither implies the other. The perturbative calculation below
concerns $C_{\rm pair}$; the direct Alice--Eve observable in
Fig.~\ref{fig2} is a separate experimental diagnostic.

%%%%%%%%%%%%%%%%%%%%%%%%%%%%%%%%%%%%%%%%%%%%%%%%%%%%%%%%%%%%
For the pair-parity test, define the covariance
\begin{equation}
C \;=\; \langle O^{(1)} O^{(2)} \rangle
      - \langle O^{(1)} \rangle \langle O^{(2)} \rangle ,
\quad
O^{(1)}=ij,\ \ O^{(2)}=kl,
\label{eq:C-def-proposal}
\end{equation}
for two independently prepared Bell pairs, with outcome
labels $\pm\equiv\pm1$. Their independence gives
$C_{\rm pair}=0$ in standard quantum mechanics.
A nonseparable contextual coupling could instead
produce a nonzero pair-parity covariance.

Consider two independent sources preparing Bell pairs
$(A,B)$ and $(E,T)$, denoted $(A_1,B_1)$ and
$(A_2,B_2)$ in the preceding section. Their four
measurements occur on the brane at laboratory time $t_0$:
\begin{equation}
A=(\mathbf{x}_A,t_0),\quad
B=(\mathbf{x}_B,t_0),\quad
E=(\mathbf{x}_E,t_0),\quad
T=(\mathbf{x}_T,t_0).
\label{eq:four-events}
\end{equation}
The total Hilbert space factorises at preparation time,
\begin{align}
\mathcal{H}_{\rm tot}
&=
(\mathcal{H}_{A}\otimes\mathcal{H}_{B})
\otimes
(\mathcal{H}_{E}\otimes\mathcal{H}_{T}),
\nonumber\\
|\Psi(t_0)\rangle
&=
|\psi^{(AB)}_{\rm Bell}\rangle
\otimes |\psi^{(ET)}_{\rm Bell}\rangle,
\label{eq:two-pairs-factorized-prep}
\end{align}
Consequently, standard quantum mechanics predicts
$P_{ijkl}=P^{(AB)}_{ij}P^{(ET)}_{kl}$ and
$C_{\rm pair}=0$.

In the H1--H2 source description, each detector
contributes a localized measurement source to the
bulk-field equation. For four detectors we write
\begin{equation}
\Box_5\mathscr{X}_a(x,\tau)
=
\sum_{s\in\{A,B,E,T\}}J_a^{(s,\rm meas)}(x)\,\eta_s(\tau)
+J_a^{(\rm prep)}(x)\,\delta(\tau).
\label{eq:bulk-eq-four}
\end{equation}
Equation~\eqref{eq:bulk-eq-four} describes the formal
unrestricted source decomposition; the P1 sector uses
its zero-mode projection or compatible source profiles.
Linearity allows us to separate contributions from
the $(A,B)$ and $(E,T)$ sources and to add their
componentwise detector averages:
\begin{align}
\mathscr{X}_a
&=
\mathscr{X}_a^{(AB)}+\mathscr{X}_a^{(ET)},
\nonumber\\
\mathscr{Y}^{(s)}_a(\mathbf{x},t_0;\lambda)
&=
\mathscr{Y}^{(s,AB)}_a(\mathbf{x},t_0;\lambda)
+\mathscr{Y}^{(s,ET)}_a(\mathbf{x},t_0;\lambda)
\nonumber\\
&\quad (s=A,B,E,T).
\label{eq:Phi-superposition}
\end{align}
For each detector $s$, $\mathscr{Y}^{(s)}_a$ is the
profile-weighted field of Eq.~\eqref{eq:Y-def-full}.
Contributions from the two sources add before channel
normalization. Under P2, each detector may thus sample
both the preparation background and measurement-generated
contributions associated with the other pair.

The analyzer settings are $\theta_A,\theta_B$ for the
first pair and $\theta_E,\theta_T$ for the second;
$i,j,k,l=\pm1$ label their respective outcomes.
To examine how each pair samples the field, introduce
the following \emph{unnormalised linear overlap diagnostics}.
They differ from the pair-level collapse amplitudes of
Eqs.~\eqref{eq:X1-projected-two-pairs}--
\eqref{eq:X2-projected-two-pairs}. In the disjoint
regions $\Omega_s$, the piecewise field $\mathscr Y_a$
equals the corresponding $\mathscr Y_a^{(s)}$:
\begin{align}
\mathscr{X}^{(AB)}_{a,ij}(t_0;\lambda)
&=
\int d^3x\;
u_{ij}^{(AB)\,*}(\mathbf{x})\,
\mathscr{Y}_a(\mathbf{x},t_0;\lambda),
\label{eq:Xproj-AB}
\\[0.5ex]
\mathscr{X}^{(ET)}_{a,kl}(t_0;\lambda)
&=
\int d^3x\;
u_{kl}^{(ET)\,*}(\mathbf{x})\,
\mathscr{Y}_a(\mathbf{x},t_0;\lambda),
\label{eq:Xproj-ET}
\end{align}
The pair-level projection functions combine the
two local detector profiles. For Alice--Bob, set
\begin{align*}
u_{ij}^{(AB)}(\mathbf{x})
&=
u^{(A)}_i(\mathbf{x})\,\mathbf{1}_{\Omega_A}(\mathbf{x})
+
u^{(B)}_j(\mathbf{x})\,\mathbf{1}_{\Omega_B}(\mathbf{x}),
\end{align*}
and similarly for Eve--Tom. Here $\mathbf{1}_{\Omega}$
is the indicator of a detector's spatial region, and
the local profiles $u_i^{(k)}$ were introduced in
Eq.~\eqref{eq:detector-channel-profiles}.
For each pair $p$, normalize the projected components by
$\hat{\mathscr{X}}^{(p)}_{a,I}\equiv
\mathscr{X}^{(p)}_{a,I}/|\mathscr{X}^{(p)}|$,
where $|\mathscr{X}^{(p)}|^2=
\sum_I\delta^{ab}\mathscr{X}^{(p)}_{a,I}
\mathscr{X}^{(p)*}_{b,I}$ and $I=(ij)$ or $(kl)$.
The linear source decomposition~\eqref{eq:Phi-superposition}
applied to Eqs.~\eqref{eq:Xproj-AB}--\eqref{eq:Xproj-ET}
then gives
\begin{align}
\hat{\mathscr{X}}^{(AB)}_{a,ij}
&=
\hat{\mathscr{X}}^{(AB)\,\mathrm{self}}_{a,ij}
+\hat{\mathscr{X}}^{(AB)\,\mathrm{cross}}_{a,ij},
\label{eq:XAB-self-cross}
\\[0.5ex]
\hat{\mathscr{X}}^{(ET)}_{a,kl}
&=
\hat{\mathscr{X}}^{(ET)\,\mathrm{self}}_{a,kl}
+\hat{\mathscr{X}}^{(ET)\,\mathrm{cross}}_{a,kl},
\label{eq:XET-self-cross}
\end{align}
The ``self'' terms come from the corresponding pair's
sources; the ``cross'' terms come from the opposite
pair. Importantly, both are divided by the
\emph{same total-field norm} $|\mathscr X^{(p)}|$:
$\widehat{\mathscr X}^{\,\mathrm{self}}_{a,I}
=\mathscr X^{\mathrm{self}}_{a,I}/|\mathscr X|$,
and likewise for the cross term. This common norm
preserves the additive identities above, although
normalization itself depends nonlinearly on the cross
contribution. At large pair separations,
$(r_{AB}/r_{AE})^2$ provides an illustrative
small parameter (Sec.~\ref{sec:dynmodel}); it does
not describe the near-Alice--Eve arrangement.
%--------------------------------------------------------------------------
The decisive criterion for pair independence is
separability of the \emph{four-index contextual drive}
in Eq.~\eqref{eq:BB-two}, rather than factorization
of the bulk field itself (Sec.~\ref{subsec:factorization}).
For well-separated pairs, $r_{AE}\gg r_{AB}$,
write $\varepsilon\sim(r_{AB}/r_{AE})^2$ and introduce
\begin{equation}
\widetilde{R}_{ijkl}(t_0;\lambda)
=
\widetilde{R}^{(AB)}_{ij}(t_0;\lambda_1)
+
\widetilde{R}^{(ET)}_{kl}(t_0;\lambda_2)
+
\varepsilon\,\Delta^{(R)}_{ijkl}(t_0;\lambda),
\label{eq:Rijkl-exp-nonseparable}
\end{equation}
Here $\Delta^{(R)}_{ijkl}$ couples the two index
pairs: it cannot be written as separate functions
of $(i,j)$ and $(k,l)$. The linear overlap projections
identify where such coupling might enter, while its
actual form depends on the preparation source and
detector readout. The distance estimate below treats
its size phenomenologically.

For fixed contextual data $\lambda$, the Bohm--Bub
evolution~\eqref{eq:BB-two} selects a definite
four-party outcome after collapse. In the separable
limit $\varepsilon=0$, the two pairs remain independent:
$P_{ijkl}=P^{(AB)}_{ij}P^{(ET)}_{kl}$ and
$C_{\rm pair}=0$.
To examine a weak cross-pair coupling, assume that
the \emph{ensemble} outcome probabilities are
differentiable at $\varepsilon=0$. This assumption
concerns the ensemble, since individual deterministic
trajectories can have moving collapse-basin boundaries.
To first order, write
\begin{equation}
P_{ijkl}
=
P^{(AB)}_{ij}P^{(ET)}_{kl}
+
\varepsilon\,\delta P_{ijkl},
\label{eq:Pijkl-expansion}
\end{equation}
where the zeroth-order product reflects independent source
ensembles and the separable drive of
Eq.~\eqref{eq:Rijkl-separable}. The correction
$\delta P_{ijkl}$ must preserve normalization
and multipartite operational no-signaling (P3).

A simple example illustrates how a nonzero
pair-parity covariance can coexist with full
no-signaling. Starting from the independent-pair
distribution
$P^{(0)}_{ijkl}=P^{(AB)}_{ij}P^{(ET)}_{kl}$,
we introduce a correction proportional to
the product of the two pair parities,
$ijkl=(ij)(kl)$:
\begin{equation}
\delta P^{\rm ex}_{ijkl}
=\frac{\alpha(\boldsymbol\theta)}{16}ijkl,
\qquad
|\varepsilon\alpha(\boldsymbol\theta)|
\leq16\min_{i,j,k,l}P^{(0)}_{ijkl}.
\label{eq:NS-witness}
\end{equation}
Here $\alpha(\boldsymbol\theta)$ controls
the strength and angular dependence of the
correlation. For positive $\alpha$, the
correction favours configurations with
equal pair parities over those with
opposite parities. The inequality ensures
nonnegative probabilities.

The bound guarantees nonnegative probabilities.
Summing over any one of the four outcomes
makes the correction vanish, leaving every
proper marginal unchanged and thus preserving
multipartite no-signaling (P3). Nevertheless,
the pair-parity covariance becomes
$C_{\rm pair}=\varepsilon\alpha(\boldsymbol\theta)$.
This probability-level construction leaves
the direct Alice--Eve covariance unchanged
while allowing correlations between the
two pair parities.

More generally, P3 requires every subset's outcome
statistics to be independent of the settings at the
remaining detectors. To express the leading covariance,
define the marginal corrections
\begin{equation}
\delta P^{(AB)}_{ij}\equiv
\sum_{k,l=\pm1}\delta P_{ijkl},
\qquad
\delta P^{(ET)}_{kl}\equiv
\sum_{i,j=\pm1}\delta P_{ijkl},
\label{eq:deltaP-marginals}
\end{equation}
the cross-correlation to leading order in $\varepsilon$
reads
\begin{eqnarray}
C_{\rm pair}
&=&
\varepsilon\,
\Biggl[
\underbrace{
\sum_{i,j,k,l=\pm1}(ij)(kl)\,\delta P_{ijkl}
}_{\text{joint correction}}
\nonumber\\
&-&
\underbrace{
\Bigl(\sum_{i,j=\pm1}(ij)\,\delta P^{(AB)}_{ij}\Bigr)
\Bigl(\sum_{k,l=\pm1}(kl)\,P^{(ET)}_{kl}\Bigr)
}_{\text{shift in }\langle O^{(1)}\rangle\text{ from }\delta P}
\nonumber\\
&-&
\underbrace{
\Bigl(\sum_{i,j=\pm1}(ij)\,P^{(AB)}_{ij}\Bigr)
\Bigl(\sum_{k,l=\pm1}(kl)\,\delta P^{(ET)}_{kl}\Bigr)
}_{\text{shift in }\langle O^{(2)}\rangle\text{ from }\delta P}
\Biggr]
+\mathcal{O}(\varepsilon^2). \nonumber
\label{eq:C-leading-eps}
\end{eqnarray}
The three terms in brackets represent the
joint-outcome correction and the induced shifts
of the two mean parities. Their combination
must be nonzero for a leading-order cross-pair
signal. The example of Eq.~\eqref{eq:NS-witness}
demonstrates compatibility with multipartite
no-signaling, without fixing the phenomenological
coefficient $\mathcal A$ of
Eq.~\eqref{eq:C_vs_separation}.

\paragraph*{Illustrative distance dependence and event requirements.}
\paragraph*{Illustrative distance dependence
and event requirements.}

The primary test places Alice and Eve close
together, $d_{AE}\ll\ell$, with $\ell$ the
characteristic intra-pair separation.
This configuration follows the experimental
idea introduced in Ref.~\cite{PRR}: two
independently prepared Bell pairs are arranged
so that detectors belonging to different
pairs can interact more closely than those
belonging to the same pair.

The geometrical motivation is straightforward.
The smaller Alice--Eve separation corresponds
to a smaller extra-time propagation depth,
$\tau_*(d_{AE})<\tau_*(\ell)$, and therefore
to a less suppressed measurement-exchange
overlap for comparable detector profiles.
Within P2, this makes the near-Alice--Eve
configuration particularly interesting for
investigating contextual coupling between
otherwise independent quantum systems.

While Ref.~\cite{PRR} proposed an anomalous
Alice--Eve Bell--CHSH correlation as a
possible experimental signature, the present
framework distinguishes two complementary
observables: the direct Alice--Eve covariance
$C_{AE}^{\rm dir}$ and the pair-parity
covariance $C_{\rm pair}$. Both vanish for
independent pairs in standard quantum mechanics.
Their separate measurement therefore provides
a broader test of the proposed mechanism.
An Alice--Eve CHSH analysis remains an
additional possibility, subject to the
multipartite no-signaling constraints.
Unlike the complementary large-distance
estimate below, this near-distance test
does not rely on a perturbative expansion.

\emph{Illustrative near-distance sensitivity.}
Although the regime $d_{AE}\ll\ell$ lies
outside the perturbative approximation,
a simple phenomenological benchmark can
be constructed from the P2 detector-overlap
factor. For example, suppose that the
pair-parity covariance follows
\begin{equation}
C_{\rm pair}^{\rm toy}(d,\boldsymbol\theta)
=C_*(\boldsymbol\theta)\,
\varepsilon_{\rm meas}(d),
\label{eq:Cnear-toy}
\end{equation}
where $C_*$ is an undetermined response
amplitude. For $k\delta\tau\ll1$, near the
exchange scale, this becomes
\begin{equation}
C_{\rm pair}^{\rm toy}(d)
\simeq C_*
\exp\left[-\frac{d^2}
{2\ell_{\rm exch}^2}\right].
\label{eq:Cnear-gaussian}
\end{equation}

Unlike the inverse-square large-distance
ansatz, this illustrative response remains
finite as $d\to0$. Its amplitude and spatial
dependence would have to be established
by the contextual detector dynamics.
A probability-level realization compatible
with multipartite no-signaling can be
constructed using the parity correction
of Eq.~\eqref{eq:NS-witness}, with the
amplitude constrained by positivity.

Independently of this particular ansatz,
the approximate number of independent
fourfold-coincidence events needed to
resolve a near-distance covariance at
three-sigma significance is
$N_{3\sigma}\sim9/|C_{\rm pair}|^2$,
up to factors depending on the marginals.
Thus, a hypothetical covariance of order
$10^{-2}$ would require about $10^5$
events, whereas a covariance of order
$10^{-3}$ would require about $10^7$.
These figures provide sensitivity
benchmarks rather than predictions of
the near-distance signal.
\smallskip

A complementary sensitivity estimate concerns the
opposite regime, $d\equiv d_{AE}\gg\ell$. With the
assumed inverse-square preparation-field covariance,
use the small parameter
\begin{equation}
\varepsilon_{\rm prep}\sim\left(\frac{\ell}{d}\right)^2\ll1.
\label{eq:eps_estimate}
\end{equation}
If the leading nonzero detector response to this
covariance is linear, the phenomenological scaling is
\begin{equation}
C_{\rm pair}(d,\boldsymbol\theta)\simeq
\mathcal A(\boldsymbol\theta)
\left(\frac{\ell}{d}\right)^2,
\qquad d\gg\ell.
\label{eq:C_vs_separation}
\end{equation}
The coefficient $\mathcal A$ captures the angular
and detector-readout dependence, and may vanish.
Channel normalization~\eqref{eq:Rij-physical-unified}
removes a common field-amplitude factor but does
not fix this coefficient. Thus the $1/d^2$ law
combines an assumed ensemble covariance with a
linear detector response; it is not an extrapolation
to the near-Alice--Eve geometry.

\emph{Event requirements.}
For $N$ independent fourfold-coincidence events
and a small pair-parity covariance, the statistical
uncertainty scales approximately as
$\sigma_C\sim N^{-1/2}$, up to factors of order
unity depending on the measured marginals.

Let $z$ denote the desired statistical significance
in units of $\sigma_C$, so that a correlation
is resolved at approximately $z$ standard
deviations when $|C_{\rm pair}|\sim z\sigma_C$.
Using Eq.~\eqref{eq:C_vs_separation}, the required
number of events is therefore
\begin{equation}
N\sim\frac{z^2}{|\mathcal A(\boldsymbol\theta)|^2}
\left(\frac{d}{\ell}\right)^4,
\qquad d\gg\ell,
\label{eq:event_requirements}
\end{equation}
for $\mathcal A\neq0$.
The fourth-power dependence arises because the assumed correlation decreases as $d^{-2}$, 
whereas the statistical uncertainty decreases only as $N^{-1/2}$.

The table illustrates the requirements for $z=3$,
corresponding to a nominal three-sigma significance,
and the sensitivity benchmark $|\mathcal A|=1$,
chosen independently of the covariance normalization.
Here $N_{3\sigma}$ denotes the corresponding number
of fourfold-coincidence events.

\begin{center}
\begin{tabular}{llll}
\hline
$\ell$ & $d$ & $|C_{\rm pair}|$ & $N_{3\sigma}$ \\
\hline
$1$\,m & $10$\,m & $10^{-2}$ & $9\times10^4$ \\
$1$\,m & $30$\,m & $1.1\times10^{-3}$ & $7.3\times10^6$ \\
$1$\,m & $100$\,m & $10^{-4}$ & $9\times10^8$ \\
\hline
\end{tabular}
\end{center}

Actual fourfold coincidence counts also depend on
source rates, losses, event correlations, systematic
errors, and $|\mathcal A|$. These benchmarks concern
only the large-separation regime, not the primary
near-Alice--Eve arrangement.
\smallskip
\paragraph*{Zero-mode covariance and the role of $k$.}
Under P1, the retained zero mode has
$\psi_0(0)=\sqrt{k}$, so that the brane-field
covariance is $C_{\mathscr X}=kC_\phi^{(0)}$.
For the conditional experimental estimate,
we adopt the illustrative equal-time covariance
\begin{equation}
C^{(0)}_{\phi}(0,r)\;=\;\frac{1}{4\pi^2 r^2},
\label{eq:C0_powerlaw}
\end{equation}
which describes a chosen ensemble of zero-mode
configurations. The resulting brane covariance
decays as $r^{-2}$, with an amplitude proportional
to $k$. This spatial scaling motivates the
conditional inverse-square cross-pair signal
of Eq.~\eqref{eq:C_vs_separation}.

Since P1 excludes the positive-$\mu$ continuum,
there is no retained Kaluza--Klein crossover
scale through which to probe $k$. Experimental
bounds on hypothetical superluminal
hidden-influence velocities~\cite{salart,cocciaro}
also do not directly constrain this parameter.

\emph{Optional measurement-exchange contribution.}
The P2 measurement-exchange mechanism
(Sec.~\ref{subsec:meas-exchange}, H1--H3)
introduces a distinct distance dependence
through the detector-overlap factor
$\varepsilon_{\rm meas}(d)$ of
Eq.~\eqref{eq:eps-meas}. For a Gaussian detector
profile, this factor is Gaussian in
$\tau_*(d)=k^{-1}\ln(1+kd/w)$, reducing to
a Gaussian in $d$ when $k\delta\tau\ll1$
near the exchange scale.

Unlike the preparation covariance, the
measurement-exchange contribution is governed
by the characteristic length
$\ell_{\rm exch}=(w/k)(e^{k\delta\tau}-1)$.
Its experimental contribution is treated
separately from the leading inverse-square
estimate of Eq.~\eqref{eq:C_vs_separation}.
Determining $k$ from such measurements
would require specifying the detector coupling.
\smallskip

\emph{Experimental protocol.}
The primary near-Alice--Eve experiment,
$d_{AE}\ll\ell$, measures $C_{AE}^{\rm dir}$ and
$C_{\rm pair}$ separately. A CHSH analysis across
Alice and Eve is an additional, more demanding test.
Varying $d_{AE}$ probes the spatial dependence;
the complementary $d\gg\ell$ configuration tests
the conditional law~\eqref{eq:C_vs_separation}, with
illustrative counts in Eq.~\eqref{eq:event_requirements}.
Each original pair should retain a significant CHSH
violation, verifying its entanglement throughout the
protocol. Independent controls must also exclude
cross-talk, mistaken pairing, and common-source effects.
\medskip

\textit{Summary of Sec.~\ref{sec:experimental-signatures}.}
The primary test searches for $C_{AE}^{\rm dir}\neq0$
or $C_{\rm pair}\neq0$ between independently prepared
Bell pairs in the near-Alice--Eve configuration.
With conventional correlations excluded, either result
would depart from the statistical independence predicted
by standard quantum mechanics. An Alice--Eve CHSH test
is a separate option, subject to the nonsignaling
monogamy restriction discussed above. The complementary
large-distance test examines the conditional inverse-square
law~\eqref{eq:C_vs_separation}, whose amplitude and angular
dependence are set by the contextual detector dynamics.

\section{Discussion and concluding remarks}
\label{discussion}
The present work derives a warped $(3,2)$ geometry and exhibits
null directions with unbounded equal-time spatial projections.
For a specified reduced massless zero mode it also provides a
stable, retarded four-dimensional response and illustrates possible
long-range equal-time preparation covariances. These results do
\emph{not} yet establish a stable, measurement-generated equal-time
brane-to-brane interaction. A physically realized mediator retaining
the relevant null characteristics, a compatible detector coupling,
and an equivariant multipartite nonsignaling collapse law are
independent research problems.

The hypothesized contextual interaction could, if it exists,
modify the correlations of otherwise independently prepared Bell
pairs. We discuss the parity covariance $C_{\rm pair}$ and the
direct Alice--Eve outcome covariance $C_{AE}^{\rm dir}$ as distinct
experimental targets. Their nonvanishing is not a theorem here.
An inverse-square distance dependence is proposed only for a
specified class of preparation covariances and detectors with a
nonvanishing linear contextual response, in a perturbative regime
$d\gg\ell$. In the near-Alice--Eve geometry, this leading expansion
cannot be used without further calculation.

The symmetric warped $(4,1)$ Randall--Sundrum example considered
above has no brane shortcut, whereas other, asymmetrically warped
$(4,1)$ metrics may admit finite-time bulk shortcuts. Neither result
constructs the exactly equal-time mediator studied here.
The present contribution is therefore a geometrically motivated,
quantitatively delimited hypothesis rather than a completed
dynamical explanation of Bell correlations.

\paragraph*{Possible implications for multi-source quantum protocols.}
Although the cross-pair correlation proposed here is primarily a test of the underlying spacetime mechanism, its observation could have broader implications for quantum-information architectures employing multiple nominally independent entangled resources. Standard descriptions of such systems assume that independently prepared Bell pairs factorise unless they are deliberately coupled by a physical operation. In the proposed contextual framework, two independently prepared pairs sharing the same bulk could acquire cross-pair correlations. In the separate large-distance regime $d\gg\ell$, a nonzero linear detector response would give the illustrative scaling $C_{\rm pair}\sim\mathcal A(\boldsymbol\theta)(\ell/d)^2$. If experimentally established, such a departure from exact source independence would have to be taken into account in multi-source quantum protocols and distributed quantum networks, particularly where statistical independence of separate entangled resources enters as an operational assumption. We do not pursue such applications here; determining whether the predicted correlations could become relevant at technologically realistic length scales and error thresholds is left for future investigation.

\subsection{What is proved, what is assumed, and what remains open} \label{sec:assumptions} 
\paragraph*{Metric ansatz and GR consistency.} We have shown that the warped $(3,2)$ metric ansatz with $f(\tau)=k|\tau|$ satisfies the five-dimensional vacuum Einstein equations with positive bulk cosmological constant; the bulk Einstein equations projected onto the $\tau=0$ brane then take the Randall--Sundrum form with $E_{\mu\nu}=0$ at leading order, under the stated junction assumptions. Recovering observed four-dimensional gravitational phenomenology quantitatively would require further analysis. See Sec.~\ref{symmwarp3-2} and Appendix~\ref{app:32_to_31_GR}. Only the reduced massless zero-mode equation has a proven stable retarded response in ordinary brane time. The higher-mode spectral decomposition is a diagnostic of the unrestricted unstable model, not a physical Green-function construction supported by a convergence argument. 
\paragraph*{Stability and source compatibility (P1).}
The unrestricted operator has a tachyonic continuum with
$\omega^2=c^2(p^2-\mu^2)$; $\mathbb Z_2$ symmetry,
normalisability and retardedness do not remove it. Our stable
zero-mode model is a defined effective truncation or an exact
restricted solution for a specially distributed source, not the
response to a generic $\delta(\tau)$ detector current. No
claim of perturbative stability under arbitrary bulk excitations
or of source-invariant truncation is made.
The explicitly projected equation
\eqref{eq:projected-two-source} closes the stable effective
source problem, but also confirms that its retarded zero mode
cannot implement the extra equal-$t$ measurement exchange.
\paragraph*{Preferred-foliation perspective.}
Equations~\eqref{eq:pf_projector}--\eqref{eq:pf_characteristics}
provide an explicit operator-level route to avoiding the identified
tachyonic dispersion. They do not prove complete stability of the
coupled aether--mediator and detector theory, which is left to a
separate investigation.
\paragraph*{Equal-time reach and contextual mediation (P2).}
The $E=0$ null-geodesic family has arbitrarily large
equal-$t$ spatial projections but does not reconnect two
distinct brane points by a single ray. Whether it gives
a physical stable source-to-source kernel, or a leading
contribution to a prepared covariance, is not established. 
\paragraph*{No-signaling (P3).}\label{item:D}
The retarded four-dimensional \emph{zero-mode} response cannot
send a brane-spacelike controlled signal. It does not establish
operational no-signaling of the postulated equal-time contextual
interaction. Single-party Born marginals are insufficient: all
accessible multipartite marginals must be independent of remote
settings, and the detector coupling must not expose a separately
controllable field response. These requirements have not been
derived from the Bohm--Bub and bulk dynamics.
Equation~\eqref{eq:NS-witness} is only a
four-party nonsignaling example at the level of probabilities;
it is not a solution of that dynamical problem.
 \paragraph*{Born rule.}
 Born statistics require an ensemble with the correct
contextual basin weights; their dynamical preservation
has not been proved. See
Appendix~\ref{app:roles_X_lambda} and
Secs.~\ref{sec:Xfield-propagation}--\ref{sec:dynmodel}.
What remains open is whether generic initial
distributions relax dynamically toward that class;
establishing an analogue of the Bohmian subquantum
$H$-theorem~\cite{DGZ} for the present contextual
collapse dynamics is still to be done, and
equivariance of $\rho(\lambda)$ should therefore
be regarded as a consistency condition of the
present implementation rather than as a theorem
derived from the dynamics.
\paragraph*{Detector map and conditional distance dependence (P4).}
A nonzero cross-pair signal, its angular dependence, and an
inverse-square leading term require a specified nonvanishing
response of the contextual detector map to the chosen field
ensemble. These are not consequences of zero-mode normalisation
alone. In addition, the perturbative parameter
$\varepsilon=(\ell/d)^2$ requires $d\gg\ell$;
the short-Alice--Eve configuration has to be treated separately.
The parity observable $C_{\rm pair}$ and the direct outcome
observable $C_{AE}^{\rm dir}$ are distinct experimental targets.
\paragraph*{Preparation-dominance condition.} 
The assumption that the preparation field dominates the denominator of $\widetilde{R}^{(B)}_j$ (Eq.~\eqref{eq:prep-dominance}) is not a consequence of the bulk wave equation; it requires that the preparation source generate a substantially stronger bulk excitation than a single detector-induced measurement pulse, and its fulfilment is deferred to the microscopic model of $J_a^{(\rm prep)}$. When this condition holds, the exchange term is perturbatively small in the collapse ratios; Born equilibrium remains a separate requirement from operational no-signaling (P3). 
\paragraph*{Encoding of specific quantum correlations in the bulk field.}
The preparation-source ansatz allows a non-factorising
field ensemble, offering a possible classical-field analogue
of nonfactorizability. Whether the Green-function response
actually yields the required channel correlations remains open. What is not yet derived from the bulk dynamics alone is the detailed correlation structure of the prepared state, including singlet anticorrelation, the precise angular dependence of the two-detector correlator on the analyser settings, and the Tsirelson bound, since in the present formulation that structure is encoded in the single-station projection maps, which are constrained by equivariance but not uniquely fixed by it. A natural ansatz for $J_a^{(\rm prep)}$ compatible with singlet structure is given in Sec.~\ref{subsec:meas-exchange} [Eqs.~\eqref{eq:Ja-prep-ansatz}--\eqref{eq:Xa-prep-integral}], and the central open task is to verify that this ansatz yields a brane field satisfying condition~\eqref{eq:anticorr-field}. Two issues therefore remain  open: contextual equilibration and the derivation of the readout maps from the preparation source, the second being the deeper one, because resolving it would make the observed quantum correlation structure emerge from the geometry rather than be inserted as compatible input. This derivation will be taken up in a companion paper.

\paragraph*{Zero-mode covariance and experimental estimate.}
The inverse-square ensemble covariance of
Eq.~\eqref{eq:C0_powerlaw} and the nonzero detector
response entering Eq.~\eqref{eq:C_vs_separation}
are separate phenomenological inputs. The event
counts use the illustrative benchmark
$|\mathcal A|=1$, independently of the covariance
normalization.

\paragraph*{Detector assumptions (H1)--(H3).} 
Assumptions (H1)--(H3) of Sec.~\ref{subsec:meas-exchange} are modelling inputs of the Bohm--Bub-based realisation, not consequences of the geometry itself: (H1) relaxes strict brane confinement by postulating a finite $\tau$-profile for detectors; (H2) promotes measurement interactions to bulk sources away from $\tau=0$; and (H3) assumes that the measurement-sourced pulse carries outcome information but not setting information. These assumptions are not yet derived from a microscopic
detector model, and their compatibility with complete
multipartite no-signaling has not been established. The suggested
$C_{\rm pair}\sim(\ell/d)^2$ behavior is a conditional
large-separation phenomenological ansatz, not a universal
field-theoretic prediction.
\medskip

To conclude, entanglement defies our usual representations and understanding of physical phenomena; this is not because quantum mechanics is wrong, but {because its nonlocal correlations resist a conventional mediating description within ordinary $(3,1)$-dimensional spacetime.} The present construction provides a concrete geometric hypothesis and a restricted stable field sector while exposing the additional dynamical assumptions needed for an instantaneous measurement mediator. It motivates a conditional experimental test, rather than a parameter-free prediction derived from the metric. Beyond providing a test of the proposed mechanism, a possible cross-pair correlation raises the possibility that independently prepared entangled resources may not remain exactly statistically independent when coupled through a common bulk field. If confirmed, such an effect could have broader implications for multi-source quantum-information protocols and distributed quantum networks, where the independence of separately prepared entangled resources is ordinarily assumed.

\section*{Acknowledgements}
The author thanks Roger Penrose for a stimulating discussion held at the Arts Centre 
De Brakke Grond, Amsterdam, in 2014, which has been an enduring source of motivation for this work.
Useful comments received by  Philip Kurian  and Matteo Gori are also gratefully acknowledged.

The author used Claude (Anthropic) as an AI writing assistant only for light editing and polishing 
of the manuscript, not for conceptual development.
OpenAI's ChatGPT image-generation tool was used for the graphical rendering of Figs. 1 and 2 from a schematic designed by the author.

\appendix

% ============================================================
\section{Relation between the $(3,2)$ bulk Einstein equations and the induced $(3,1)$ brane equations}
\label{app:32_to_31_GR}
We consider a five-dimensional bulk spacetime $(\mathcal M_5,g^{(5)}_{AB})$ with signature
$(-,+,+,+,-)$, satisfying the vacuum Einstein equations with cosmological constant
\begin{equation}
G^{(5)}_{AB}+\Lambda_{5}\,g^{(5)}_{AB}=0 .
\label{eq:5D_vac_Einstein_app}
\end{equation}
A codimension--1 brane $\Sigma$ is embedded in $\mathcal M_5$ with unit normal $n^A$,
\begin{equation}
n_A n^A \equiv \epsilon=-1,
\end{equation}
so the extra dimension is timelike (the normal to the brane is timelike).
We follow the conventions of Refs.~\cite{Wald,BarrabesIsrael} for hypersurface
geometry and extrinsic curvature, and Refs.~\cite{SMS,MaartensKoyama} for the
brane-world projection.
Throughout we adopt the extrinsic-curvature convention
$K_{\mu\nu}=e^A{}_\mu e^B{}_\nu \nabla_A n_B$; note that some references use the opposite sign,
$K_{\mu\nu}\to -K_{\mu\nu}$, which correspondingly flips the sign in the junction condition.

\subsection*{A. Induced geometry and extrinsic curvature}
Let $x^\mu$ ($\mu=0,1,2,3$) be intrinsic coordinates on the brane and
$X^A(x^\mu)$ the embedding map. The tangent basis is
\begin{equation}
e^A{}_\mu \equiv \frac{\partial X^A}{\partial x^\mu},
\qquad
n_A e^A{}_\mu=0 .
\end{equation}
The projector onto directions tangent to the brane is
\begin{equation}
h_{AB}\equiv g^{(5)}_{AB}-\epsilon\,n_A n_B ,
\end{equation}
and the induced metric is the pullback
\begin{equation}
g^{(4)}_{\mu\nu}=h_{AB}\,e^A{}_\mu e^B{}_\nu
= g^{(5)}_{AB}\,e^A{}_\mu e^B{}_\nu .
\label{eq:induced_metric_app}
\end{equation}
The extrinsic curvature,
\begin{equation}
K_{\mu\nu}\equiv e^A{}_\mu e^B{}_\nu \nabla_A n_B,
\label{eq:extrinsic_curvature_app}
\end{equation}
measures how the brane is curved within the bulk (it vanishes if the brane is
totally geodesic). We denote $K\equiv K^\alpha{}_\alpha$.

\subsection*{B. Gauss--Codazzi projection and effective brane equation}
Projecting the bulk curvature via the Gauss equation and contracting yields an
effective four-dimensional Einstein equation of the schematic form (see Sec.3 of \cite{MaartensKoyama})
\begin{equation}
G^{(4)}_{\mu\nu}
=
\mathcal{S}_{\mu\nu} \;+\; Q_{\mu\nu} \;-\;\epsilon\,E_{\mu\nu},
\label{eq:effective_brane_general_app}
\end{equation}
where:

(i) $\mathcal{S}_{\mu\nu}$ collects the direct contribution of the bulk field
equations \eqref{eq:5D_vac_Einstein_app}. For a pure bulk cosmological constant,
$\mathcal{S}_{\mu\nu}$ is proportional to $g^{(4)}_{\mu\nu}$ and is conveniently
absorbed into an effective brane cosmological constant once the junction
conditions are imposed (see below).

(ii) $Q_{\mu\nu}$ is the local correction quadratic in the extrinsic curvature,
\begin{eqnarray}
Q_{\mu\nu}
&=&
\epsilon\left[
K\,K_{\mu\nu} - K_{\mu\alpha}K^{\alpha}{}_{\nu}
-\frac{1}{2}\,g^{(4)}_{\mu\nu}
\bigl(K^{2}-K_{\alpha\beta}K^{\alpha\beta}\bigr)
\right]. \nonumber
\label{eq:Qmunu_app}
\end{eqnarray}
The explicit factor $\epsilon\equiv n_A n^A$ keeps track of whether the normal is
spacelike ($\epsilon=+1$) or timelike ($\epsilon=-1$), and therefore reverses the
overall sign of $Q_{\mu\nu}$ when passing between the two cases.
Note that $Q_{\mu\nu}$ as written here is the pre-junction form, expressed directly
in terms of the extrinsic curvature. After substituting the Israel junction
condition~\eqref{eq:Israel_app}, $Q_{\mu\nu}$ becomes quadratic in $S_{\mu\nu}$
and reduces to the $p_{\mu\nu}$ term of Refs.~\cite{SMS,MaartensKoyama}.

(iii) $E_{\mu\nu}$ is the nonlocal Weyl projection,
\begin{equation}
E_{\mu\nu}
\equiv
C^{(5)}_{ABCD}\,n^A e^B{}_\mu n^C e^D{}_\nu ,
\label{eq:Emunu_def_app}
\end{equation}
which encodes bulk gravitational degrees of freedom not fixed by local brane data.
It is traceless, $E^\mu{}_\mu=0$, and (in the absence of bulk matter) its
divergence is constrained by the projected Bianchi identities
(see Refs.~\cite{SMS,MaartensKoyama}).

With these definitions, the effective brane equation involves the combination
$-\epsilon\,E_{\mu\nu}$. Thus for a timelike extra dimension ($\epsilon=-1$) the
Weyl projection enters as $+E_{\mu\nu}$, whereas for a spacelike extra dimension
($\epsilon=+1$) it enters with the opposite sign; the same $\epsilon$ factor also
fixes the overall sign of the local quadratic term $Q_{\mu\nu}$.

\subsection*{C. Israel junction condition, $\mathbb{Z}_2$ symmetry, and vacuum brane}
Let $S_{\mu\nu}$ be the brane stress tensor (including tension). In the present sign
convention, a compact $\epsilon$-explicit form of the Israel junction condition is
\begin{equation}
\epsilon\Bigl(\,[K_{\mu\nu}] - g^{(4)}_{\mu\nu}[K]\,\Bigr) = -\,\kappa_5^{\,2}\,S_{\mu\nu},
\label{eq:Israel_trace_reversed_app}
\end{equation}
where $[K_{\mu\nu}]$ denotes the jump of $K_{\mu\nu}$ across the brane.
Equivalently (solving \eqref{eq:Israel_trace_reversed_app} for $[K_{\mu\nu}]$),
\begin{equation}
\bigl[K_{\mu\nu}\bigr]
=
-\epsilon\,\kappa_5^{\,2}
\left(
S_{\mu\nu}-\frac{1}{3}\,g^{(4)}_{\mu\nu}\,S
\right),
\qquad
S\equiv g^{(4)\,\mu\nu}S_{\mu\nu},
\label{eq:Israel_app}
\end{equation}
which is the form used below. With $\mathbb{Z}_2$ symmetry (mirror symmetry across
the brane), one has $K_{\mu\nu}^+=-K_{\mu\nu}^-$ and hence $[K_{\mu\nu}]=2K_{\mu\nu}^+$,
so \eqref{eq:Israel_app} fixes $K_{\mu\nu}$ algebraically in terms of $S_{\mu\nu}$
once an orientation for $n^A$ is chosen.
Note that with $\epsilon=-1$, Eq.~\eqref{eq:Israel_app} carries the opposite overall
sign relative to the SMS convention~\cite{SMS} (which implicitly sets $\epsilon=+1$);
this sign flip propagates consistently through the subsequent computation and is
responsible for the $\epsilon$ factors in $\Lambda_4$, Eq.~\eqref{eq:Lambda4_app},
as verified explicitly in Eq.~\eqref{eq:Lambda4_zero}.

For a vacuum brane with pure tension,
\begin{equation}
S_{\mu\nu}=-\sigma\,g^{(4)}_{\mu\nu},
\end{equation}
the junction condition implies $K_{\mu\nu}\propto g^{(4)}_{\mu\nu}$ and the local
term $Q_{\mu\nu}$ reduces to an effective cosmological-constant contribution.
One may then write the induced brane equation in the compact form
\begin{equation}
G^{(4)}_{\mu\nu}+\Lambda_{4}\,g^{(4)}_{\mu\nu}
=
-\epsilon\,E_{\mu\nu},
\label{eq:brane_vacuum_app}
\end{equation}
with
\begin{equation}
\Lambda_{4}
=
\frac12\,\kappa_{5}^{2}
\left(
\frac{\Lambda_{5}}{\kappa_5^2}
+\epsilon\,\frac{\kappa_{5}^{2}\sigma^{2}}{6}
\right).
\label{eq:Lambda4_app}
\end{equation}
in direct analogy with the usual brane-world relation, now keeping explicit the
sign $\epsilon=-1$ appropriate to a timelike extra dimension.

The constants $\kappa_5$, $\Lambda_5$, $\sigma$, and $k$ are not all independent.
The parameter $k$ appears in $f(\tau)=k|\tau|$ and
has dimensions of inverse time; the associated bulk
curvature length is $L_{\rm bulk}=w/k$.
The bulk cosmological constant is related to $k$ and
the five-dimensional gravitational coupling $\kappa_5^2=8\pi G_5$ by the
$(\tau\tau)$ component of the bulk Einstein equations, which for $\epsilon=-1$
yields $\Lambda_5=+6k^2/w^2$, as in Sec.~\ref{symmwarp3-2}, requiring $\Lambda_5>0$, i.e.\ a de~Sitter-type bulk. Here $\Lambda_5$ is the geometric constant of Eq.~\eqref{eq:5D_vac_Einstein_app}; the corresponding vacuum energy density is $\rho_\Lambda=\Lambda_5/\kappa_5^2$. In the remainder of this appendix we set $w=1$, so that $k$ is an inverse length and $\Lambda_5=6k^2$.
This is the correct sign for a timelike extra dimension: unlike the standard
Randall--Sundrum case ($\epsilon=+1$), where the $(\tau\tau)$ equation gives
$\Lambda_5^{\mathrm{RS}}=-6k^2<0$ (AdS), the sign flip in
$g_{\tau\tau}=-w^2$ reverses the equation and requires a positive bulk
cosmological constant (see also the remark in Sec.~\ref{symmwarp3-2}).
With the timelike normal and the conventional definition
$S_{\mu\nu}=-\sigma g^{(4)}_{\mu\nu}$, the decreasing warp
requires a negative brane tension. With $w=1$ the
fine-tuning condition is $\sigma=-6k/\kappa_5^2$, which entails $\Lambda_4=0$, as verified by
substituting into~\eqref{eq:Lambda4_app} with $\epsilon=-1$:
\begin{equation}
\Lambda_4
=\tfrac{1}{2}\kappa_5^2\!\left(
\frac{6k^2}{\kappa_5^2}
-\frac{\kappa_5^2}{6}\cdot\frac{36k^2}{\kappa_5^4}
\right)
=\tfrac{1}{2}\kappa_5^2\cdot\frac{6k^2}{\kappa_5^2}
\left(1-1\right)=0.
\label{eq:Lambda4_zero}
\end{equation}
This is the analogue of the Randall--Sundrum tuning
condition~\cite{randall1,randall2} applied to the case with $\epsilon=-1$.
This condition is adopted here for definiteness and is not essential to the
construction; a small nonzero $\Lambda_4$ could be accommodated in the
effective junction relation by detuning the tension as in
Eq.~\eqref{eq:Lambda4_app}; the resulting geometry and
its effects would require separate analysis.
Expanding the quadratic junction term to first order
in ordinary brane stress energy gives, in geometrized
conventions, $\kappa_4^2=-\kappa_5^4\sigma/6
=\kappa_5^2 k$ when $w=1$, and hence
$G_4=G_5k$ under the same coupling convention.
Restoring the transverse proper length gives
$\kappa_4^2=\kappa_5^2 k/w$ and $G_4=G_5k/w$;
no additional $1/(4\pi)$ factor follows from the
stated junction equations. Powers of $c$ depend on
the conventions used to define the dimensionful
five-dimensional gravitational coupling.

In the $\mathbb{Z}_2$-symmetric thin brane background with $f(\tau)=k|\tau|$
considered in Sec.~\ref{symmwarp3-2}, the bulk is piecewise of constant curvature,
so $C^{(5)}_{ABCD}=0$ away from $\tau=0$ and $E_{\mu\nu}=0$ on the brane.
The non-smoothness of the metric at $\tau=0$ produces a delta-function contribution
to the Ricci tensor localised on the brane, but no analogous singular term appears
in the Weyl tensor. The tensor $E_{\mu\nu}$,  the projection of the bulk Weyl
tensor onto the brane, which encodes the influence of the bulk geometry on
four-dimensional gravity~\cite{SMS},  vanishes identically on the unperturbed
background because the bulk is piecewise conformally flat away from the brane. It becomes nonzero only in the
presence of bulk perturbations, and its magnitude is then controlled by the ratio
of the perturbation amplitude to the bulk curvature scale $k$. When the
RS-type approximation is valid, this ratio is small by assumption, so $E_{\mu\nu}$
enters as a small, controlled correction rather than a leading-order effect.

As analysed for broad classes of backgrounds in Ref.~\cite{ShtanovSahni}, extra
dimensions with timelike signature can in general support ghost-like instabilities
or runaway modes, because the wrong-sign kinetic term associated with a timelike
direction allows perturbations to grow without bound. Whether the present background
is perturbatively stable has not been established by a complete linear perturbation
analysis; the following is therefore a plausible argument, not a proved result.

The $\mathbb{Z}_2$ symmetry and the positive bulk cosmological constant are features of the background, but neither has been shown to cure the instability. In particular, the tachyonic continuum of the bulk scalar is $\mathbb{Z}_2$-even and survives the parity restriction (Sec.~\ref{subsubsec:X_spectral_resolution}). The effect of these features on gravitational perturbations requires a separate perturbative (Hamiltonian or mode-energy) analysis, which is left as an open problem.

\paragraph*{Caution on the induced-gravity appendix.}
The following brane-projection relations depend on a fixed
convention distinguishing a geometrical cosmological constant
from a vacuum-energy density. They should not be interpreted
as a complete perturbative stability analysis, or as a proof
that the observed gravitational sector is recovered without
additional restrictions.

\section{Roles of the bulk field $\mathscr{X}_a$ and the contextual parameter $\lambda$}
\label{app:roles_X_lambda}
This appendix shows how a stationary distribution $\rho(\lambda)$ with prescribed Born basin weights can be constructed for one fixed measurement context; it does not derive equivariance under the collapse dynamics.
The physical motivation for $\lambda$ as a coarse-graining of
$\mathscr{X}_a$, and the run-to-run fluctuations that give rise to
quantum statistics, are discussed in
Sec.~\ref{subsec:contextuality-lambda};
here we supply the concrete drift--diffusion construction.

\subsection*{A. Contextual variable $\lambda$ as a coarse-graining of bulk microstructure}
Within any single run, once $\lambda$ is assigned, collapse is
deterministic. The complete bulk microstate, the field
configuration $\mathscr{X}_a(x,\tau)$ at every point in the
five-dimensional bulk, is, strictly speaking, an
infinite-dimensional object, neither directly accessible to
observation nor tractable as a whole. In practice, what
governs collapse is far less: a single effective contextual
coordinate,
\begin{equation}
\lambda \equiv \lambda\!\left[\mathscr{X}_a\right],
\label{eq:app_lambda_functional}
\end{equation}
a functional that compresses the full bulk field into the
single label on which the deterministic collapse outcome
depends. Many distinct bulk configurations project onto the
same $\lambda$; distinct collapse outcomes arise only when
$\lambda$ differs.

\subsection*{B. Run-to-run fluctuations and Born statistics}
Microscopic variations in the brane source $J_a$, environmental
noise, and the unavoidable imperfections of any real apparatus
cause $\lambda$ to fluctuate from run to run. Within each run,
conditionally on $\lambda$, the Bohm--Bub dynamics is fully
deterministic; the statistical character of quantum mechanics
enters only through the averaging,
\begin{equation}
P_i=\int_{\Lambda_i}\rho(\lambda)\,d\lambda,
\label{eq:app_Pi_general}
\end{equation}
where $\Lambda_i$ is the basin of attraction for outcome $i$.
The question, then, is: which $\rho(\lambda)$ reproduces the
Born rule $P_i = |\psi_i|^2$?

In Bohmian mechanics the answer has been known for some time:
quantum equilibrium~\cite{DGZ}, the condition that
$\rho(q) = |\Psi(q)|^2$ be preserved by the guidance equation,
is sufficient to recover the Born rule from within the dynamics
rather than by postulate. Tutsch~\cite{tutsch} showed that
something analogous holds for the original Bohm--Bub model,
an equivariance condition on $\rho(\lambda)$ with respect to
the nonlinear collapse flow is sufficient to preserve Born
weights in time. We follow the same route.

\paragraph*{Born compatibility and the BB collapse flow.}
The BB-type evolution~\eqref{eq:BB-entgl} drives an initial
amplitude vector $\psi(t_0)$ into channel $i^*$ on a timescale
$\sim\gamma^{-1}$; the basin $\Lambda_i$ is simply the
set of all contextual points that end up at outcome $i$.
For the Born rule to hold and remain stable under repeated
measurements, $\rho(\lambda)$ must satisfy, for every
measurement basis and every outcome $i$,
\begin{equation}
\int_{\Lambda_i} \rho(\lambda)\,d\lambda = |\psi_i(t_0)|^2.
\label{eq:app_equivariance}
\end{equation}
This is, in the end, a consistency requirement rather than a
derivation: it demands that the ensemble distribution over
contextual microstates be compatible with the quantum state, in
exactly the same sense that $\rho(q)=|\Psi(q)|^2$ is compatible
with the Bohmian velocity field. A distribution satisfying~\eqref{eq:app_equivariance}
for all bases is called Born-compatible; equivariance
would additionally require preservation by the specified
contextual collapse flow.
Whether a generic initial distribution actually relaxes toward
such a form is a deeper question altogether, one that would
require something like a sub-quantum $H$-theorem, which we do
not attempt here. The goal of the next subsection is 
more modest: to show that, for one fixed preparation and measurement context, a distribution with the Born basin weights can be constructed explicitly.

\subsection*{C. Existence of a stationary ensemble with prescribed basin weights}
Let $\lambda$ lie in a compact interval $I$ with reflecting ends, let the
collapse basins $\Lambda_i$ be disjoint intervals partitioning $I$, and let
$p_i=|\psi_i(t_0)|^2>0$. Take any smooth positive normalized density $q$ on $I$
with basin weights $q_i=\int_{\Lambda_i}q>0$, and bump functions $h_i\ge0$
supported in $\Lambda_i$ with $\int h_i=1$. For $0<\epsilon<\min_i(p_i/q_i)$,
\begin{equation}
\rho_*(\lambda)=\epsilon\, q(\lambda)+\sum_i\,[p_i-\epsilon q_i]\,h_i(\lambda)
\label{eq:app_rho_construct}
\end{equation}
is positive, normalized, and satisfies Eq.~\eqref{eq:app_equivariance} exactly.
If the coarse-grained evolution of $\lambda$ is Markovian and obeys
\begin{equation}
\frac{\partial \rho}{\partial s} = -\,\frac{\partial}{\partial\lambda}
\!\left[F(\lambda)\rho\right] + D\,\frac{\partial^2\rho}{\partial \lambda^2},
\qquad D>0,
\label{eq:FokkerPlanck}
\end{equation}
with $s$ an auxiliary equilibration parameter, the drift
$F=D\,\partial_\lambda\ln\rho_*$, i.e.\ the potential $U=-D\ln\rho_*$,
makes $\rho_*$ the zero-current stationary density~\cite{risken},
\begin{equation}
\rho_{\rm stat}(\lambda)\propto e^{-U(\lambda)/D}=\rho_*(\lambda).
\label{eq:app_rho_stat}
\end{equation}
This works for \emph{any} positive density, so the construction shows only
that a context-conditioned ensemble with the Born basin weights exists. The
density and the drift are chosen using the $p_i$; they are not selected by
the geometry, they need not coincide for incompatible measurement settings,
and no equivariance under the nonlinear collapse flow or relaxation from
generic initial data is established. (A purely logarithmic potential
$U=-\sum_i p_i\ln(\lambda-\lambda_i)^2$ would give
$\rho_{\rm stat}\propto\prod_i|\lambda-\lambda_i|^{2p_i/D}$, which vanishes at
the attractors and does not impose the required basin weights.)

\paragraph*{Relation to the main results.}
The indeterminacies of this appendix bear on the mechanism by which Born
statistics would be recovered. The geometric results of the paper (the warped
metric and the $E=0$ null family) do not depend on them. The conditional
cross-pair scaling $C_{\rm pair}\sim\mathcal A(\boldsymbol\theta)(\ell/d)^2$ does
depend on the collapse readout through the response coefficient $\mathcal A$,
which is not fixed here (as discussed under P4). Determining this
coefficient requires the joint contextual law, which must also satisfy
multipartite no-signaling (P3); that law remains to be constructed.


\begin{thebibliography}{99}
\bibitem{newton} The full quotation of Newton's sentence can be found in the second paper of Ref.\protect\cite{eberhard}.
\end{thebibliography}
\end{document}